%% file: main.tex
\documentclass[10pt,conference]{IEEEtran}
\usepackage{cite}
\usepackage{amsmath,amssymb,amsfonts}
\usepackage{algorithmic}
\usepackage{graphicx}
\usepackage{textcomp}
\usepackage[hyphens]{url}
\usepackage{fancyhdr}
\usepackage{hyperref}
\usepackage{xspace}
\usepackage{float}

\usepackage{comment}
\usepackage{multirow}
\usepackage[table,xcdraw]{xcolor}
\usepackage{colortbl}
\usepackage{soul}
\usepackage{subcaption} 
\usepackage{booktabs}
\usepackage{pifont}
\newcommand{\cmark}{\ding{51}}
\newcommand{\xmark}{\ding{55}}

\newcommand{\acceleratorname}{\textit{RescuCore}\xspace}

\newcommand{\ignore}[1]{}

\newcommand{\labeltext}[3][]{%
    \@bsphack%
    \csname phantomsection\endcsname
    \def\tst{#1}%
    \def\labelmarkup{\emph}
    \def\refmarkup{}%
    \ifx\tst\empty\def\@currentlabel{\refmarkup{#2}}{\label{#3}}%
    \else\def\@currentlabel{\refmarkup{#1}}{\label{#3}}\fi%
    \@esphack%
    \labelmarkup{#2}
}

\title{Integrated GPU Core for Mamba-3}

\author{
  Shiva Raja, William  Desueza, Ajay Joshi     \\
\\
  Department of Electrical and Computer Engineering \\
  Boston University, Boston, MA  \\
  {\{shivar,willjdes,joshi\}@bu.edu}
}

\begin{document}

\maketitle
\pagestyle{plain}

\input{abstract}

\input{introduction}

\input{background}

\input{profiling}

\input{architecture}

\input{evaluation}

\input{related_work}

\input{conclusion}

\bibliographystyle{IEEEtranS}
\bibliography{refs}


\end{document}

%% file: abstract.tex
\begin{abstract}

Mamba-3, the latest Mamba state-space model (SSM), enables efficient long-context sequence modeling, matching Mamba-2's perplexity with half the state size and raising decoding arithmetic intensity to improve GPU utilization.
However, Mamba-3's algorithmic innovations, including multi-input, multi-output (MIMO) computation, exponential-trapezoidal discretization, and rotary positional embeddings (RoPE), introduce complex execution patterns that interleave dense projections, element-wise operations, and sequential state updates.
Despite software-level kernel fusion, these operations remain inefficient on conventional GPUs.
Our characterization of Mamba-3 hybrid models shows that SSM execution remains fragmented across many kernels, incurring substantial kernel launch and data movement overheads.
Furthermore, increased register pressure compared with Mamba-2 reduces warp occupancy and limits memory-latency hiding.

We propose \acceleratorname, a specialized GPU execution unit that accelerates Mamba-3 inference without compromising the execution of the non-SSM layers of hybrid models on existing GPU resources, eliminating the need for a separate dedicated accelerator for Mamba-3.
Integrated alongside CUDA Cores and Tensor Cores within each GPU Streaming Multiprocessor (SM) subpartition, \acceleratorname{} employs a dedicated local register file while leveraging the existing GPU memory hierarchy and warp scheduling infrastructure.
While Tensor Cores handle dense projections, attention, and MLP layers, and CUDA Cores execute remaining scalar and control operations, \acceleratorname handles discretization, normalization, RoPE, MIMO computation, recurrent state updates, and output gating using a deeply pipelined datapath, reducing execution fragmentation and intermediate data movement. 
\acceleratorname reduces intermediate memory traffic, kernel-launch overhead, and dependent-operation latency, improving the execution efficiency of Mamba-3.

We evaluate \acceleratorname on representative Mamba-3 hybrid models using cycle-accurate Accel-Sim simulation, RTL synthesis for area estimation, and synthesis-informed AccelWattch power modeling.
Compared with conventional GPU execution, \acceleratorname achieves \(1.4\times\)--\(9.4\times\) speedup and \(1.4\times\)--\(5.0\times\) reduction in energy consumption during prefill, and \(1.3\times\)--\(2.8\times\) speedup and \(1.2\times\)--\(2.6\times\) reduction in energy consumption during autoregressive decoding.
These results demonstrate the potential of tightly integrated, specialized GPU execution units for accelerating emerging SSM-based hybrid models.
\end{abstract}

%% file: introduction.tex
\section{Introduction}\label{INTRODUCTION}

Recent advances in sequence modeling increasingly adopt structured state-space models (SSMs), including Mamba~\cite{gu2023mamba,dao2024mamba2,lahoti2026mamba3}, for efficient long-context processing.
Hybrid architectures combine SSMs with attention~\cite{Attention}, multilayer perceptrons (MLPs), and mixture of experts (MoE) for language, vision-language, and molecular workloads~\cite{Lenz2025Jamba,Li2025MaTVLM,Hu2025MOLMamba}.
Unlike quadratic-cost attention~\cite{Attention}, SSMs~\cite{SSM1} process sequences with linear complexity and fixed-size recurrent states.
Mamba-3~\cite{lahoti2026mamba3} further improves expressiveness and arithmetic intensity through exponential-trapezoidal discretization, rotary embeddings (RoPE)~\cite{su2024roformer}, and multi-input multi-output (MIMO) computation.
However, these features introduce fine-grained transformations and dependent state updates alongside dense operations, creating a mismatch with conventional GPU execution.
Consequently, algorithmic efficiency does not necessarily translate into GPU performance.

\begin{figure}[t]
    \centering
    \includegraphics[width=0.49\textwidth]{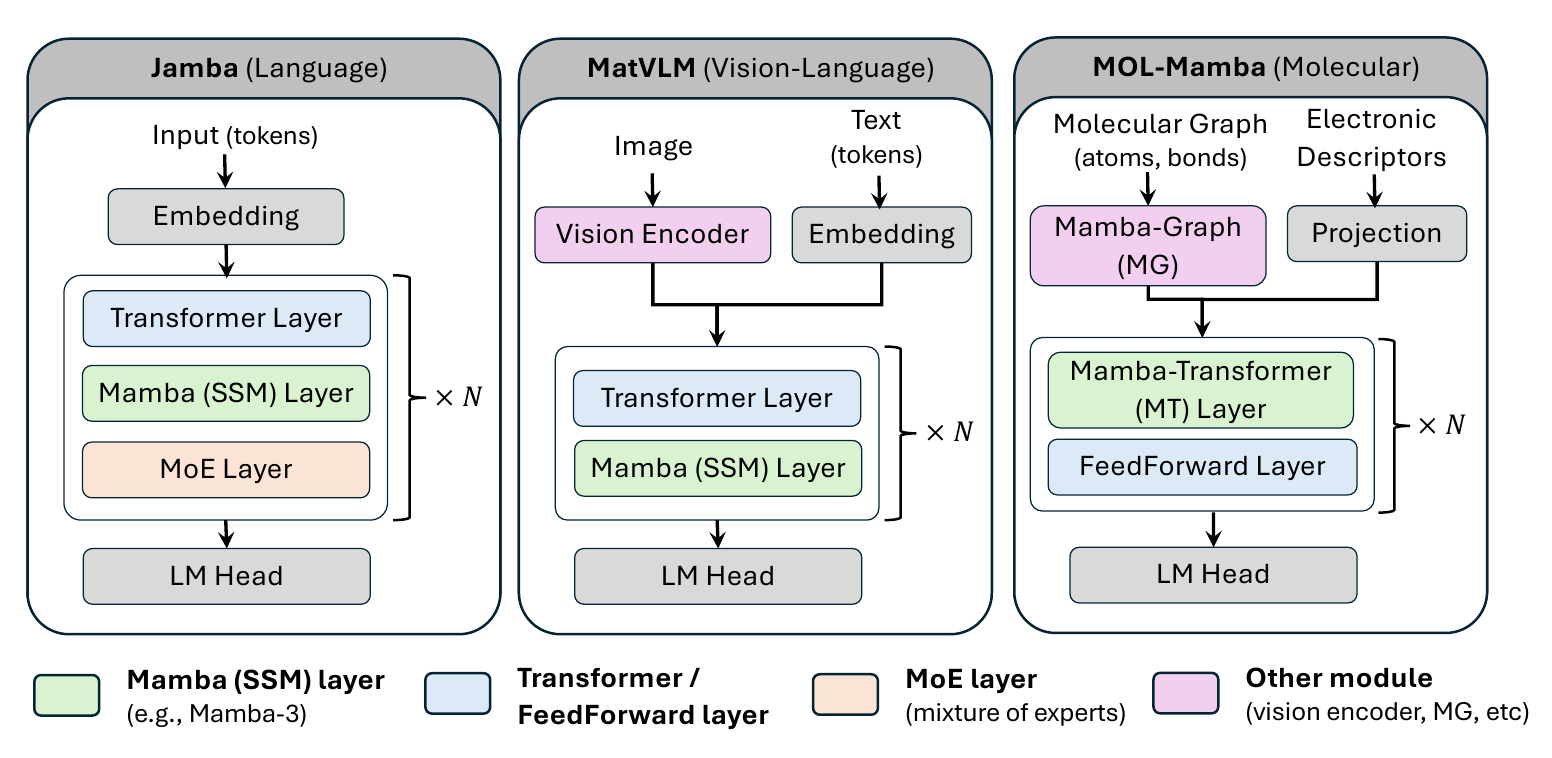}
    \caption{Simplified hybrid models of Jamba~\cite{Lenz2025Jamba}, MaTVLM~\cite{Li2025MaTVLM}, and MOL-Mamba~\cite{Hu2025MOLMamba}, highlighting SSM and non-SSM components.}
    \label{fig:hybrid_model_archs}
\end{figure}

\begin{figure*}[ht]
    \centering
    \includegraphics[width=1\textwidth]{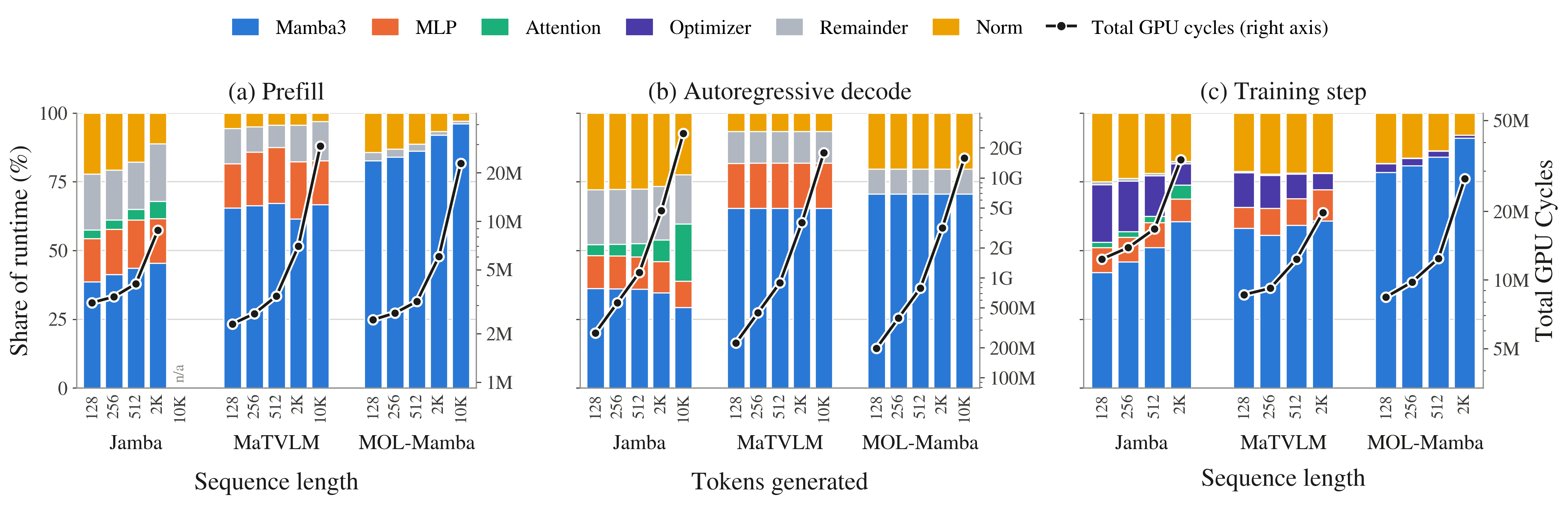}
    \caption{Runtime breakdown of three Mamba-3 hybrid models during (a) prefill, (b) autoregressive decoding, and (c) training across sequence lengths, simulated with Accel-Sim (A100, batch size 1, state size $N=32$, MIMO rank 2).
    Mamba-3 is the largest single contributor to runtime across the evaluated configurations.}
    \label{fig:intro_hyrbid_runtime}
\end{figure*}

Prior work has accelerated SSMs through algorithmic reformulation, software optimization, and specialized hardware.
Mamba introduced hardware-aware selective scans, while Mamba-2 employed structured state-space duality (SSD) to exploit matrix multiplication and Tensor Core parallelism~\cite{gu2023mamba,dao2024mamba2}.
Specialized accelerators such as MARCA, FastMamba, EpochCore, and RAIL improve SSM execution through tailored datapaths and dataflows~\cite{li2024marca,FastMamba,raja2025epochcore,zhu2026rail}.
PIMBA and NOVA explore memory-centric execution~\cite{kim2025pimba,jung2026nova}, whereas HLX introduces pipelined hardware for attention and Mamba-2 SSD in hybrid models~\cite{li2025hlx}.
GPU-integrated specialization has also been explored for other workloads~\cite{kim2025virgo,fhecore}.
These approaches establish the value of specialized execution, but do not directly resolve the combined kernel-fragmentation and dependent-operation bottlenecks of Mamba-3 within conventional GPU Streaming Multiprocessors (SMs).

Despite these advances, efficiently executing Mamba-3 within GPU-resident hybrid models presents two distinct challenges.
First, Mamba-3's discretization, normalization, RoPE, MIMO computation, and recurrent updates execute across multiple GPU kernels, introducing repeated launches, synchronization, and intermediate data movement.
Second, removing kernel boundaries does not eliminate the sequential arithmetic dependencies of recurrent computation.
Software fusion can reduce launch overhead, but long dependency chains, register pressure, and limited independent work still constrain latency hiding.
Moreover, standalone accelerators may require additional coordination with GPU-resident attention and feed-forward layers.
These limitations motivate a GPU-integrated architecture that specializes dependent recurrent execution while preserving conventional GPU resources for dense and non-SSM operations.

To quantify these limitations, we characterize three representative hybrid-model architectures: Jamba~\cite{Lenz2025Jamba}, MaTVLM~\cite{Li2025MaTVLM}, and MOL-Mamba~\cite{Hu2025MOLMamba}.
We replace their original SSM layers with Mamba-3 while preserving the surrounding model architecture, as illustrated in Figure~\ref{fig:hybrid_model_archs}.
Figure~\ref{fig:intro_hyrbid_runtime} shows that Mamba-3 kernels account for $33$--$96\%$ of simulated GPU execution cycles across the evaluated configurations.
Our detailed characterization using Accel-Sim~\cite{AccelSim} reveals that each Mamba-3 layer executes 26 GPU kernels during prefill and 25 during decoding at $B=32$, of which 24 and 22, respectively, are middle kernels targeted by \acceleratorname{}.
During autoregressive decoding, kernel-launch overhead accounts for $42\%$ of simulated execution cycles in one characterized configuration under our launch-latency model.
Importantly, software fusion alone does not eliminate the underlying execution bottlenecks.
On an NVIDIA A100, our hand-fused CUDA implementation reduces batch-1 decoding latency from $49.2\,\mu$s to $14.4\,\mu$s, yet approximately $10\,\mu$s remains in dependent operations.
Increasing register-file or cache capacity provides at most $6\%$ improvement in the evaluated configurations.
Furthermore, Mamba-3 exhibits substantially higher register pressure than Mamba-2, reducing achievable warp occupancy and limiting latency hiding.
These observations indicate that Mamba-3 performance is constrained not simply by insufficient GPU compute or memory capacity, but by fragmented execution and dependent recurrent operations that remain expensive after software optimization.

To address these challenges, we propose \acceleratorname, to the best of our knowledge, the first specialized Mamba-3 execution unit integrated within GPU\footnote{Although evaluated on NVIDIA GPUs, the proposed architecture is applicable to other GPUs.} SM subpartitions to accelerate Mamba-3 inference.
Rather than replacing conventional GPU resources, \acceleratorname complements CUDA Cores and Tensor Cores by specializing the recurrent computation between Mamba-3's input and output projections.
Existing GPU resources continue to execute dense projections, attention, feed-forward layers, and other non-SSM operations.
Within each SM subpartition, \acceleratorname combines exponential-trapezoidal discretization, normalization, RoPE, MIMO computation, recurrent state updates, and output gating into a deeply pipelined datapath.
Dedicated arithmetic stages and direct operand forwarding reduce intermediate data movement and avoid repeatedly transferring dependent values through the GPU memory hierarchy.
The architecture exploits parallelism across independent state channels while respecting dependencies between consecutive updates to the same recurrent state.
By integrating specialized recurrent execution with existing GPU scheduling and memory resources, \acceleratorname avoids the need for a separate accelerator and the associated inter-device data transfers.

Our key contributions are:
\begin{itemize}
    \item We characterize Mamba-3 execution across representative hybrid-model workloads, identifying kernel fragmentation, dependent-operation latency, and register pressure as major GPU performance limitations, including those remaining after software fusion.

    \item We propose \acceleratorname,  to the best of our knowledge, the first GPU-integrated execution architecture that specializes Mamba-3 recurrent computation within SM subpartitions while retaining conventional GPU execution for dense and non-SSM operations.

    \item We design and evaluate a deeply pipelined recurrent datapath with dedicated arithmetic stages, local state reuse, and direct operand forwarding, examining scheduling, numerical precision, performance, energy efficiency, and hardware cost.
\end{itemize}

We evaluate \acceleratorname using cycle-accurate Accel-Sim~\cite{AccelSim} simulation, RTL synthesis, and synthesis-informed AccelWattch~\cite{AccelWattch} power modeling.
Across Mamba-3 hybrid workloads, simulation projects $1.4\times$--$9.4\times$ speedup and $1.4\times$--$5.0\times$ energy reduction during prefill, and $1.3\times$--$2.8\times$ speedup and $1.2\times$--$2.6\times$ energy reduction during decoding.
The selected BF16 configuration ($N=32$, $R=4$) incurs an estimated $5.9\%$ A100 die-area overhead.
These results demonstrate the potential of GPU-integrated recurrent specialization while retaining general-purpose execution flexibility.


%% file: background.tex
\section{Background}
\label{BACKGROUND}

This section introduces the role of structured state-space models (SSMs) in sequence modeling, the Mamba-3 recurrence and its computational stages, the parallelism available during inference, and the numerical representations relevant to recurrent-state computation. 
Unlike Transformer attention, which directly combines information from tokens in the input sequence, SSMs maintain a compact recurrent state that summarizes previously processed information. 
SSMs can be used as standalone sequence-modeling layers or combined with attention and feed-forward layers in hybrid architectures, where they provide efficient sequence processing while attention retains direct access to token representations. 
During inference, prefill processes the input sequence to establish the recurrent state, while autoregressive decoding subsequently generates one token at a time, updating this state as each new token is processed. 
These two phases expose different computational parallelism and execution characteristics, which are central to the acceleration opportunities examined in this work.

\subsection{State-Space Models and Mamba-3 Recurrence}
\label{BG_SSM}

Structured SSMs model dependencies across a sequence by maintaining a compact hidden state that evolves as new inputs are processed. 
Rather than explicitly revisiting all preceding tokens, the model updates this state to retain information relevant to subsequent computation. 
The state evolution and output generation are described by the following continuous-time linear system:
\begin{equation}
\begin{aligned}
    \dot{\mathbf{h}}(t)
        &= \mathbf{A}(t)\mathbf{h}(t)
        + \mathbf{B}(t)x(t), \\
    y(t)
        &= \mathbf{C}(t)^{\top}\mathbf{h}(t),
\end{aligned}
\label{eq:ssm_continuous}
\end{equation}
where $\mathbf{h}(t)\in\mathbb{R}^{N}$ is the recurrent state, $x(t),y(t)\in\mathbb{R}$ are the input and output, $\mathbf{A}(t)\in\mathbb{R}^{N\times N}$ is the state-transition matrix, and $\mathbf{B}(t),\mathbf{C}(t)\in\mathbb{R}^{N}$ are input and output projections. Structured SSMs constrain $\mathbf{A}(t)$ to enable efficient recurrent computation~\cite{SSM2,gu2023mamba}.

Mamba~\cite{gu2023mamba} and Mamba-2~\cite{dao2024mamba2}, two structured SSM models, employ exponential-Euler discretization. 
While Mamba uses a real diagonal state-transition matrix, Mamba-2 restricts the transition within each SSM head to a scalar multiple of the identity:
\begin{equation}
    \mathbf{A}_{t}=a_t\mathbf{I}_{N},
    \qquad a_t<0.
\label{eq:mamba2_A}
\end{equation}
The resulting discrete recurrence is
\begin{equation}
\begin{aligned}
    \mathbf{h}_{t}
        &= \alpha_t\mathbf{h}_{t-1}
        + \gamma_t\mathbf{B}_{t}x_t, \\
    y_t
        &= \mathbf{C}_{t}^{\top}\mathbf{h}_{t}
        + D x_t,
\end{aligned}
\label{eq:mamba12_recurrence}
\end{equation}
where
\begin{equation}
    \alpha_t=e^{\Delta_t a_t}\in(0,1),
    \qquad
    \gamma_t=\Delta_t,
    \qquad
    \Delta_t>0.
\label{eq:mamba12_coeff}
\end{equation}
Here, $D$ is the learned skip coefficient, and $\Delta_t$ is the input-dependent discretization step. 
Mamba-3~\cite{lahoti2026mamba3} extends this formulation through data-dependent state decay, exponential-trapezoidal discretization, complex-valued state dynamics implemented using rotary transformations, and optional multi-input multi-output (MIMO) recurrent updates.

Conceptually, the recurrence combines information retained from previous tokens with a contribution from the current input. 
The state-transition coefficient controls how much previous information persists, while the input-dependent terms determine how new information enters the state. 
The resulting output therefore reflects both the current token and the sequence history encoded in the recurrent state.

\subsubsection{Data-Dependent State Decay}
Mamba-3 controls the retention of information in its recurrent state through the decay coefficient
\[
\alpha_t = e^{\Delta_t a_t},
\]
where \(\Delta_t\) is the input-dependent discretization step and \(a_t < 0\) is the state-transition parameter. 
The resulting coefficient satisfies \(0 < \alpha_t < 1\), determining how strongly the previous state contributes to the current update. 
A value of \(\alpha_t\) near one preserves more of the previous state, whereas a smaller value causes previously accumulated information to decay more rapidly. 

Conceptually, this mechanism allows the recurrent state to retain or forget information at different rates as the input sequence evolves.

\subsubsection{Exponential-Trapezoidal Discretization}
Unlike exponential-Euler discretization, which incorporates only the current input, Mamba-3's exponential-trapezoidal recurrence incorporates both current and previous inputs:
\begin{equation}
\begin{aligned}
    \mathbf{h}_{t}
    ={}& \alpha_t\mathbf{h}_{t-1}
       + \beta_t\mathbf{B}_{t-1}x_{t-1}
       + \gamma_t\mathbf{B}_{t}x_t,
\end{aligned}
\label{eq:mamba3_trap}
\end{equation}
with coefficients
\begin{equation}
\begin{aligned}
    \alpha_t &= e^{\Delta_t a_t}, \\
    \beta_t  &= (1-\lambda_t)\Delta_t\alpha_t, \\
    \gamma_t &= \lambda_t\Delta_t,
    \qquad \lambda_t\in[0,1].
\end{aligned}
\label{eq:mamba3_coeff}
\end{equation}
Here, $\lambda_t$ controls the relative contributions of the previous and current input terms. 
Setting $\lambda_t=\tfrac{1}{2}$ yields trapezoidal weighting, whereas $\lambda_t=1$ recovers exponential-Euler discretization~\cite{lahoti2026mamba3}. 
Computing these coefficients requires data-dependent exponential and arithmetic operations, while the recurrence additionally depends on the previous timestep's projected input.

By incorporating both the current and previous input contributions, exponential-trapezoidal discretization provides a more expressive approximation of how the recurrent state evolves between timesteps. 
The resulting coefficients determine how much prior state is retained and how information from adjacent tokens contributes to the updated state.

\begin{figure*}[h]
    \centering
    \includegraphics[width=0.95\textwidth]{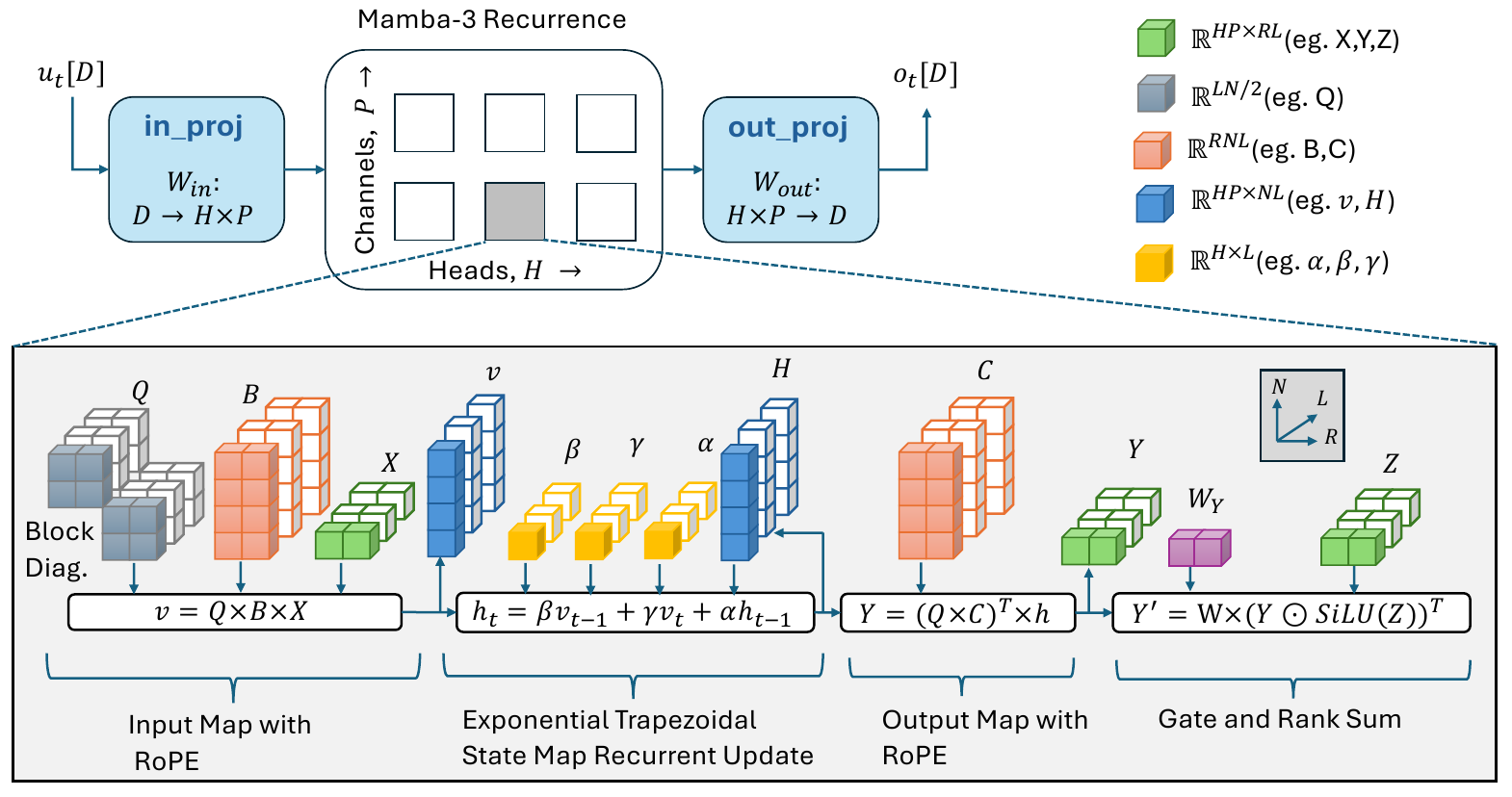}
    \caption{
    Mamba-3 layer organization and recurrent dataflow. The input projection (\texttt{in\_proj}) maps each \(D\)-dimensional token \(u_t\) into \(H\) heads with \(P\) channels per head. Within each head, channels share the state projections \(B,C\) and recurrence coefficients \(\alpha,\beta,\gamma\), but maintain independent \(N\)-dimensional states \(h_t\). From left to right, the recurrence applies RoPE (\(Q\)) to map input features \(X\) into \(v\), updates \(h_t=\alpha h_{t-1}+\beta v_{t-1}+\gamma v_t\), reads out rank-\(R\) features \(Y\) using the rotated projection \(QC\), and applies gating (\(Z\)) and rank reduction to produce channel outputs. The output projection (\texttt{out\_proj}) combines the \(HP\) channel outputs into a \(D\)-dimensional token \(o_t\). Heads and channels are independent, while state updates are sequential across timesteps.}
    \label{fig:mamba3_full_dataflow}
\end{figure*}

\subsubsection{Complex-Valued State Dynamics Using RoPE}

Mamba-3 introduces complex-valued state transitions that combine real-valued decay with rotational dynamics. 
Rather than explicitly performing complex arithmetic, the implementation represents the state using real-valued tensors and applies two-dimensional rotations to pairs of state dimensions~\cite{lahoti2026mamba3}. 
Each rotation is defined as
\begin{equation}
    \mathbf{R}(\phi)=
    \begin{bmatrix}
        \cos\phi & -\sin\phi \\
        \sin\phi &  \cos\phi
    \end{bmatrix},
\label{eq:mamba3_rotation_operator}
\end{equation}
where $\phi$ is a data-dependent rotation angle.

The cumulative rotation (\(\mathbf{Q}_t\)) is applied to the input and output state projections as follows:
\begin{equation}
\begin{aligned}
    \mathbf{Q}_t
        &:= \prod_{i=0}^{t}\mathbf{R}_{i}^{\top}, \\
    \widetilde{\mathbf{B}}_t
        &:= \mathbf{Q}_t\mathbf{B}_t, \\
    \widetilde{\mathbf{C}}_t
        &:= \mathbf{Q}_t\mathbf{C}_t.
\end{aligned}
\label{eq:mamba3_rope}
\end{equation}
The additive composition of two-dimensional rotations allows the product of rotation matrices to be expressed as a single rotation:
\begin{equation}
    \prod_{i=0}^{t}\mathbf{R}^{\top}(\phi_i)
        =\mathbf{R}^{\top}\left(\sum_{i=0}^{t}\phi_i\right).
\label{eq:mamba3_cumm_prod}
\end{equation}
Consequently, the cumulative transformation requires only angle accumulation rather than repeated matrix multiplication. 
Applying the resulting rotation to a pair of adjacent even- and odd-indexed elements yields
\begin{equation}
    \begin{bmatrix}
        y_e \\ y_o
    \end{bmatrix}
    =
    \begin{bmatrix}
        \cos\phi & \sin\phi \\
        -\sin\phi & \cos\phi
    \end{bmatrix}
    \begin{bmatrix}
        u_e \\ u_o
    \end{bmatrix}.
\label{eq:mamba3_rotation}
\end{equation}
Thus, RoPE introduces angle accumulation, trigonometric evaluation, and pairwise arithmetic before the recurrent state update.

While state decay controls how strongly previous information is retained, rotary transformations allow the state dynamics to represent phase changes across timesteps. 
Applying these rotations to the input and output state projections introduces position-dependent interactions into the recurrent computation without explicitly storing complex-valued states.

\subsubsection{MIMO State Update}

Mamba-3 generalizes the single-input single-output (SISO) recurrence to rank-$R$ MIMO updates, increasing computation per recurrent-state access~\cite{lahoti2026mamba3}. 
For one head and channel, let $\mathbf{h}_t\in\mathbb{R}^{N}$ denote the recurrent state, $\mathbf{X}_t\in\mathbb{R}^{R}$ the input features, and $\widetilde{\mathbf{B}}_t,\widetilde{\mathbf{C}}_t\in\mathbb{R}^{N\times R}$ the RoPE-transformed projections. 
The recurrence and state readout are
\begin{equation}
\begin{aligned}
    \mathbf{h}_t
      ={}& \alpha_t\mathbf{h}_{t-1}
       + \beta_t\widetilde{\mathbf{B}}_{t-1}\mathbf{X}_{t-1}
       + \gamma_t\widetilde{\mathbf{B}}_t\mathbf{X}_t, \\
    \mathbf{Y}_t
      ={}& \widetilde{\mathbf{C}}_t^{\top}\mathbf{h}_t,
\end{aligned}
\label{eq:mamba3_mimo}
\end{equation}
where $\mathbf{Y}_t\in\mathbb{R}^{R}$ is the rank-$R$ state readout. 
The MIMO formulation increases arithmetic intensity by performing additional input-state and state-output computation without proportionally increasing the recurrent-state size. 
However, the update at timestep $t$ still depends on the state produced at timestep $t-1$.

Conceptually, MIMO allows each recurrent state to incorporate multiple input features and produce multiple output features, increasing the information processed through a shared state representation. 
This increases the computation performed per state access without proportionally increasing state storage, improving the arithmetic intensity of the recurrence.

\subsubsection{Output Gating}

Following the MIMO state readout and skip connection, Mamba-3 applies element-wise gating and reduces the rank-$R$ features to the channel output. Given gating features $\mathbf{Z}_t\in\mathbb{R}^{R}$ and learned output weights $\mathbf{W}_o\in\mathbb{R}^{R}$, the gated output is
\begin{equation}
\begin{aligned}
    y_t'
    &= \sum_{r=1}^{R}
       W_{o,r}Y_{t,r}\operatorname{SiLU}(Z_{t,r}), \\
    \operatorname{SiLU}(z)
    &= z\sigma(z)=\frac{z}{1+e^{-z}}.
\end{aligned}
\label{eq:output_gating}
\end{equation}

The gating operation selectively weights the features read from the recurrent state, allowing the model to control which information contributes to the channel output. 
The rank-reduced result is then passed through the block's output projection to produce the token representation consumed by subsequent model layers. 
Together, discretization, rotary transformations, MIMO updates, exponential trapezoidal recurrence, and gating determine how sequence information is accumulated, transformed, and exposed to the rest of the model.

\subsection{Mamba-3 Dimensions and Execution Parallelism}

As illustrated in Fig.~\ref{fig:mamba3_full_dataflow}, a Mamba-3 layer first applies an input projection (\texttt{in\_proj}) to transform each \(D\)-dimensional token embedding into features organized across \(H\) heads, each containing \(P\) recurrent channels. 
Each channel maintains an independent \(N\)-dimensional recurrent state, while channels within the same head share the state projections and recurrence coefficients. 
This organization exposes parallelism across heads and channels, although successive updates to each channel's state remain dependent across timesteps.
Thus, the recurrent representation contains $HP$ channels, which need not equal the model embedding dimension $D$.

Within each head, the $P$ channels share head-level discretization coefficients and state projections, while maintaining independent recurrent states. 
Each channel $(h,p)$ maintains an $N$-dimensional state $\mathbf{h}_{h,p,t}\in\mathbb{R}^{N}$, giving $HPN$ recurrent state elements across the layer. 
The MIMO rank $R$ determines the number of input contributions and output features associated with each channel. 
Per-channel MIMO weights map the projected input into $R$ features, which interact with the shared head-level state projections to update the channel's recurrent state.

The recurrent computation follows the dataflow in Eqs.~\ref{eq:mamba3_coeff}--\ref{eq:mamba3_mimo}. 
Input-dependent coefficients and normalized projections are generated from the input projection outputs, followed by the rotary transformations and MIMO input mapping. 
Each channel then combines the decayed previous state with the current and previous input contributions. 
The updated state is read out through the output projection vectors, followed by gating and reduction across the MIMO rank. 
The resulting $HP$ channel outputs are combined by the learned dense output projection (\texttt{out\_proj}) to recover a $D$-dimensional output token.

This organization exposes parallelism across the \(H\) heads and \(P\) channels, since each channel maintains an independent recurrent state. 
However, successive updates to the same state remain dependent across timesteps. 
During \textit{prefill}, the model processes the input sequence and computes the recurrent states needed for subsequent generation, exploiting parallelism across heads and channels as well as parallel scan algorithms along the sequence dimension. 
During autoregressive decoding, each newly generated token is processed using the states retained from preceding tokens, and the updated states are carried forward to the next decoding step. 
Thus, \textit{prefill} offers additional sequence-level parallelism, whereas \textit{decoding} primarily exploits parallelism across independent heads and channels while preserving sequential state dependencies.

\subsection{GPU Execution of Hybrid Transformer-SSM Models}
\label{BG_GPU}

Hybrid Transformer–SSM models combine attention and feed-forward layers with Mamba-3 layers, introducing distinct computational characteristics within the same inference workload. 
Attention and feed-forward computations primarily involve dense matrix operations that map well to existing GPU execution resources. 
Mamba-3 layers also perform dense input and output projections, but their intermediate computations consist of discretization, normalization, rotary transformations, MIMO processing, recurrent state updates, and output gating.

Unlike dense projections, these intermediate operations involve fine-grained arithmetic and dependencies between successive computational stages. 
Conventional GPU implementations distribute them across multiple kernels, potentially requiring intermediate results to be transferred through the GPU memory hierarchy. 
Although kernel fusion can reduce some of this overhead, dependencies within the recurrent computation remain.

Consequently, the execution efficiency of a hybrid model depends not only on accelerating its dense operations but also on efficiently processing its Mamba-3 layers. 
Section~\ref{sec:mamba3_bottlenecks} characterizes these execution bottlenecks and motivates the GPU-integrated specialization proposed in this work.

\subsection{Numerical Representation for Recurrent State}
\label{BG_POSIT}

Numerical precision affects both the accuracy and hardware efficiency of hybrid Transformer–SSM models. 
These models combine attention and feed-forward computations with Mamba-3 layers, whose operations include dense projections and recurrent processing. 
Within Mamba-3, numerical errors are particularly important for the recurrent state because it is repeatedly updated and reused across timesteps, allowing rounding errors to propagate through subsequent computations.

In this work, we investigate numerical representations for all operations implemented in \acceleratorname, including discretization, normalization, rotary transformations, MIMO processing, recurrent state updates, and output gating. 
Dense input and output projections, as well as attention and feed-forward computations, remain outside the scope of the proposed hardware. 
However, our numerical accuracy evaluation applies each candidate precision uniformly across the entire hybrid model, including these surrounding operations, to assess its impact on end-to-end model accuracy. 
The hardware evaluation separately examines the implementation cost of each precision within \acceleratorname.

To examine the trade-off between numerical accuracy and hardware efficiency, we consider five number formats: FP32, TF32, BF16, Posit16, and INT8, spanning conventional floating-point, posit, and integer representations.
BF16 retains FP32's eight-bit exponent but uses a seven-bit fraction, preserving a wide dynamic range with reduced significand precision~\cite{kalamkar2019bfloat16}. 
FP16 provides an alternative 16-bit floating-point representation with a different range-precision tradeoff.

Posit is another numerical representation that uses variable-length regime and fraction fields~\cite{Gustafson2017Posit,Lu2020PositDNN,Mallasen2025PositSurvey}. 
A Posit$(16,e_s)$ value comprises a sign bit, a variable-length regime, up to $e_s$ exponent bits, and a fraction occupying the remaining bits. For a finite, nonzero value,
\begin{equation}
\begin{aligned}
    x &= (-1)^s\mathrm{useed}^{\,k}2^e(1+f), \\
    \mathrm{useed} &= 2^{\,2^{e_s}},
\end{aligned}
\label{eq:posit}
\end{equation}
where $s$, $k$, $e$, and $f$ denote the sign, regime value, exponent, and fractional component, respectively. 
We evaluate the accuracy implications of these representations for Mamba-3 in Section~III, while considering their hardware costs in the proposed architecture.

Unlike intermediate values that are used only within a single operation, recurrent states are repeatedly updated and reused across tokens, making their numerical representation important for preserving information over long sequences. 
Consequently, selecting a compact arithmetic format requires balancing hardware efficiency against the accumulation of numerical error, motivating the accuracy and hardware-cost evaluation presented in the following sections.

%% file: profiling.tex
\section{Mamba-3 Bottleneck Analysis}\label{sec:profiling}

To motivate the integration of \acceleratorname into the GPU, we characterize Mamba-3~\cite{lahoti2026mamba3} execution across representative hybrid models during prefill and autoregressive decoding. 
Using PyTorch Profiler~\cite{pytorchProfiler}, Nsight Systems~\cite{NVIDIA_NsightSystems}, Nsight Compute~\cite{NVIDIA_NsightCompute}, and Accel-Sim~\cite{AccelSim}, we investigate three questions: (1) What limits prefill and decoding? (2) How much performance can software kernel fusion recover? and (3) Can conventional GPU architectural enhancements eliminate the remaining inefficiencies?
We also evaluate the accuracy of Mamba-3-based hybrid models and reduced-precision arithmetic to identify opportunities for efficient specialized execution within existing GPUs.



\subsection{Hybrid Workloads and Mamba-3 Accuracy}
\label{HybridModels}

To evaluate Mamba-3 acceleration in realistic inference workloads, we consider hybrid models that originally incorporate different variants of Mamba-based state-space models (SSMs) alongside computational components such as Transformer attention, mixture-of-experts (MoE), and graph processing. 
These workloads exhibit different patterns of Mamba integration, allowing us to assess the applicability of GPU-integrated specialization beyond standalone Mamba models. 
We study three representative models: Jamba~\cite{Lenz2025Jamba}, MaTVLM~\cite{Li2025MaTVLM}, and MOL-Mamba~\cite{Hu2025MOLMamba}. 
Simplified architectures of these hybrid models are shown in Figure~\ref{fig:hybrid_model_archs}.
In each model, we replace the original Mamba or Mamba-2 components with Mamba-3 while preserving the surrounding architecture. 
This allows us to evaluate Mamba-3 acceleration across diverse hybrid workloads while isolating the effects of the SSM replacement from changes to other model components.
\acceleratorname is specifically designed for Mamba-3 operations; although earlier Mamba variants share some computational primitives, supporting their original recurrence formulations would require additional compatibility considerations and is beyond the scope of this work.

\begin{enumerate}
\item \textit{Jamba:} Jamba is a hybrid language model that interleaves Transformer attention and Mamba layers at a 1:7 ratio, with MoE modules replacing selected MLPs. 
This design combines periodic global attention with efficient recurrent sequence processing. 
We substitute Mamba-3 for the original Mamba layers while retaining the attention, MoE, and normalization components. 
Mamba-3 accounts for approximately \(45\%\) of the total GPU execution cycles during inference, compared with approximately \(50\%\) for the original Mamba-1 layers.
\item \textit{MaTVLM:} MaTVLM is a hybrid vision-language model that replaces selected Transformer self-attention operations with Mamba-2 while preserving the vision encoder, multimodal connector, and remaining decoder components. 
We replace these Mamba-2 operations with Mamba-3 without modifying the surrounding Transformer structure. 
Mamba-3 accounts for approximately \(62\%\) of the total GPU execution cycles during inference, compared with approximately \(53\%\) for the original Mamba-2 layers.
\item \textit{MOL-Mamba:} MOL-Mamba combines graph neural networks, Mamba-based sequence modeling, and Transformer processing for molecular representation learning. 
Its Mamba-Graph module captures long-range dependencies among graph-derived atom representations using graph-conditioned discretization, while a Mamba-Transformer fuser integrates structural and electronic features. 
We replace the Mamba operations with Mamba-3 while retaining graph conditioning and the surrounding GNN and Transformer components. 
Mamba-3 accounts for approximately \(92\%\) of the total GPU execution cycles during inference, compared with approximately \(80\%\) for the original Mamba-1 layers.
\end{enumerate}

\begin{figure}[t]
    \centering
    \includegraphics[width=0.48\textwidth]{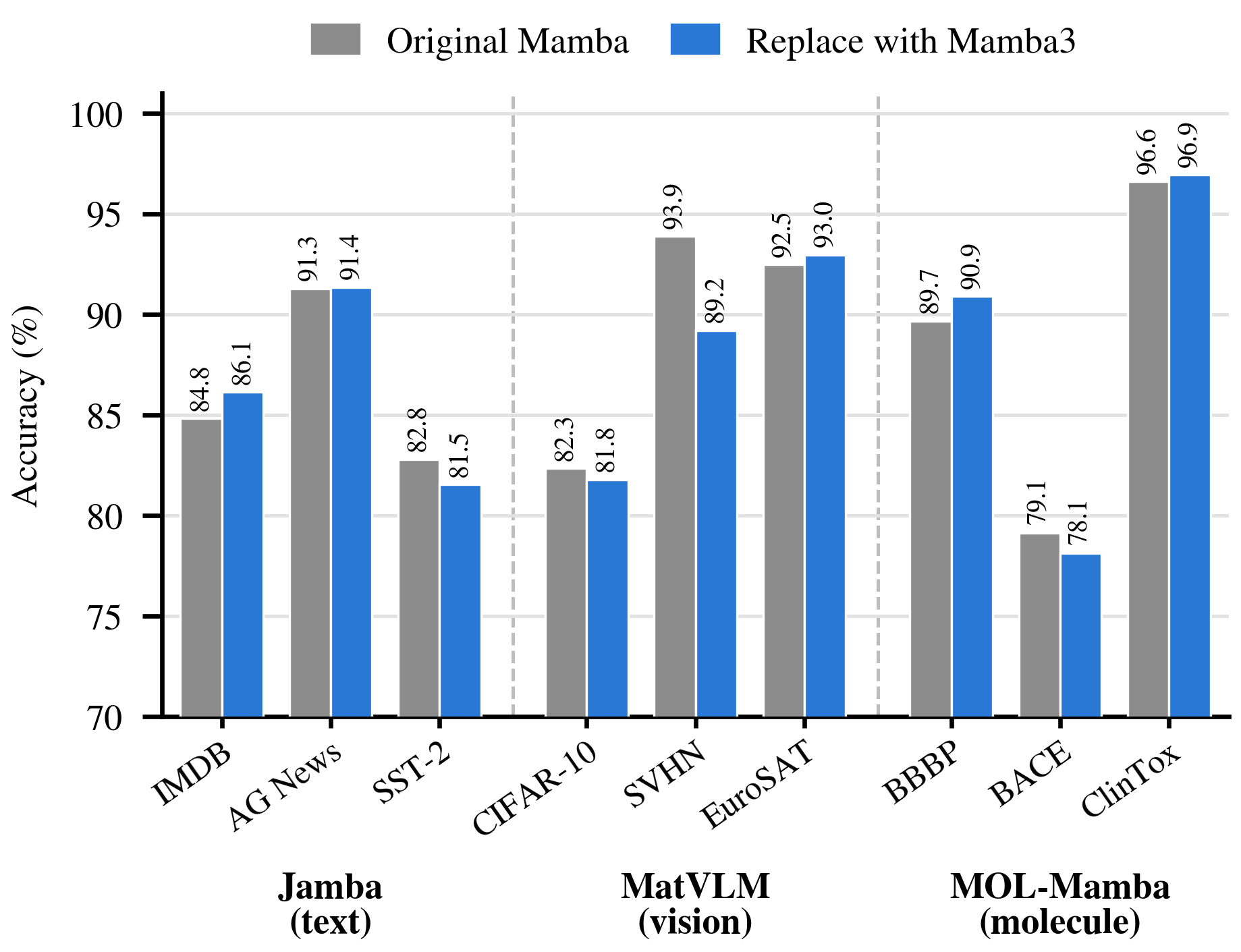}
    \caption{Classification accuracy of three hybrid models with the original Mamba mixer and with the mixer replaced by Mamba-3. Each pair uses the same architecture and training budget.}
    \label{fig:class_accuracy_v1_vs_v3_all_models}
\end{figure}



To study the impact of Mamba-3 execution in representative hybrid architectures, we replace the original Mamba or Mamba-2 layers with Mamba-3 while preserving the surrounding model components. 
Although the original configurations exhibit slightly better execution performance, our objective is not to establish Mamba-3 as a superior replacement, but to evaluate the acceleration opportunities for Mamba-3 within established hybrid workloads. 
Since this substitution changes the recurrent computation, we also assess its effect on model-level accuracy. 
Fig.~\ref{fig:class_accuracy_v1_vs_v3_all_models} compares the original and Mamba-3-based configurations across nine benchmarks spanning language, vision, and molecular workloads. 
Accuracy improves on five benchmarks and decreases on four, with an average change of \(-0.46\) percentage points and an average absolute change of \(1.21\) percentage points. 
The largest reduction occurs on SVHN~\cite{netzer2011svhn} at \(4.7\) percentage points, while the remaining changes are within \(1.3\) percentage points. 
These results indicate that the Mamba-3-based configurations retain broadly comparable accuracy on most evaluated benchmarks, supporting their use as representative workloads for evaluating \acceleratorname, while highlighting that the substitution may not be suitable for all accuracy-sensitive applications.

Across these representative hybrid workloads, Mamba-3 layers account for approximately \(33\%\) to \(96\%\) of total GPU execution cycles, despite being integrated with other computational components. 
This substantial contribution motivates accelerating Mamba-3 through specialized execution units integrated into the GPU, while preserving existing GPU resources for attention, MoE, graph processing, and other non-SSM operations.

\begin{figure}[t]
    \centering
    \includegraphics[width=0.5\linewidth]{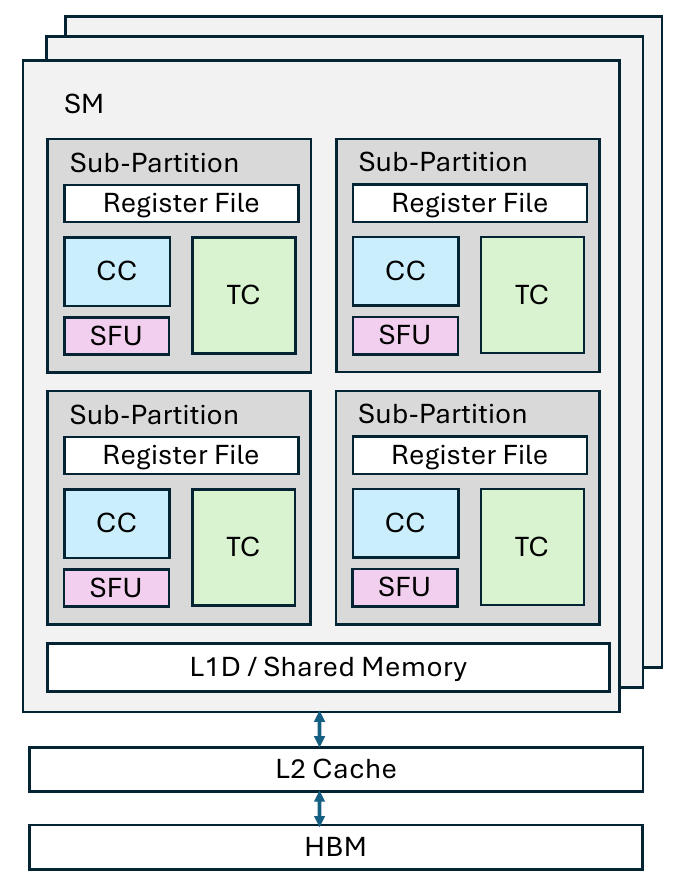}
    \caption{NVIDIA A100 SM with CUDA cores (CC), special function units (SFU), Tensor Cores (TC) and memory interfaces}
    \label{fig:SM_Cores}
\end{figure}

\begin{figure}[t]
    \centering
    \includegraphics[width=\linewidth]{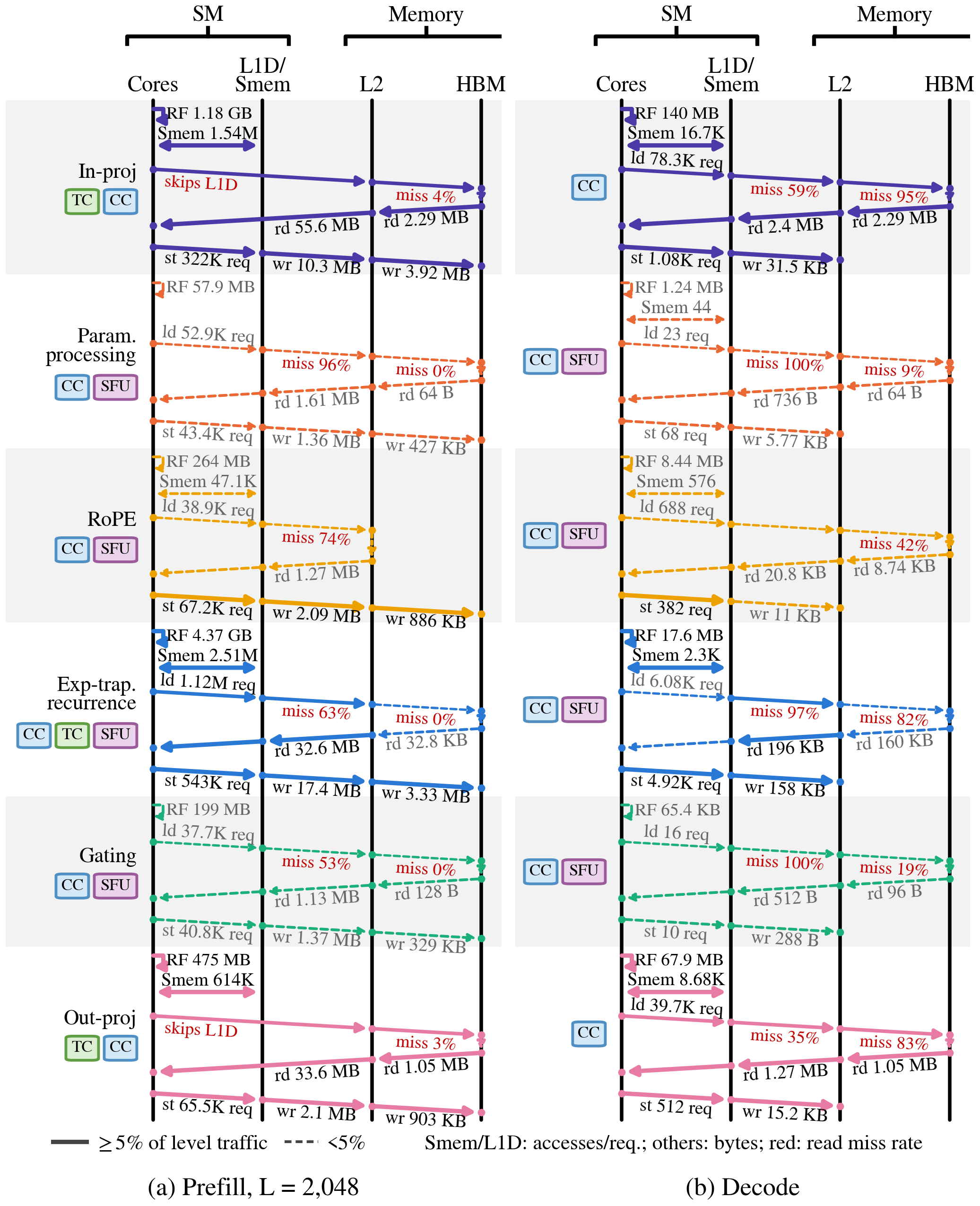}
    \caption{Memory handshakes of the Mamba-3 kernel groups on NVIDIA A100 GPU cores (corresponding to Figure~\ref{fig:SM_Cores}). (a) Prefill ($L=2,048$) and (b) Decode. Each group shows register-file (RF) and shared-memory (Smem) traffic, then loads (bytes read, rd) and stores (bytes written, wr) through L1D, L2 and HBM; red text gives read miss rates, and solid arrows carry at least 5\% of a level's traffic.}
    \label{fig:mamba3_sm_memory_1col}
\end{figure}

\subsection{Execution Bottlenecks in Prefill and Decoding}
\label{sec:mamba3_bottlenecks}

We characterize the execution of Mamba-3 on NVIDIA GPUs to identify bottlenecks in prefill and autoregressive decoding. 
As described in Section~\ref{BACKGROUND}, Mamba-3 extends the recurrent computation with discretization, rotary positional embeddings (RoPE), and rank-$R$ exponential trapezoidal recurrent state updates, increasing the computation performed per token.
On conventional GPUs, these operations are distributed across multiple kernels and exhibit dependencies between successive computational stages, introducing execution overheads that we characterize in this section.

\subsubsection{Kernel Execution and Bottleneck Analysis}


The input projection (\textit{In-Proj}) first generates features for the recurrent computation. The \textit{Parameter Processing} block performs normalization and computes the discretization coefficients $\alpha_t$, $\beta_t$, and $\gamma_t$ in Equation~\ref{eq:mamba3_coeff}, using data-dependent exponential and arithmetic operations. 
Subsequently, the \textit{RoPE} block applies rotary transformations to the input and output state projections, as described in Equations~\ref{eq:mamba3_cumm_prod} and~\ref{eq:mamba3_rotation}.
Next, the \textit{Exp-Trap. Recurrence} block performs the exponential-trapezoidal recurrent state update and readout in Equation~\ref{eq:mamba3_mimo}, followed by output gating and rank reduction in Equation~\ref{eq:output_gating}. 
Finally, the output projection (\textit{Out-Proj}) produces the representation passed to the next layer. 

On conventional GPUs, these operations execute as separate kernels, requiring intermediate results to be transferred through the GPU memory hierarchy between stages. 
Figure~\ref{fig:mamba3_sm_memory_1col} illustrates the associated memory accesses for the four operation groups measured on NVIDIA A100 GPU, shown in Figure~\ref{fig:SM_Cores}. 
Each group stages operands through registers and shared memory, with global memory loads frequently missing in L1D and being served by L2 or, when necessary, HBM. 
Kernel outputs generate additional writes through the memory hierarchy. 
For a Mamba-3 layer with $L=2,048$, these groups collectively generate 1.9 million L1D requests, 2.6 million shared-memory accesses, 58.8 MB of L2 traffic, and 5.0 MB of HBM traffic, accounting for approximately 37\% and 38\% of the layer's total L2 and HBM traffic, respectively. 
During prefill, L1D read miss rates range from 53\% to 96\% across the four groups, with 4.97 MB of HBM writes, including 3.33 MB from recurrent computation. 
During decoding, all four groups generate memory requests that reach HBM. 
These measurements highlight the memory traffic associated with executing dependent Mamba-3 operations across separate GPU kernels.



\begin{table}[!t]
\centering
\caption{Mamba-3 kernels per layer and number of GPU cycles on A100 NVIDIA GPU ($L$=$2K$,$N$=$32$,$R$=$2$,$B$=$32$), grouped by function.}
\label{tab:mamba3-kernel-groups}
\begin{tabular}{lrrrr}
\hline
\rowcolor[HTML]{EFEFEF} 
\textbf{} & \multicolumn{2}{c}{\cellcolor[HTML]{EFEFEF}\textbf{Prefill}} & \multicolumn{2}{c}{\cellcolor[HTML]{EFEFEF}\textbf{Decode}} \\ \hline
\rowcolor[HTML]{EFEFEF} 
\textbf{Group} & \multicolumn{1}{r}{\cellcolor[HTML]{EFEFEF}\textbf{Kernels}} & \multicolumn{1}{r}{\cellcolor[HTML]{EFEFEF}\textbf{Cycles}} & \multicolumn{1}{r}{\cellcolor[HTML]{EFEFEF}\textbf{Kernels}} & \multicolumn{1}{r}{\cellcolor[HTML]{EFEFEF}\textbf{Cycles}} \\ \hline
In projections & 1 & 27.0\% & 1 & 12.7\% \\
Discretization & 11 & 1.8\% & 11 & 31.3\% \\
Normalization & 4 & 2.7\% & 4 & 12.8\% \\
RoPE & 2 & 4.0\% & 3 & 12.0\% \\
Exp-Trap. recurrence & 7 & 50.1\% & 4 & 20.7\% \\
Out projections & 1 & 14.4\% & 2 & 10.6\% \\
\hline
Total & 26 & 100\% & 25 & 100\% \\
\hline
\end{tabular}
\end{table}

Table~\ref{tab:mamba3-kernel-groups} shows that at $B=32$, one Mamba-3 layer executes 26 kernels during prefill and 25 during decoding.
Excluding the dense input and output projections, 24 prefill kernels and 22 decoding kernels perform discretization, normalization, RoPE, and exponential-trapezoidal recurrence, exchanging intermediate values through the GPU memory hierarchy.
These middle kernels are the target of \acceleratorname{}, which consolidates their operations into a single specialized execution pipeline.
Their individual execution times differ substantially between the two inference phases.
During prefill, recurrent computation is dominant and accounts for $50.1\%$ of layer cycles, whereas during decoding parameter processing such as discretization and normalization is dominant and accounts for $44.1\%$ of the time.
The separate kernels exchange intermediate values, such as discretization coefficients and transformed state projections, through the GPU memory hierarchy rather than retaining all values within a single kernel. 
Figure~\ref{fig:mamba3_sm_memory_1col} quantifies the associated L2 traffic: during prefill, recurrent kernels generate 32.6~MB of reads and 17.4~MB of writes, while projection kernels generate 85.8~MB of reads and 12.4~MB of writes. 
These results reveal a shift from recurrence-dominated prefill to parameter-processing-dominated decoding, alongside substantial L2 traffic and a large number of parameter-processing kernel launches.
These measurements characterize the A100 GPU; the relative execution costs may differ across GPU architectures because of variations in compute throughput, memory hierarchy, and kernel execution overhead.

\begin{figure*}[t!]
    \centering
    \includegraphics[width=0.95\textwidth]{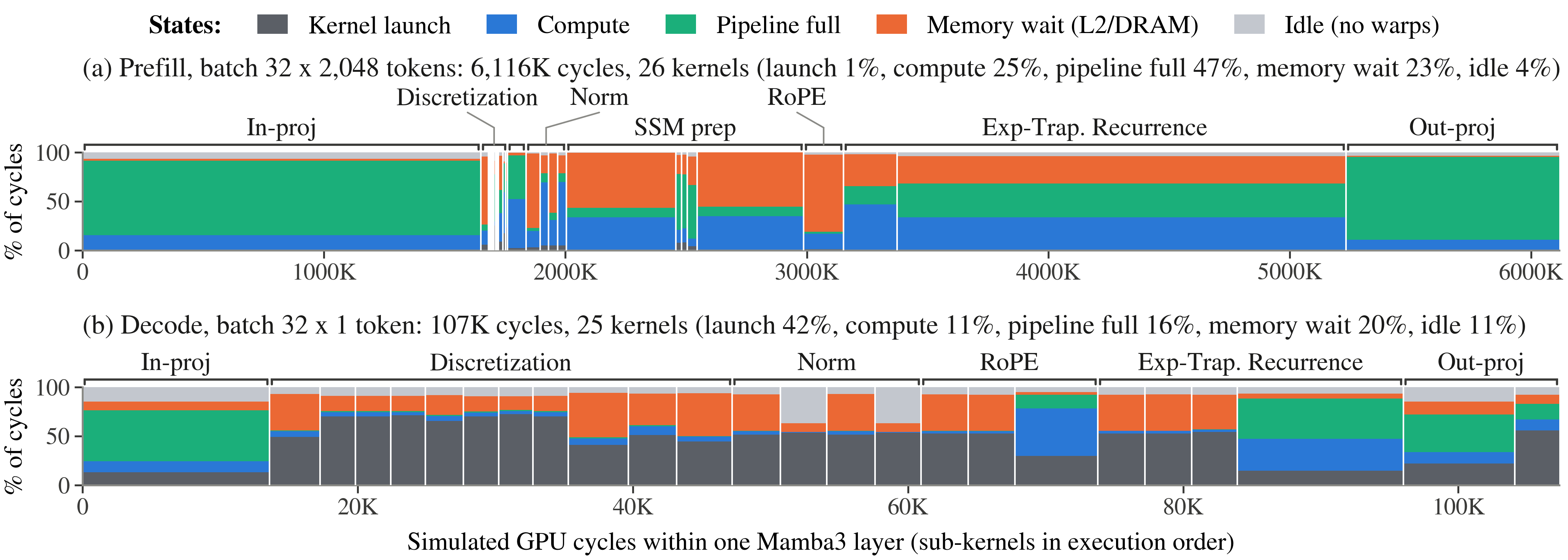}
    \caption{Cycle breakdown of one Mamba-3 layer at 2K tokens for (a) prefill and (b) single-token decode. Each column is one sub-kernel in execution order; its width is the sub-kernel's cycles and its height splits those cycles into kernel launch and warp-scheduler states. Simulated with Accel-Sim (A100, 1.8K-cycle kernel launch latency, batch size 32, N=32, MIMO rank 2)}
    \label{fig:intro_mamba3_kernel_breakdown}
\end{figure*}

Motivated by prior work on top-down performance analysis~\cite{topdownperformanceanalysis}, GPU instruction-level stall attribution~\cite{gpuperformanceadvisor}, and GPU profiling methodologies~\cite{NVIDIA_NsightCompute}, we classify Mamba-3 execution cycles into five categories: \textit{kernel launch}, \textit{compute}, \textit{pipeline full}, \textit{memory wait}, and \textit{idle}, as shown in Figure~\ref{fig:intro_mamba3_kernel_breakdown}. 
Using Accel-Sim~\cite{AccelSim}, we apply this classification to distinguish modeled kernel launch overhead, useful computation, execution-pipeline contention, memory-related stalls, and idle cycles, respectively. These categories are tailored to our Mamba-3 analysis rather than adopted directly from any single profiling methodology.

During \textit{prefill}, the Mamba-3 layer executes 26 kernels over $6.116M$ cycles. 
The fused Exponential Trapezoidal Recurrence scan dominates execution, consuming approximately 50.1\% of layer cycles, with time distributed across computation, pipeline stalls, and memory waits. 
Although the input and output projection kernels spend approximately 75\% of their cycles in the pipeline-full state, their GEMMs achieve only 11\%--20\% of peak Tensor Core throughput.
The RoPE angle kernel spends 64\% of its cycles waiting on memory, while launch overhead accounts for 69\%--79\% of cycles in small discretization and normalization kernels. 
Thus, prefill performance is primarily constrained by the recurrent scan, with additional inefficiencies from memory stalls and small-kernel execution.

During \textit{decoding}, the Mamba-3 layer executes 25 kernels over $107K$ cycles. 
Each decoding step processes only one token per sequence, substantially reducing useful work per kernel launch compared with prefill.
Consequently, kernel launch latency accounts for 42\% of the total simulated Mamba-3 layer execution cycles during decoding. 
For individual discretization kernels, which perform relatively little computation, launch latency can account for as much as 88\% of the cycles attributed to launching and executing the kernel. 
In contrast, instruction issue accounts for only 3\%--6\% of total layer cycles. Memory stalls are concentrated in RoPE, exponential trapezoidal recurrence, and projection operations, which consume intermediate results produced by preceding kernels.


\begin{figure*}[t!]
    \centering
    \begin{subfigure}[b]{0.48\textwidth}
        \centering
        \includegraphics[width=\linewidth]{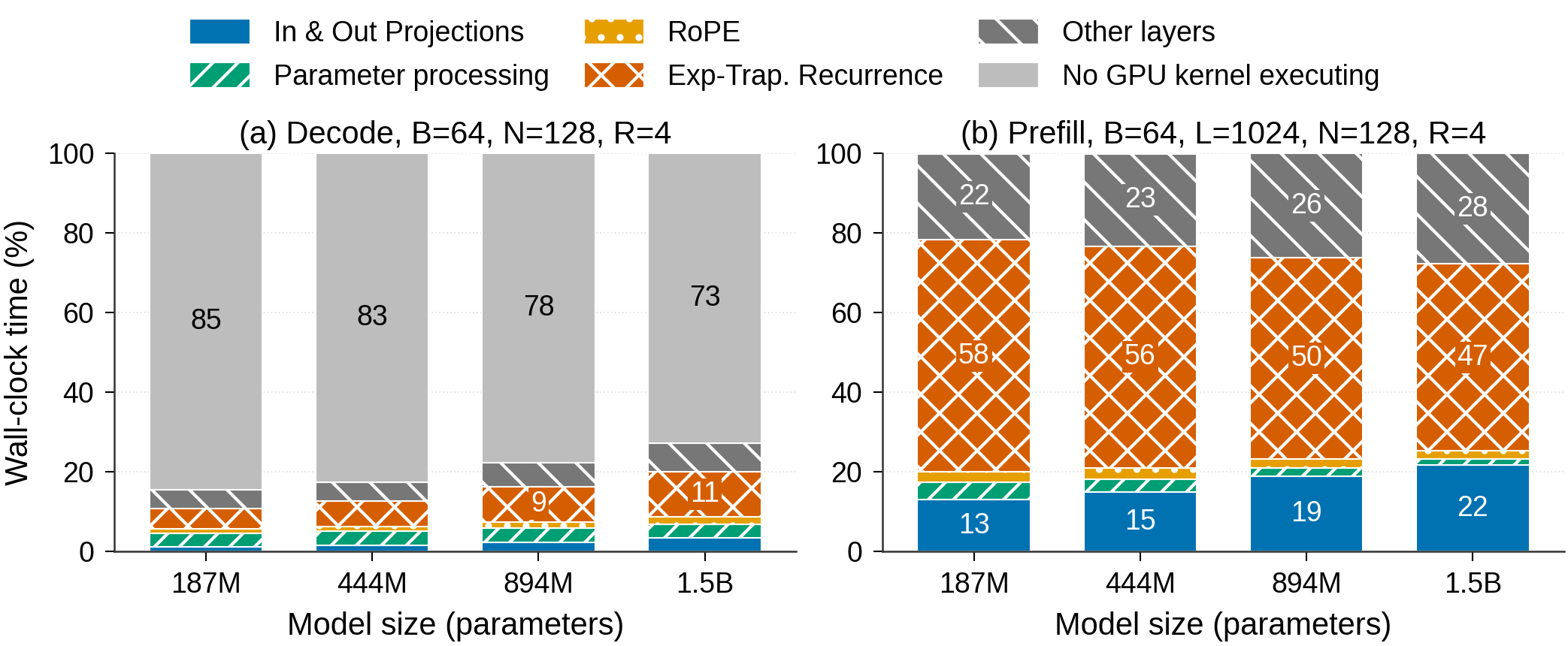}
        \caption{Fused SSM kernel share of wall-clock time under parameter scaling.}
        \label{fig:ssm_sensitivity_param}
    \end{subfigure}
    \hfill
    \begin{subfigure}[b]{0.48\textwidth}
        \centering
        \includegraphics[width=\linewidth]{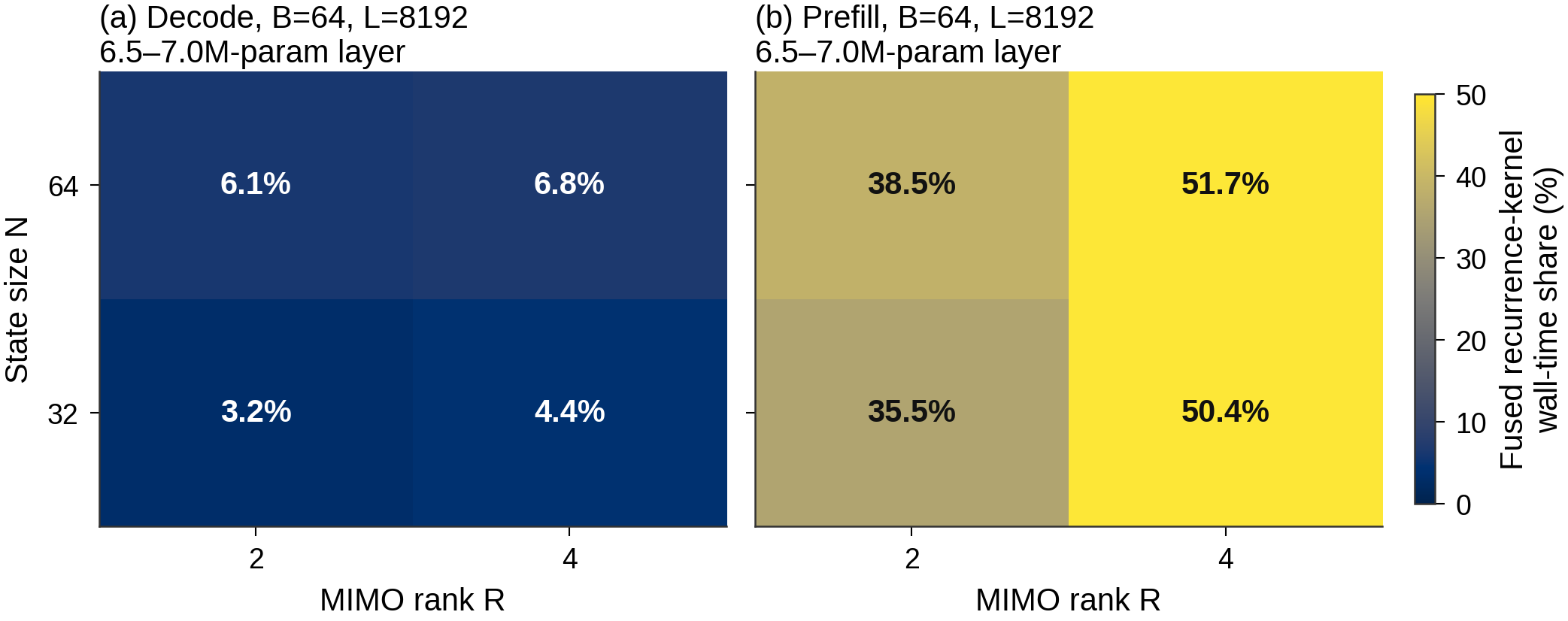}
        \caption{Exp-Trap. Recurrence kernel share across scaled $R$ and $N$.}
        \label{fig:ssm_sensitivity_heatmap}
    \end{subfigure}
    \caption{Sensitivity analysis of Mamba-3 as fraction of total wall-clock time spent in various kernels, for prefill and decode.}
    \label{fig:ssm_sensitivity}
\end{figure*}

\textit{Sensitivity to Model Parameters:}
Figure~\ref{fig:ssm_sensitivity} examines how the execution distribution changes with the recurrent state size $N$ and MIMO rank $R$ during prefill and decoding. 
We distinguish the fused SSM kernel, which implements the recurrent computation in the baseline GPU implementation, from the surrounding parameter processing, RoPE, and projection kernels. 
Increasing $N$ and $R$ changes the computational workload but does not eliminate the kernel boundaries and data dependencies between these stages. 
Across the evaluated configurations, the fused SSM kernel accounts for only 3.2\%--6.8\% of total decoding wall time, compared with up to 51.7\% during prefill. 
This difference reflects the two execution regimes: decoding incurs substantial overhead from launching short-running kernels, whereas prefill amortizes launch overhead over larger computational workloads. 
Consequently, substantial execution time remains outside the fused SSM kernel in both phases, particularly during decoding, motivating acceleration of the surrounding Mamba-3 operations.


\textit{Architectural Implications:}
These results identify two complementary acceleration opportunities. 
During decoding, the large number of short kernels and repeated intermediate transfers motivate consolidating dependent operations into a locally connected execution pipeline. 
During prefill, the dominant recurrent scan motivates a datapath that exploits parallelism across independent heads and channels while efficiently processing sequential state updates. 
Accordingly, we propose integrating a specialized execution unit within the GPU to pipeline discretization, normalization, RoPE, MIMO computation, and recurrent updates with local intermediate forwarding, while retaining existing GPU execution resources for dense projections and other hybrid-model operations.

\subsection{Software Kernel Fusion}
\label{sec:sw_fusion}

Our profiling in the previous section identifies kernel fragmentation and launch overhead as dominant bottlenecks in Mamba-3 decoding, motivating software fusion to reduce kernel launches and intermediate data movement.
To assess how effectively fusion addresses these bottlenecks, we implement hand-fused CUDA kernels for the complete Mamba-3 decoding path on an NVIDIA A100 80GB GPU.

\paragraph{Single-sequence decoding.}
At $B=1$, $N=32$, $R=2$, the publicly released Mamba-3 CUDA implementation~\cite{lahoti2026mamba3} executes 25 GPU kernels per decoding step and requires $49.2\,\mu$s, even with CUDA Graphs~\cite{nvidiaCudaGraphs}, which reduce CPU-side submission overhead by capturing and replaying sequences of GPU operations.
Our optimized implementation reduces execution to two GPU kernels, achieving $14.4\,\mu$s latency ($3.4\times$ speedup) while matching reference outputs within a relative error of $2\times10^{-4}$.
To assess whether the remaining kernel boundary can be eliminated, we evaluate redundant input-projection computation and software grid-wide synchronization.
These alternatives increase latency to $20.7\,\mu$s and $16.5\,\mu$s, respectively, indicating that additional fusion introduces more overhead than it removes.
In the two-kernel implementation, approximately $10\,\mu$s is spent on dependent activation, normalization, RoPE, and recurrent-update operations, whereas the theoretical DRAM transfer lower bound is only $2.2\,\mu$s per decoding step, and measured SM and DRAM utilization remain low at $9\%$ and $15\%$, respectively.
Our evaluated software-fusion implementations substantially reduce launch overhead at $B=1$ but leave significant dependent-operation latency and resource-utilization bottlenecks.
\paragraph{Impact of batch size}
Single-sequence decoding is the most favorable case for fusion, because almost all of the decoding step is fixed per-kernel cost.
Since Section~\ref{sec:gpu_arch} compares fusion with architectural changes at $B=32$ and $B=64$, we repeat the measurement for $B\in\{1,32,64,128\}$  and for six configurations ($N\in\{32,64\}$, $R\in\{1,2,4\}$; $d_\text{model}=512$).
At each point we report the faster of two fused designs: the two-kernel design above, and a three-kernel variant that keeps the fused middle kernel but leaves both projections to cuBLAS, because the two-kernel design's input projection supports at most eight sequences and its split-K output projection, which accumulates partial sums with atomics, scales poorly with $B$.
All fused variants match the reference within a relative error of $1.3\times10^{-3}$ at every batch size.
Table~\ref{tab:sw_fusion_batch} summarizes the results.

\begin{table}[t]
\centering
\caption{Mamba-3 decoding latency per layer ($\mu$s) of the shipped kernels (CUDA Graph) and of the fastest fused implementation on an A100, and the resulting speedup (Spd., $\times$), at increasing batch size $B$.}
\label{tab:sw_fusion_batch}
\footnotesize
\setlength{\tabcolsep}{2.5pt}
\begin{tabular}{@{}lccccccccc@{}}
\toprule
& \multicolumn{3}{c}{$B=1$} & \multicolumn{3}{c}{$B=32$} & \multicolumn{3}{c}{$B=64$} \\
\cmidrule(lr){2-4}\cmidrule(lr){5-7}\cmidrule(l){8-10}
$N$, $R$ & Ship. & Fused & Spd. & Ship. & Fused & Spd. & Ship. & Fused & Spd. \\
\midrule
32, 1 & 39.3 & 14.3 & 2.75 & 61.1 & 40.1 & 1.52 & 69.5 & 61.3 & 1.13 \\
32, 2 & 49.2 & 14.4 & 3.42 & 77.7 & 45.0 & 1.73 & 85.8 & 70.5 & 1.22 \\
32, 4 & 50.1 & 15.7 & 3.19 & 80.4 & 63.5 & 1.27 & 90.0 & 105.4 & 0.85 \\
64, 1 & 39.4 & 14.6 & 2.71 & 66.3 & 53.0 & 1.25 & 79.8 & 85.7 & 0.93 \\
64, 2 & 49.6 & 15.3 & 3.24 & 81.7 & 61.1 & 1.34 & 97.8 & 100.6 & 0.97 \\
64, 4 & 50.7 & 16.5 & 3.07 & 84.2 & 77.9 & 1.08 & 101.6 & 133.1 & 0.76 \\
\bottomrule
\end{tabular}
\end{table}

The benefit of fusion shrinks steadily as the batch grows.
For $N=32$ and $R=2$, the speedup falls from 3.42 at $B=1$ to 1.73 at $B=32$, 1.22 at $B=64$, and 0.83 at $B=128$.
Across the six configurations, fusion achieves $2.75$--$3.42$ at $B=1$ but only $1.08$--$1.73$ at $B=32$ and $0.76$--$1.22$ at $B=64$, where the fused implementation is slower than the shipped kernels for four of the six configurations; at $B=128$ it is slower for all six.
The decline results from the two implementations scaling very differently with $B$.

\emph{The shipped kernels scale sublinearly.}
Increasing the batch from 1 to 64 multiplies the work by $64\times$ but the shipped latency by only $1.75$--$2.02$.
At $B=1$, the kernels between the two projections take $30.6$--$43.1\,\mu$s in total, roughly $1.5$--$2\,\mu$s each, a cost set by kernel boundaries (an empty kernel alone occupies $0.92\,\mu$s in a CUDA Graph) and by short, latency-bound kernels that occupy only a few SMs, not by the amount of work.
This time grows by at most $14\%$ from $B=1$ to $B=4$.
Additional sequences are absorbed by SMs that are idle at small batch sizes, so between $B=32$ and $B=64$ each additional sequence adds only $0.25$--$0.54\,\mu$s.
The projections behave similarly: their cost is dominated by reading the weights once per step, so for $N=32$, $R=2$ they grow from $8.2\,\mu$s at $B=1$ to only $14.0\,\mu$s at $B=64$.
The fixed per-kernel cost that fusion removes therefore stays constant in absolute terms while its share of the decoding step declines.

\emph{The fused kernel scales almost linearly.}
From $B=1$ to $B=64$, the fused latency grows by $4.3$--$8.1\times$, and each additional sequence between $B=32$ and $B=64$ adds $0.66$--$1.72\,\mu$s, $2.4$--$4.4\times$ more than in the shipped kernels.
The fused kernel was designed for single-sequence latency and trades extra computation for fewer synchronizations.
Each thread block, which processes one head of one sequence, recomputes the RMSNorm and RoPE of the $B$ and $C$ projections that all heads of the sequence share, so this work is repeated 32 times per sequence, and the state update runs as scalar FP32 loops.
At $B=1$ the kernel launches only a few dozen thread blocks on the 108 SMs, so the redundant work runs on otherwise idle SMs and costs no latency.
At $B=64$ its 2,048 blocks fill the GPU, and the redundancy costs throughput directly: Nsight Compute shows that the fused kernel becomes instruction-issue bound, with $57$--$66\%$ SM throughput and an IPC of $2.4$--$2.7$, while DRAM throughput is only $4$--$10\%$ and occupancy is limited to $35$--$46\%$ by shared memory.
The shipped implementation, in contrast, already performs the recurrence in a single fused CuTe DSL kernel that takes $9$--$19\,\mu$s at $B=64$, compared with $48$--$118\,\mu$s for the recurrence in our fused kernel.

\emph{Larger state and rank move the crossover to smaller batches.}
The per-sequence work in the fused kernel grows with $N$ and $R$ (its executed instructions increase by about $2.5\times$ from $N=32$, $R=1$ to $N=64$, $R=4$), so fusion stops paying off between $B=64$ and $B=128$ for $N=32$ with $R\le2$, but already between $B=32$ and $B=64$ for $N=32$, $R=4$ and for all $N=64$ configurations.

In summary, kernel fusion removes a fixed cost per decoding step, which dominates at $B=1$ but not at the batch sizes used in serving.
At $B=32$ and $B=64$, the decoding step is dominated by per-sequence work in the recurrence and its dependent element-wise operations, which fusion does not reduce, and software fusion yields at most 1.73 and 1.22, respectively.
Where the fused implementation is slower, the shipped kernels remain the fastest software option, so we count software fusion as $1.0\times$ in those configurations in Section~\ref{sec:gpu_arch}.

\subsection{GPU Resource Utilization and Scaling Limits}

\begin{figure}[!t]
    \centering
    \includegraphics[width=\linewidth]{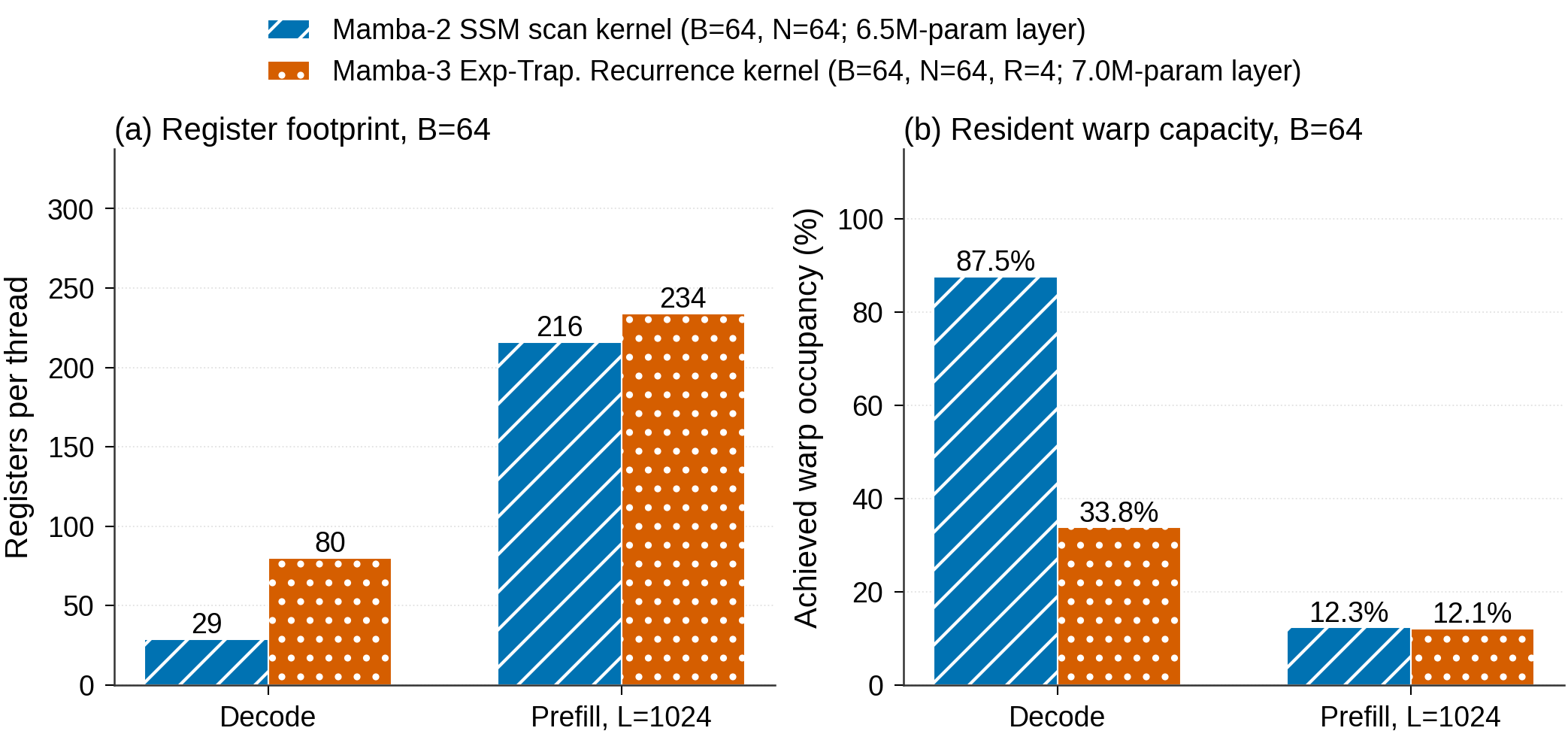}
    \caption{Register usage per thread and achieved warp occupancy for Mamba-2 and Mamba-3 SSM kernels for decoding/prefill.}
    \label{fig:nsight_register}
\end{figure}

\begin{figure*}[htbp]
    \centering
    \includegraphics[width=0.75\textwidth]{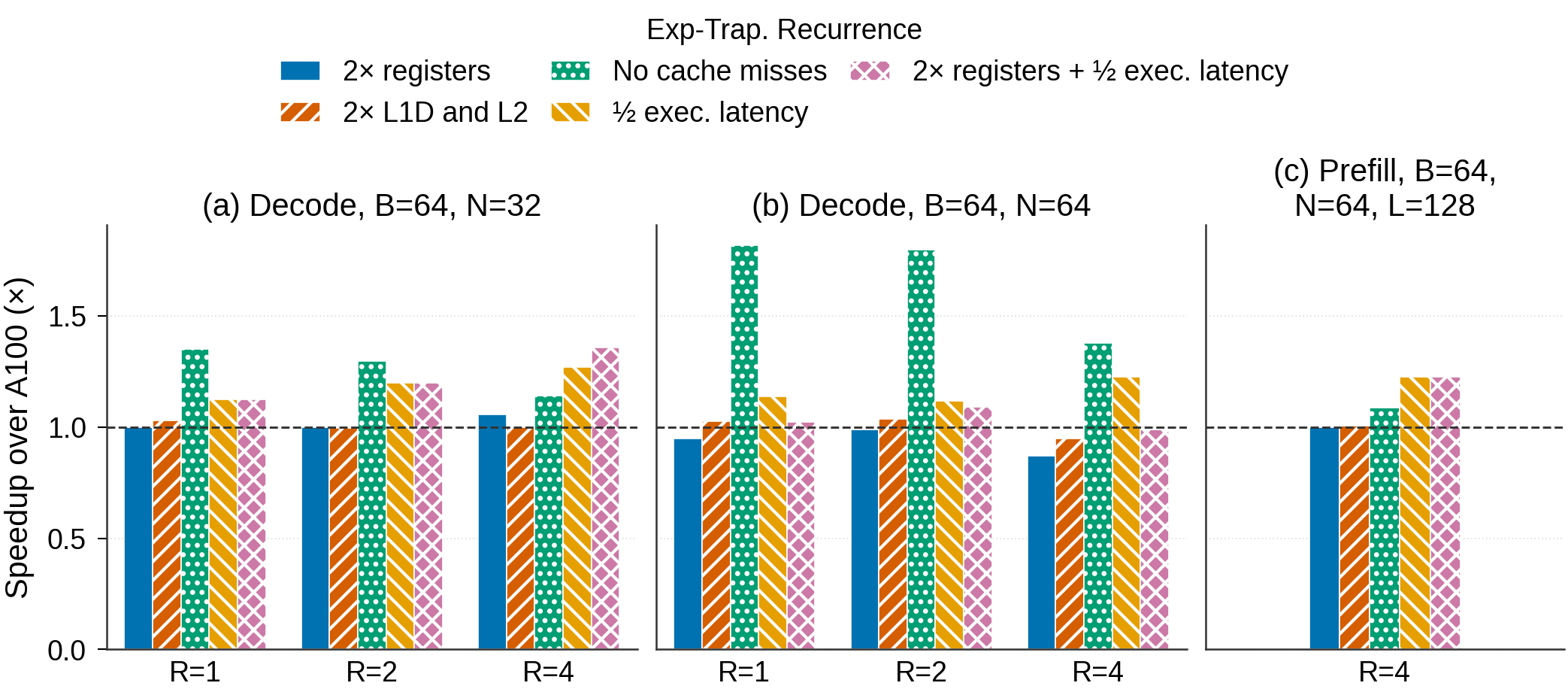}
    \caption{Impact of simulated microarchitectural modifications to the NVIDIA A100 GPU on Mamba-3 Exp-Trap. Recurrence kernel execution time (Accel-Sim).}
    \label{fig:hardware_scaling}
\end{figure*}

To determine whether conventional GPU resource scaling can alleviate Mamba-3's execution bottlenecks, we first characterize its microarchitectural utilization using Nsight Compute~\cite{NVIDIA_NsightCompute} and compare it against Mamba-2~\cite{dao2024mamba2}. 
We then evaluate the performance sensitivity of Mamba-3 to changes in GPU compute and memory resources using Accel-Sim~\cite{AccelSim}.

Figure~\ref{fig:nsight_register} shows that Mamba-3's additional recurrent operations substantially increase register pressure. 
During decoding ($B=64$, $N=64$), the fused Mamba-3 kernel requires 80 registers per thread, compared with 29 for Mamba-2, a $2.8\times$ increase. 
Consequently, achieved warp occupancy decreases from 87.5\% to 33.8\%, reducing the number of resident warps available to hide instruction and memory latencies. 
Thus, despite its higher arithmetic intensity, Mamba-3 exposes less thread-level parallelism to the GPU.

To assess whether additional hardware resources can recover this lost efficiency, we simulate several modifications to the A100 architecture: doubling the SM register-file capacity, doubling the L1 and L2 cache capacities, halving instruction latencies, and combining a doubled register file with halved instruction latencies. 
We also evaluate an idealized memory hierarchy with no cache or DRAM misses to estimate the upper-bound benefit of eliminating memory-access stalls.

As shown in Figure~\ref{fig:hardware_scaling}, doubling the register file or cache capacities improves execution time by at most 6\%, indicating that additional capacity alone does not substantially alleviate the observed bottlenecks. 
An ideal memory hierarchy provides up to $1.8\times$ speedup during decoding but only $1.09\times$ during prefill, highlighting the greater sensitivity of decoding to memory stalls. 
Halving instruction latencies achieves up to $1.27\times$ and $1.23\times$ speedups during decoding and prefill, respectively, while combining reduced instruction latency with a doubled register file yields at most $1.36\times$ speedup. 
These results indicate that neither additional storage capacity nor faster execution units alone can fully address the dependencies and execution overheads of Mamba-3.

Overall, the increased register demand of Mamba-3 limits warp occupancy, while its dependent recurrent operations restrict the benefits of conventional GPU resource scaling. These findings motivate a GPU-integrated specialized datapath that pipelines dependent operations and retains intermediate values locally, rather than relying solely on larger register files, caches, or faster general-purpose execution units.

\begin{figure}[t]
    \centering
    \includegraphics[width=\linewidth]{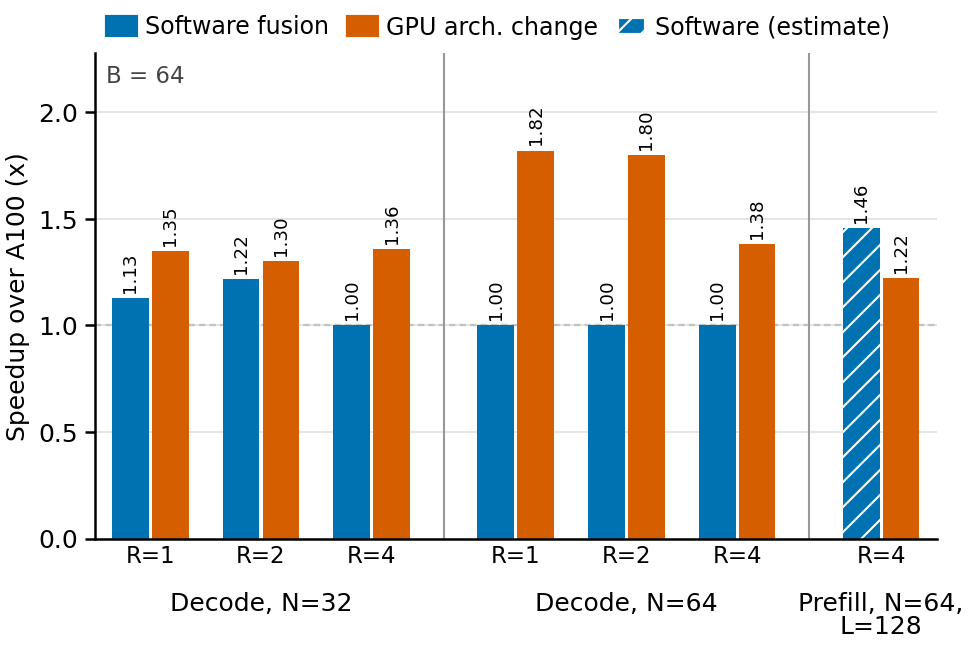}
    \caption{Speedup over the released Mamba-3 CUDA implementation~\cite{lahoti2026mamba3} on an NVIDIA A100 ($B=64$).
    Software fusion reports the best measured CUDA fusion result during decoding, while GPU architectural changes show the best-performing evaluated GPU modification.
    The hatched prefill software-fusion bar is an estimate rather than a measured implementation.
    Results span state sizes $N=32,64$ and MIMO ranks $R=1,2,4$.}
    \label{fig:gpu_optimization_limits}
\end{figure}

\subsection{Comparison across configurations}
\label{sec:gpu_arch}
Figure~\ref{fig:gpu_optimization_limits} compares measured software fusion against GPU architectural modifications simulated in Accel-Sim at $B=64$, including increased register-file and cache capacities and reduced instruction latency. 
Unlike the $3.42\times$ improvement at $B=1$, software fusion achieves at most $1.22\times$ decoding speedup, as larger batches better amortize kernel launches. 
The simulated architectural modifications achieve $1.30\times$--$1.82\times$ decoding speedup, but neither approach consistently eliminates recurrent bottlenecks. 
For prefill, software fusion is estimated to achieve $1.46\times$ speedup, compared with $1.22\times$ for the best simulated GPU modification. 
These results highlight the limitations of conventional optimizations.

\subsection{Mamba-3 Quantization}\label{BACKG_POSIT}

To assess the feasibility of reduced-precision arithmetic in the proposed GPU-integrated execution unit, we evaluate the sensitivity of Mamba-3 accuracy to different numerical representations. 
Using the Language Model Evaluation Harness~\cite{gao2024framework}, we benchmark the released 187M-parameter Mamba-3 SISO and MIMO ($R=4$) checkpoints across seven zero-shot NLP tasks. 
We compare BF16~\cite{kalamkar2019bfloat16} and FP16 baselines against INT8 weight-only quantization and the experimental Posit16 format~\cite{Gustafson2017Posit,Lu2020PositDNN,Mallasen2025PositSurvey}. 
As defined in Equation~\ref{eq:posit}, the Posit representation encodes a value using sign, regime, exponent, and fraction fields, providing a tapered-precision alternative to conventional floating-point formats. 
We evaluate Posit16 with an exponent-size parameter of $es=1$.

\begin{figure*}[!t]
    \centering
    \includegraphics[width=0.8\linewidth]{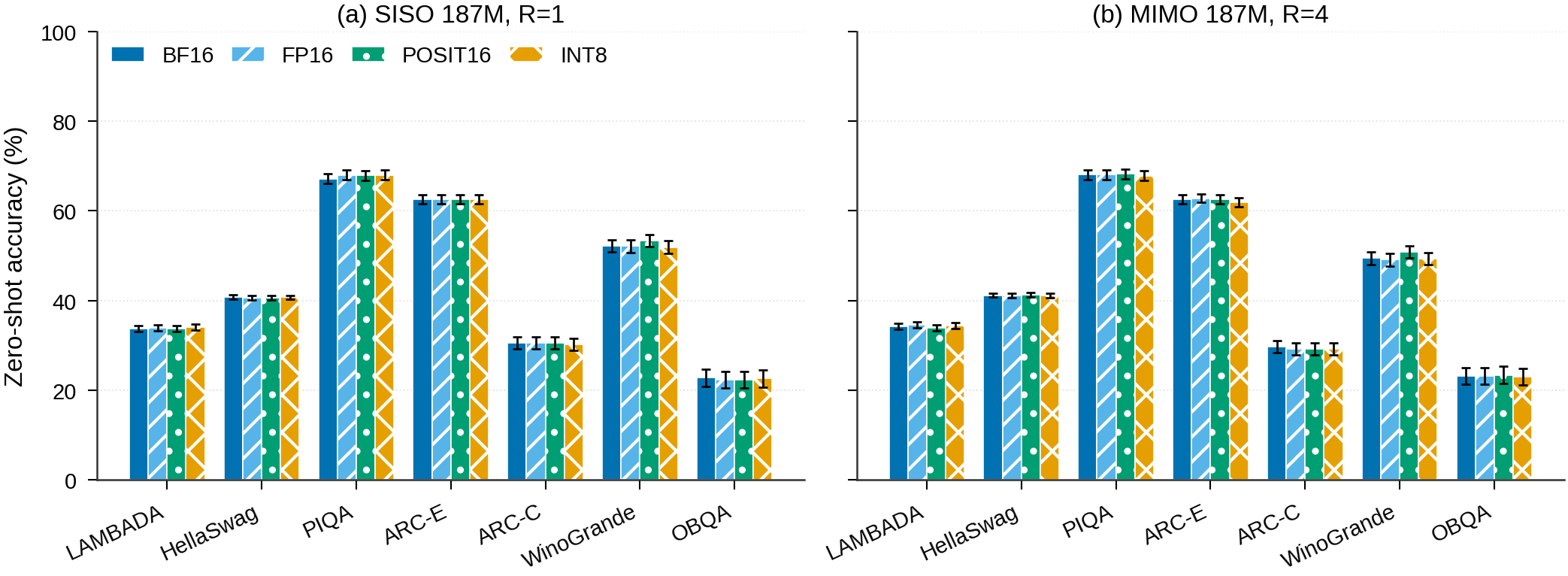}
    \caption{Seven-task zero-shot accuracy of 187M-parameter Mamba-3 SISO and MIMO ($R=4$) models under BF16, FP16, INT8 weight-only quantization, and Posit16 ($es=1$).}
    \label{fig:accuracy_lm_harness}
\end{figure*}

Figure~\ref{fig:accuracy_lm_harness} compares the zero-shot accuracy across these representations. 
Relative to BF16, Posit16 and INT8 change the seven-task average accuracy by at most 0.24 percentage points across the evaluated Mamba-3 configurations. 
These results indicate that the evaluated models tolerate reduced numerical precision with minimal accuracy degradation, supporting the feasibility of compact arithmetic representations for Mamba-3 acceleration. 
However, the choice of hardware representation must also account for arithmetic complexity, area, and power, which we examine in the subsequent architecture design.

\subsection{Architectural Implications}

Our characterization identifies complementary bottlenecks in Mamba-3 execution: decoding is dominated by kernel-launch overhead and dependent-operation latency, while prefill suffers from limited recurrent-scan parallelism and inefficient GPU utilization. 
Software kernel fusion substantially reduces dispatch and intermediate-memory overheads, but sequential dependencies remain on the critical path. 
Similarly, increasing GPU register and cache capacities yields limited benefits, while reduced-precision arithmetic offers opportunities to improve hardware efficiency without significant accuracy degradation.

Together, these findings establish a performance target for specialized hardware. 
An effective Mamba-3 accelerator must outperform not only the released GPU implementation but also a stronger baseline incorporating additional software fusion and reduced memory-access overhead. 
Our experiments show that even aggressive GPU optimizations provide limited gains, with a combined register-file expansion and instruction-latency reduction achieving at most $1.36\times$ speedup, while idealized memory eliminates stalls to provide up to $1.8\times$ during decoding. 
Accordingly, we target Mamba-3 layer speedups exceeding these optimized GPU baselines, rather than improvements attributable solely to kernel fusion or additional GPU resources.

To achieve this, we propose a specialized execution unit integrated within the GPU that pipelines discretization, normalization, RoPE, MIMO computation, and recurrent state updates with local intermediate forwarding, while retaining existing GPU resources for dense projections and other hybrid-model operations. The resulting architecture is described in Section~\ref{ARCHITECTURE}.

%% file: architecture.tex
\section{RescuCore Architecture}
\label{ARCHITECTURE}

\begin{figure*}[!t]
    \centering
    \includegraphics[width=0.7\linewidth]{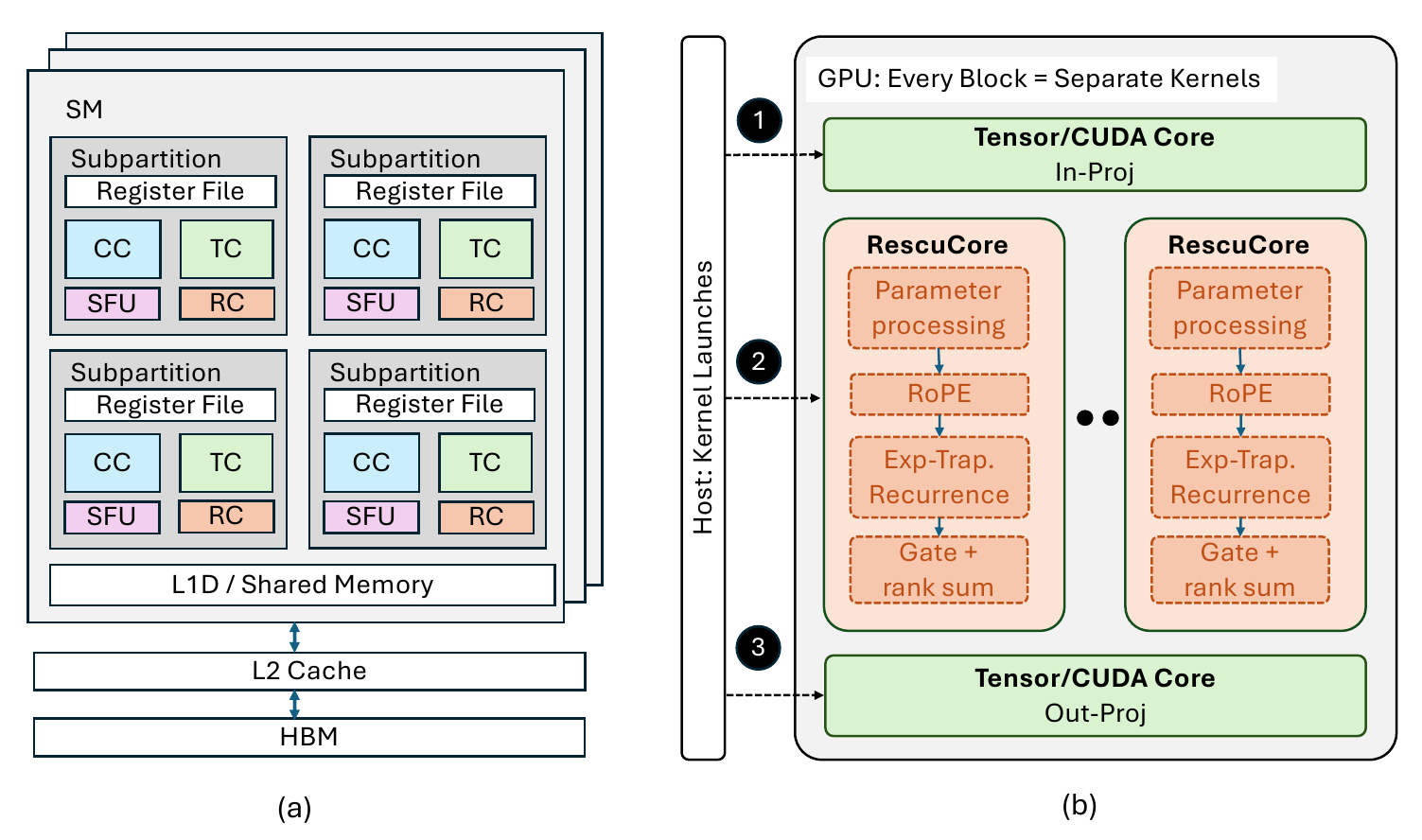}
    \caption{(a) Integration of \acceleratorname (RC) within GPU SM subpartitions alongside CUDA Cores (CC), Tensor Cores (TC), and special-function units (SFUs), with dedicated local registers for recurrent states and intermediate operands. (b) Mamba-3 execution with conventional GPU resources handling dense projections, while \acceleratorname executes parameter processing with MIMO, RoPE, exponential-trapezoidal recurrence, and output gating in a fused kernel. Independent heads and channels are distributed across RC units.}
    \label{fig:SM_Cores_RC_Combined}
\end{figure*}

Our bottleneck analysis shows that software fusion reduces kernel-launch overhead but leaves dependent Mamba-3 arithmetic constrained by conventional GPU execution resources.
We propose \acceleratorname, an SM-integrated specialized execution unit that combines a fused recurrent datapath, dedicated local registers, and direct operand forwarding.
Unlike CUDA kernel fusion, \acceleratorname retains recurrent states and intermediate operands within the specialized unit, avoiding repeated general-purpose register-file accesses and intermediate memory transfers.
Its pipelined execution exploits independent head/channel operations while preserving the GPU's existing resources for dense projections and non-SSM computation.

\subsection{GPU Integration and Execution Model}

\begin{figure*}[!t]
    \centering
    \includegraphics[width=0.7\linewidth]{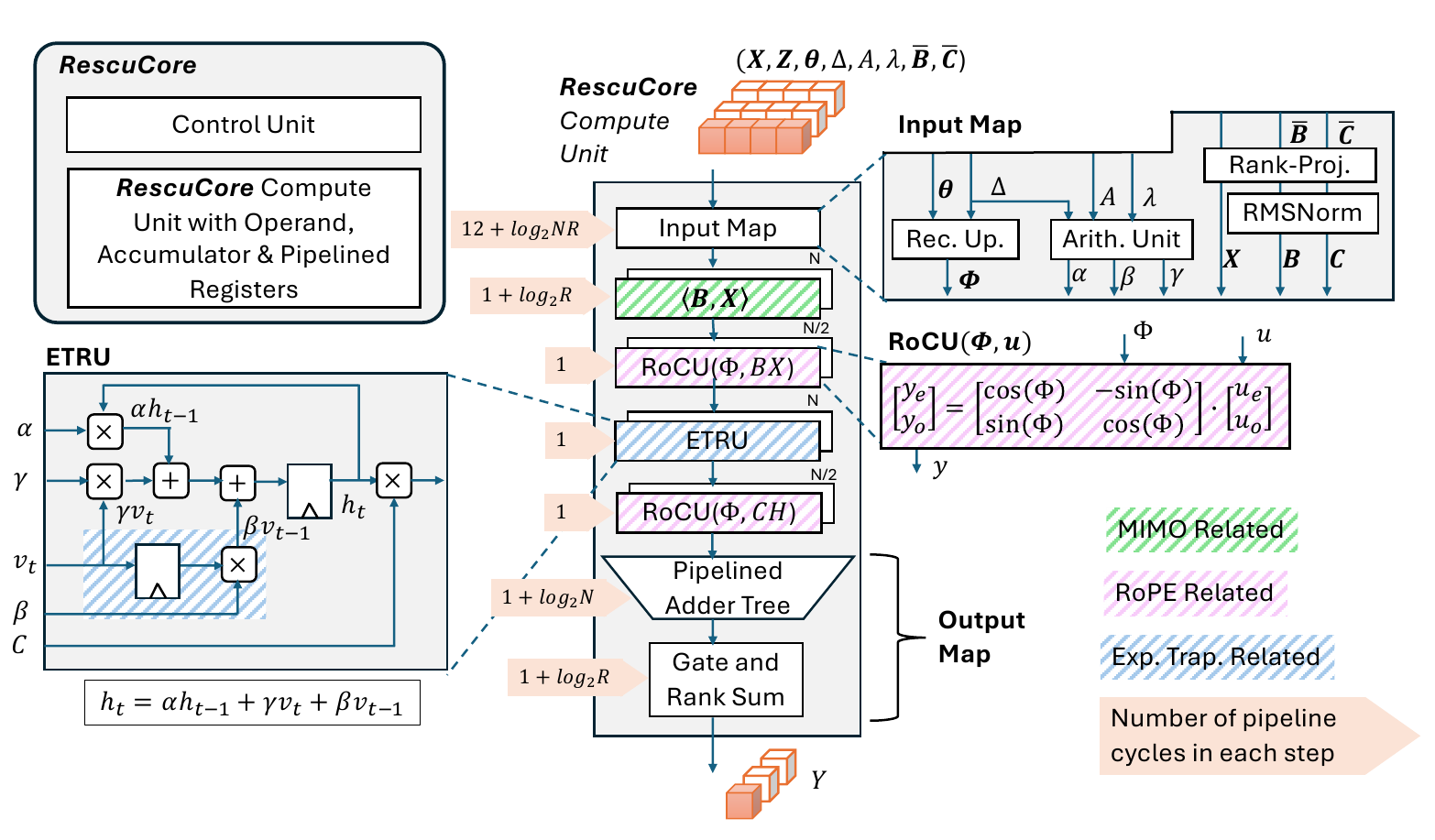}
    \caption{\acceleratorname microarchitecture with a dedicated local operand, accumulator, and pipelined registers, control logic, and a seven-stage recurrent datapath. Input mapping generates coefficients and normalized operands; MIMO mapping computes input-to-state products; two RoCUs perform rotations; ETRU updates recurrent states; and output mapping performs reduction and gating. The local registers store recurrent states and reusable intermediate operands, while pipeline registers between stages support direct forwarding and pipelined execution.}
    \label{fig:rescucore_micro_arch}
\end{figure*}

Figure~\ref{fig:SM_Cores_RC_Combined}b illustrates the division of work between conventional GPU resources and \acceleratorname.
Input/output projections execute on existing GPU compute units, whereas parameter processing, normalization, RoPE, MIMO recurrence, and gating execute within a single specialized kernel.
Attention, feed-forward layers, and other hybrid-model components continue using conventional GPU resources.

As shown in Fig.~\ref{fig:SM_Cores_RC_Combined}a, we integrate one \acceleratorname unit per SM subpartition, providing 432 units across the modeled A100's 108 SMs.
Each unit operates alongside existing CUDA Cores, Tensor Cores, and special-function units, using the GPU's scheduling and memory-access infrastructure.
Crucially, \acceleratorname employs dedicated local operand, accumulator, and pipeline registers to manage recurrent states and intermediate operands, independently of the SM subpartition register file.
Discretization coefficients, MIMO products, rotated operands, and recurrent states remain within this local storage and are forwarded directly between arithmetic stages.
This organization avoids repeated accesses to conventional GPU registers and memory for intermediate recurrent computation, enabling tightly coupled, pipelined execution.

The input projection produces per-head and per-channel activations, which are distributed to the corresponding \acceleratorname units
for recurrent computation.
Completed recurrent outputs return to conventional GPU resources for the output projection.
By retaining intermediate values locally, \acceleratorname reduces L2 traffic for the specialized recurrent path by $4.4\times$ during prefill and approximately $60\times$ during decoding in the characterized configurations.
These reductions are only achieved in the recurrent operations; dense projections continue using the conventional GPU memory hierarchy.

\subsection{\acceleratorname Microarchitecture}

Figure~\ref{fig:rescucore_micro_arch} illustrates the seven-stage datapath, which incorporates dedicated local operand, accumulator, and pipeline registers to hold recurrent states, discretization coefficients, MIMO products, rotated operands, and other intermediate values. 
Dedicated control logic schedules independent channel
operations and orchestrates operand forwarding across pipeline
stages, minimizing accesses to the SM subpartition register file
during recurrent computation.

\textit{Input mapping and MIMO computation.}
The input mapping stage receives projected features $(\mathbf{X},\mathbf{Z},\boldsymbol{\theta},\boldsymbol{\Delta},\mathbf{A},\boldsymbol{\lambda},\overline{\mathbf{B}},\overline{\mathbf{C}})$ and prepares the operands required by subsequent stages.
Dedicated arithmetic performs RMS normalization~\cite{zhang2019rmsnorm}, rank projection, exponential-trapezoidal coefficient generation $(\alpha_t,\beta_t,\gamma_t)$, and rotation-angle accumulation.
The MIMO mapping stage then computes the input-to-state products using $N$ parallel length-$R$ dot-product circuits, exploiting spatial parallelism across state dimensions.
The resulting operands are stored in pipeline registers and forwarded directly to the first rotation stage.

\textit{Rotation Compute Unit (RoCU).}
RoCU implements Mamba-3's complex-valued state dynamics using real-valued pairwise rotations as in Eq.~\ref{eq:mamba3_rotation_operator}.
Each pair of even- and odd-indexed operands represents the real and imaginary components of one complex value.
For an $N$-element input vector, RoCU instantiates $N/2$ rotation circuits operating concurrently, each computing
\begin{equation}
\begin{bmatrix}
y_{2i}\\
y_{2i+1}
\end{bmatrix}
=
\begin{bmatrix}
\cos\Phi_t & \sin\Phi_t\\
-\sin\Phi_t & \cos\Phi_t
\end{bmatrix}
\begin{bmatrix}
u_{2i}\\
u_{2i+1}
\end{bmatrix},
\quad 0\leq i<N/2.
\label{eq:rescucore_rocu}
\end{equation}
The trigonometric values are retrieved from LUTs integrated within \acceleratorname, avoiding repeated execution of sine and cosine instructions on GPU special-function units.
As shown in Fig.~\ref{fig:rescucore_micro_arch}, two RoCU stages apply these rotations to the MIMO input contribution before recurrence and to the state-readout contribution after recurrence.
Their outputs are retained in local registers and forwarded directly to the next arithmetic stage.

\textit{Exponential-Trapezoidal Recurrence Unit (ETRU).}
ETRU implements the recurrent update
\begin{equation}
\mathbf{h}_t
=
\alpha_t\mathbf{h}_{t-1}
+\gamma_t\mathbf{v}_t
+\beta_t\mathbf{v}_{t-1},
\label{eq:rescucore_etru}
\end{equation}
where $\mathbf{v}_t$ is the rotated MIMO input contribution.
ETRU contains $N$ parallel state-update circuits, each computing one element of $\mathbf{h}_t$:
\begin{equation}
h_{t,i}
=
\alpha_t h_{t-1,i}
+\gamma_t v_{t,i}
+\beta_t v_{t-1,i},
\quad 0\leq i<N.
\label{eq:rescucore_etru_element}
\end{equation}
Each circuit includes multipliers and adders to compute the decayed-state, current-input, and previous-input contributions.
The previous state and previous-input operands are read from dedicated local registers, while updated values are written back through an internal feedback path.
All $N$ state elements are updated in parallel, reducing the arithmetic latency of each recurrent step while preserving inter-timestep dependencies.
Pipeline registers capture intermediate results at stage boundaries, ensuring that data propagates through the datapath without bypassing pipeline stages.
Each \acceleratorname maintains its own local registers and feedback path, with no register sharing or operand forwarding across \acceleratorname.
The updated recurrent states are made available to subsequent timesteps only after the corresponding computations complete.

\textit{Output mapping and gating.}
The output mapping stage computes the state-readout product using a pipelined adder tree that reduces the $N$ state-element contributions.
The reduction tree has a logarithmic depth of $\lceil\log_2 N\rceil$, allowing partial sums from different independent operations to occupy successive pipeline stages.
Dedicated gating circuitry applies the SiLU~\cite{elfwing2018sigmoid} activation and element-wise gating to the reduced outputs, followed by rank-summation circuitry that combines the $R$ MIMO outputs into the channel result.
The completed output is forwarded to conventional GPU resources for the dense output projection.
Together, the pipelined reduction, gating, and rank summation avoid additional GPU kernels and intermediate memory transfers.

\textit{Local pipelining and forwarding.}
Dedicated local operand, accumulator, and pipeline registers retain intermediate values throughout the recurrent datapath.
The $N/2$ parallel RoCU circuits, $N$ parallel ETRU circuits, and pipelined output-reduction circuitry exploit spatial parallelism to accelerate recurrent computation.
As shown in Fig.~\ref{fig:rescucore_micro_arch}, each datapath stage either completes in a single cycle or employs internal pipelining for multi-cycle operations, targeting an initiation interval of one cycle ($\mathrm{II}=1$) across the datapath, subject to recurrence-feedback dependencies.
This organization combines spatially parallel arithmetic with local operand forwarding and pipelined execution, reducing the need for repeated GPU instruction issue.

\subsection{Dependency-Aware Pipelining}

\acceleratorname{} retains recurrent states and intermediate operands in dedicated local storage, enabling pipelined execution with operand forwarding.
Each unit holds one channel's state and processes consecutive tokens at $\mathrm{II}=1$, since the single-register recurrence feedback ($L_{\mathrm{fb}}=1$) makes each updated state available before the next token requires it.

The seven-stage datapath comprises input mapping, RMSNorm, input MIMO product, recurrent update, output MIMO product, state reduction, and rank reduction.
Delay-matched operands ensure token alignment, yielding a pipeline depth of
\begin{equation}
D_{\mathrm{pipe}}=18+2\log_2 N'+3\log_2 R',
\label{eq:rescucore_depth}
\end{equation}
where $N'$ and $R'$ are the native state size and MIMO rank.
RTL simulation confirms this depth and $\mathrm{II}=1$ (see Section~\ref{sec:eval_ii1} for more details).

For $H$ heads, $P$ channels per head, and $N_{\mathrm{RC}}$ units, each unit sequentially processes
\begin{equation}
c=\left\lceil\frac{HP}{N_{\mathrm{RC}}}\right\rceil
\label{eq:rescucore_ops}
\end{equation}
channels, streaming $T=L$ tokens during prefill or $T=1$ during decoding.
Switching channels requires pipeline draining, reset, and weight loading over $W=2(2N'R'+2R'+N'/2+3)$ cycles.
Saving and restoring recurrent state adds $\sigma=2N'+N'/2$ cycles during prefill and twice that during decoding.
The estimated execution cycles are
\begin{equation}
\begin{aligned}
C_{\mathrm{RC}} &= c \left(
    \kappa T + W + D_{\mathrm{pipe}} + \sigma
    \right), \\[6pt]
\kappa &= \max \left(
    1, \frac{B_{\mathrm{tok}}}{\mathrm{BW}_{\mathrm{SM}}}
    \right).
\end{aligned}
\label{eq:rescucore_cycles}
\end{equation}
where $B_{\mathrm{tok}}$ denotes per-token operand traffic for four units per SM and $\mathrm{BW}_{\mathrm{SM}}=64$\,B/clk is the A100 L2-to-SM bandwidth~\cite{abdelkhalik2022demystifying}.
Although RTL sustains $\kappa=1$, estimated bandwidth limits yield $\kappa=5.1$, $2.9$, and $4.7$ for Jamba, MaTVLM, and MOL-Mamba, respectively.
Figures~\ref{fig:speedup_sw_arc_rescucore}--\ref{fig:runtime_energy_rescucore} assume $\kappa=2$, with bandwidth sensitivity evaluated in Section~\ref{sec:eval_freq}.

For $B=1$, $H=16$, $P=64$, $N_{\mathrm{RC}}=432$, and $(N',R')=(32,4)$, each unit processes $c=3$ channels with $D_{\mathrm{pipe}}=34$ and $W=566$.
At $\kappa=2$ and $1.41$\,GHz, the estimated \acceleratorname execution requires $14{,}328$ cycles ($10.2\,\mu$s) for $L=2048$ prefill and $2{,}286$ cycles ($1.6\,\mu$s) for decoding.
In comparison, execution on A100 NVIDIA GPU without \acceleratorname requires $504{,}730$ cycles ($358\,\mu$s) for prefill and $58{,}866$ cycles ($41.7\,\mu$s) for decoding, corresponding to speedups of $35.2\times$ and $25.8\times$, respectively.
Prefill is dominated by operand delivery, whereas decoding is dominated by channel-switching overhead.
These estimates exclude GPU projections and interface overhead, which are accounted for in the complete-layer evaluation in Section~\ref{EVALUATION}.

\subsection{Architectural Implications}

\acceleratorname addresses Mamba-3's execution bottlenecks, which stem from memory accesses between dependent operations, through three complementary mechanisms: a unified datapath that removes intermediate kernel boundaries and their global-memory round trips, dedicated local registers that retain recurrent states and intermediate operands, and pipelined arithmetic with direct feedback and forwarding.
For one Mamba-3 layer at $L=2048$, the 24 GPU kernels that \acceleratorname replaces issue 22.2\,MB of L2 writes, 5.0\,MB of DRAM writes, and 4.9\,GB of register-file traffic, whereas \acceleratorname writes only its 4\,MB output and keeps all other intermediates on chip.
Its SM subpartition integration preserves conventional GPU execution for dense projections and heterogeneous model components.
Together, these mechanisms target the dependent-operation latency that remains after software fusion while exploiting independent channel parallelism for recurrent throughput.

%% file: evaluation.tex
\section{Evaluation}\label{EVALUATION}

In this section, we evaluate the \acceleratorname microarchitecture. 
We first investigate the hardware implications of different numerical formats and select a suitable format for \acceleratorname. 
We then explore the design space across different model sizes, evaluating key hardware metrics to identify the best configuration. 
Finally, using the selected numerical format and model size, we evaluate the end-to-end performance of \acceleratorname, including throughput, energy efficiency, and other performance and efficiency metrics. 
We further quantify the performance and energy benefits of integrating \acceleratorname into an NVIDIA A100 GPU for inference on hybrid models.

\begin{figure*}[!ht]
    \centering
    \includegraphics[width=0.9\textwidth]{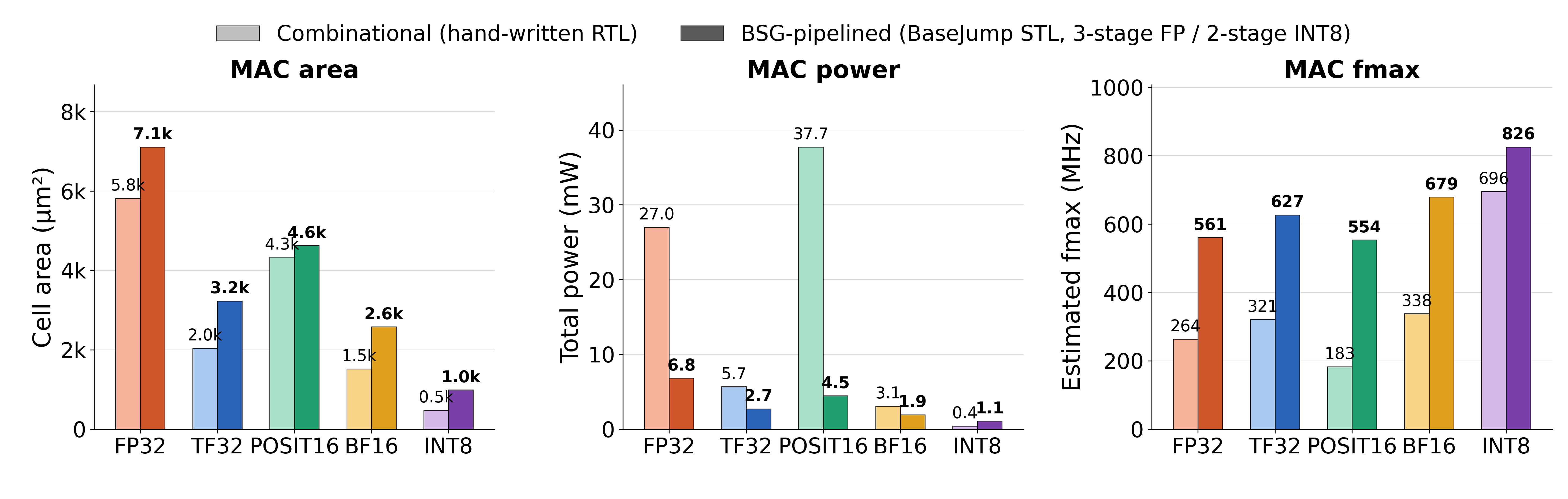}
    \caption{Area, power, and maximum synthesis frequency of combinational and pipelined MAC primitives across FP32, TF32, Posit16 ($es=2$), BF16~\cite{kalamkar2019bfloat16}, and INT8 in FreePDK45~\cite{FreePDK45}. Combinational BF16 provides a favorable floating-point area--power tradeoff, motivating its use within the functionally pipelined \acceleratorname datapath.}
    \label{fig:mac_comb_vs_pipelined}
\end{figure*}

\begin{figure*}[!ht]
    \centering
    \includegraphics[width=0.8\textwidth]{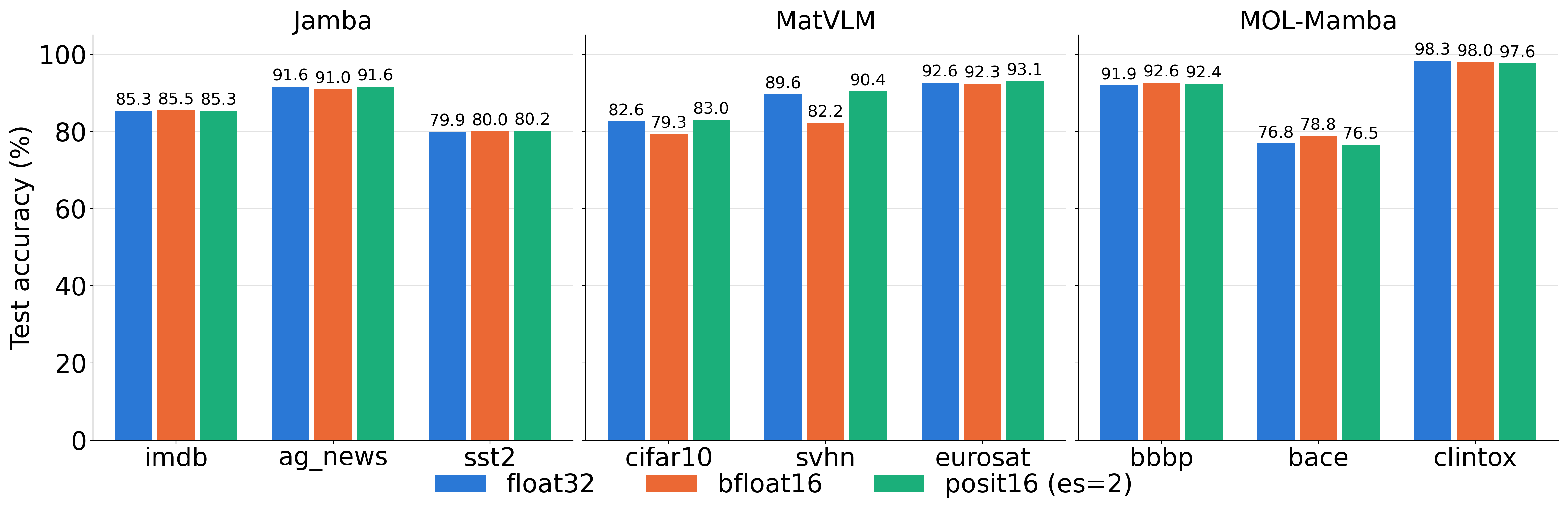}
    \caption{Inference accuracy of modified Mamba-3 hybrid models using FP32, BF16, and Posit16 ($es=2$) arithmetic across the evaluated datasets.}
    \label{fig:accuracy_mactypes}
\end{figure*}

\subsection{Experimental Methodology}

\textit{Workloads and model configurations.}
We evaluate \acceleratorname using three representative hybrid models: Jamba~\cite{Lenz2025Jamba}, a hybrid language model combining Mamba, Transformer attention, and mixture-of-experts layers; MaTVLM~\cite{Li2025MaTVLM}, a hybrid vision-language model; and MOL-Mamba~\cite{Hu2025MOLMamba}, a molecular representation learning model. 
For each hybrid model, we replace the original Mamba or Mamba-2 layers with Mamba-3 while preserving the remaining model components.
We select Mamba-3 configurations that achieve accuracy comparable to the original model, as demonstrated in Figure~\ref{fig:class_accuracy_v1_vs_v3_all_models}.
Table~\ref{tab:hyrid-model-params} summarizes the evaluated model configurations and datasets, including the recurrent state dimension $N$, MIMO rank $R$, number of heads $H$, and channels per head $P$. 
We additionally evaluate synthetic model configurations to characterize the effects of $N$, $R$, and sequence length independently.


\begin{table}[t]
\centering
\caption{Mamba-3 model dimensions used for each dataset across the three hybrid Mamba-3 models: sequence length ($L$), state map size ($N$), MIMO rank ($R$), batch size ($B$), number of heads ($H=8$), number of channels ($P=64$) and model dimension ($D=256$).}
\label{tab:hyrid-model-params}
\begin{tabular}{llllll}
\hline
\rowcolor[HTML]{C0C0C0} 
\textbf{Model} & \textbf{Dataset} & \textbf{L} & \textbf{N} & \textbf{R} & \textbf{B} \\ \hline
 & IMDB~\cite{maas2011imdb} & 256 & 128 & 4 & 32 \\ \cline{2-6} 
 & AG-News~\cite{zhang2015agnews} & 256 & 128 & 4 & 32 \\ \cline{2-6} 
\multirow{-3}{*}{Jamba} & SST2~\cite{socher2013sst} & 256 & 128 & 4 & 32 \\ \hline
 & CIFAR-10~\cite{krizhevsky2009cifar} & 64 & 64 & 1 & 64 \\ \cline{2-6} 
 & SVHN~\cite{netzer2011svhn} & 64 & 64 & 1 & 64 \\ \cline{2-6} 
\multirow{-3}{*}{MatVLM} & EuroSAT~\cite{helber2019eurosat} & 64 & 64 & 1 & 64 \\ \hline
 & BBBP~\cite{martins2012bbbp} & 200 & 64 & 1 & 32 \\ \cline{2-6} 
 & BACE~\cite{subramanian2016bace} & 200 & 64 & 1 & 32 \\ \cline{2-6} 
\multirow{-3}{*}{MOL-Mamba} & ClinTox~\cite{gayvert2016clintox} & 200 & 64 & 1 & 32  \\ \hline
\end{tabular}
\end{table}

\textit{Hardware characterization.}
We develop a parameterized RTL implementation of \acceleratorname in Verilog HDL and synthesize it using Cadence\textregistered{}  Genus\texttrademark{} with the FreePDK45~\cite{FreePDK45} 45-nm process and NanGate~\cite{NangateOpenCellLibrary} standard-cell library. 
The implementation includes the input parameter processing, MIMO mapping, rotation, exponential trapezoidal recurrent update, and output gating datapaths, together with registers for intermediate values and lookup tables for normalization and trigonometric functions.
We characterize the synthesized area, power, and timing across different hardware configurations and numerical representations. 
Area estimates are scaled to a 7-nm technology node~\cite{STILLMAKER} for comparison with the NVIDIA A100 GPU. 
These values are technology-scaled synthesis estimates rather than measurements from fabricated hardware.

\textit{GPU performance and energy modeling.}
We estimate the execution cycles of \acceleratorname for different model dimensions $(N,R)$ and workload dimensions $(L,B,H)$. 
To evaluate system-level performance, we integrate these execution estimates into Accel-Sim~\cite{AccelSim}, configured to model an NVIDIA A100 GPU. 
In the augmented GPU, \acceleratorname executes the intermediate Mamba-3 operations, while the dense input and output projections remain on conventional GPU execution resources. 
Non-Mamba operations, including attention and feed-forward computation, also continue to execute on the GPU.
We use AccelWattch~\cite{AccelWattch} to estimate system-level power and energy consumption, incorporating the synthesized power characteristics of \acceleratorname. 
The resulting performance and energy estimates account for the execution of both the specialized recurrent datapath and the remaining GPU workload.

\begin{figure}[!ht]
    \centering
    \includegraphics[width=0.46\textwidth]{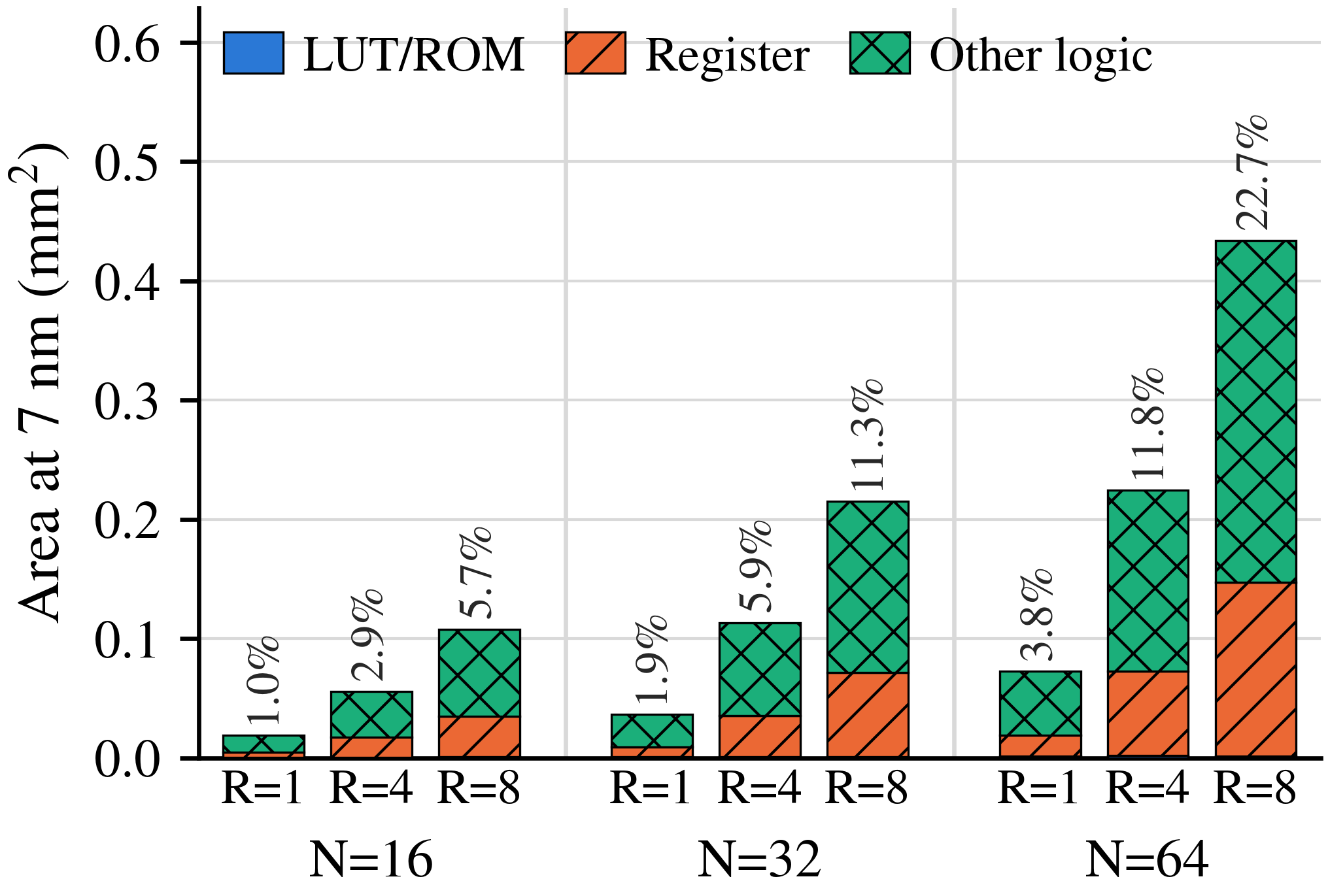}
    \caption{Area of \acceleratorname across \(N\) and \(R\) with BF16 datatype scaled to 7nm technology. Relative area increase when including a \acceleratorname in every SM sub-cluster of NVIDIA A100 GPU (total area \(826mm^{2}\)~\cite{nvidia2020a100}).}
    \label{fig:bf16_area}
\end{figure}

\begin{figure}[!ht]
    \centering
    \includegraphics[width=0.46\textwidth]{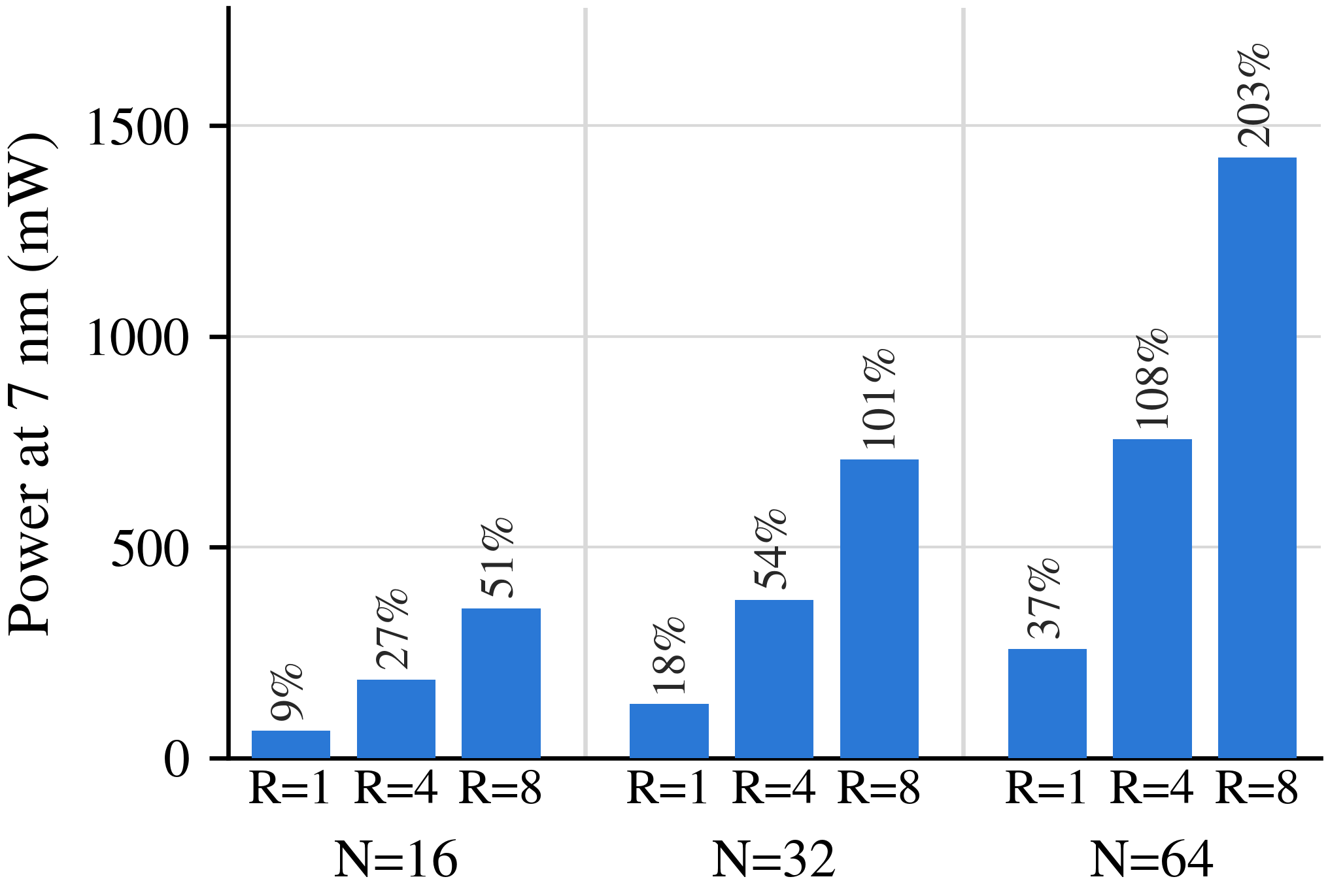}
    \caption{Power of \acceleratorname across \(N\) and \(R\) with BF16 datatype compared against power breakdown of NVIDIA SM sub-cluster A100 (\(700mW\)), B200 (\(385mW\)).}
    \label{fig:bf16_power}
\end{figure}

\subsection{Numerical Precision and Datapath Selection}

Numerical precision is particularly important for recurrent state-space models because rounding and quantization errors can propagate across successive state updates. 
At the same time, arithmetic representation affects the area, power, and critical-path delay of the replicated \acceleratorname datapath. 
We therefore evaluate both the hardware characteristics and inference accuracy of candidate numerical formats.

\textit{Arithmetic cost.}
We evaluate five numerical formats for \acceleratorname: IEEE 754 single-precision floating point (FP32)~\cite{ieee754_2019}, TensorFloat-32 (TF32) a hybrid 19-bit format from NVIDIA~\cite{nvidiaa100} for Tensor Core, Brain Floating Point (BF16)~\cite{kalamkar2019bfloat16}, 16-bit Posit (Posit-16)~\cite{Gustafson2017Posit} and 8-bit integer type (INT8). 
Using BaseJump STL~\cite{taylor2018basejump}, we implement combinational and pipelined multiply-accumulate (MAC) primitives in Verilog HDL and synthesize them with Cadence\textregistered{} Genus\texttrademark{} using the FreePDK45~\cite{FreePDK45} process and NanGate~\cite{NangateOpenCellLibrary} standard-cell library. 
Figure~\ref{fig:mac_comb_vs_pipelined} compares their post-synthesis area, power, and maximum operating frequency ($f_{\mathrm{max}}$).

Among the evaluated floating-point-class combinational MACs, BF16 provides the lowest area and power. The BF16 MAC occupies $1,521\mu\mathrm{m}^{2}$ and consumes $3,078\mu\mathrm{W}$, compared with $2,042\mu\mathrm{m}^{2}$ and $5,689\mu\mathrm{W}$ for TF32, $5,820\mu\mathrm{m}^{2}$ and $27,017\mu\mathrm{W}$ for FP32, and $4,338\mu\mathrm{m}^{2}$ and $37,746\mu\mathrm{W}$ for Posit16. 
BF16 also achieves the highest combinational operating frequency among these floating-point-class implementations at $338$MHz, compared with $321$MHz for TF32, $264$MHz for FP32, and $183$MHz for Posit16.

INT8 provides the lowest arithmetic cost, with an area of $479\mu\mathrm{m}^{2}$, power of $429\mu\mathrm{W}$, and maximum combinational frequency of $696$MHz. 
However, its limited dynamic range and integer representation make it unsuitable as a direct replacement for the floating-point recurrent datapath without additional quantization mechanisms. 
We therefore use INT8 as a hardware reference rather than selecting it for \acceleratorname.

\textit{Effect of arithmetic pipelining.}
Pipelining improves the maximum operating frequency of the MAC primitives but increases their area. 
For BF16, the MAC area increases from $1,521$ to $2,579\mu\mathrm{m}^{2}$, corresponding to approximately $+70\%$ additional area. FP32 increases from $5,820$ to $7,112\mu\mathrm{m}^{2}$, or approximately $+22\%$. 
Pipelining also reduces the estimated power of most evaluated floating-point formats, although the additional registers introduce an area cost.

Because \acceleratorname replicates arithmetic operators throughout its datapath, the cost of individual MAC primitives directly affects the complete unit. 
We therefore select combinational BF16 arithmetic primitives and pipeline the overall \acceleratorname datapath at the functional-unit level, avoiding the additional register overhead of individually pipelining every MAC.

\textit{Inference accuracy.}
Figure~\ref{fig:accuracy_mactypes} compares the accuracy of Mamba-3 hybrid models using FP32, BF16, and Posit16 with exponent size $es=2$. 
The results show that BF16 maintains accuracy close to the higher-precision baseline across the evaluated workloads, supporting its use in the recurrent datapath.

This experiment complements the Mamba-3 numerical-sensitivity analysis in Section~\ref{BACKG_POSIT}. 
The two studies use different model configurations and Posit exponent settings; consequently, their accuracy results are considered separately.

\subsection{Impact of Model Dimensions on Hardware Cost}

We next develop a parameterized Verilog HDL implementation of \acceleratorname to explore the hardware design space across different state-map sizes, $N$, and MIMO ranks, $R$. 
The implementation includes all components of the \acceleratorname microarchitecture shown in Figure~\ref{fig:rescucore_micro_arch}. 
The design incorporates registers for storing intermediate values and lookup tables (LUTs) for evaluating RMSNorm and trigonometric functions, including $\cos(\Phi)$ and $\sin(\Phi)$.

We synthesize parameterized \acceleratorname implementations across multiple combinations of $N$ and $R$. Figure~\ref{fig:bf16_area} compares the resulting hardware area, scaled to $7nm$, and expresses the overhead relative to the NVIDIA A100 die area of $826,\mathrm{mm}^{2}$. 
Figure~\ref{fig:bf16_power} presents the corresponding power estimates and compares them with the reference power budgets of GPU SM subpartitions.

Both figures illustrate the cost of increasing the dimensions supported by a single \acceleratorname unit.
A larger native configuration can process more recurrent-state elements and MIMO contributions without partitioning the computation, but requires additional arithmetic logic, intermediate storage, and associated circuitry. 
Conversely, a smaller implementation reduces the cost of GPU-wide replication at the expense of additional execution when model dimensions exceed the hardware capacity.

These trends motivate a fixed-size \acceleratorname design that balances native execution capability against the area and power overhead of integrating specialized units throughout the GPU. 

\textit{Selected \acceleratorname configuration.}
Combining the model-dimension hardware study with the numerical-format evaluation, we select a fixed \acceleratorname implementation supporting state dimension $N'=32$, MIMO rank $R'=4$, and BF16 arithmetic. 
This design provides a compact native datapath while supporting larger model configurations through the execution mapping described in the next subsection.

The selected \acceleratorname occupies an estimated $0.11\mathrm{mm}^{2}$ after scaling to $7nm$. Registers account for $29.6\%$ of the synthesized area, combinational logic for $66.7\%$, and lookup tables for $3.7\%$. 
Replicating this design across the A100 SM subpartitions incurs an estimated $+5.9\%$ die-area overhead. 
We use this same fixed configuration for all subsequent system-level evaluations.

\begin{figure}[t]
    \centering
    \includegraphics[width=0.48\textwidth]{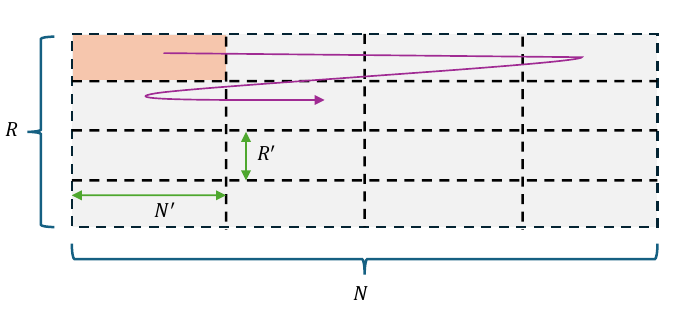}
    \caption{Scale-up mapping using techniques from~\cite{ScaleSim} to evaluate models of dimensions $(N,R)$ on a fixed-dimension \acceleratorname accelerator $(N',R')$.}
    \label{fig:scale_up}
\end{figure}

\begin{figure}[t]
    \centering
    \includegraphics[width=0.45\textwidth]{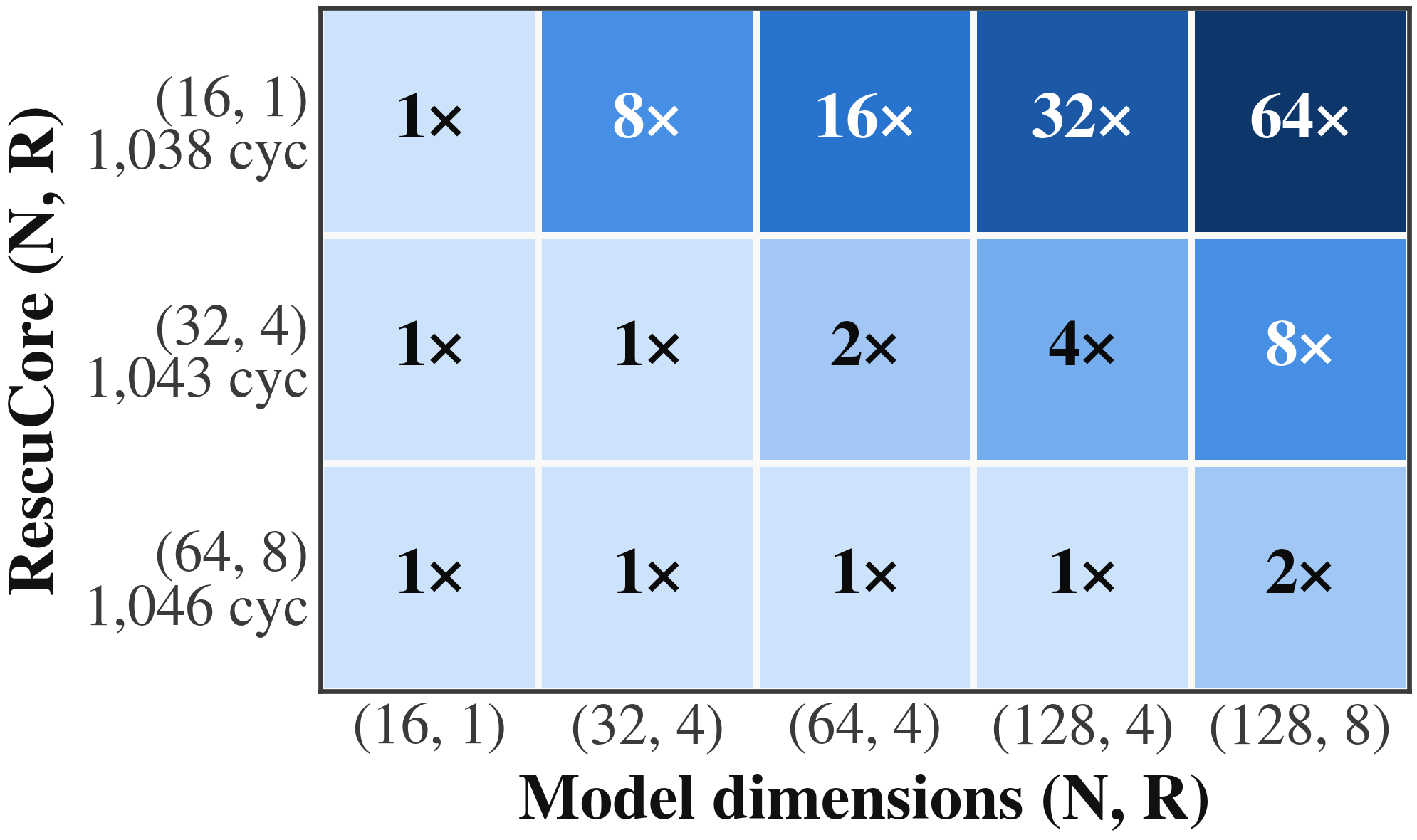}
    \caption{Inference cycles to decode L=1024 tokens across various \acceleratorname dimensions and model dimensions showing the impact of scale-up. For example, selecting $N^{'}=32,R^{'}=4$ \acceleratorname dimension will cause a \(4\times\) latency increase to evaluate model with $N=128,R=4$ using a single \acceleratorname or require four \acceleratorname units operating in parallel.}
    \label{fig:scale_out_heatmap}
\end{figure}

\textit{Register storage requirements.}

\begin{figure}[h]
    \centering
    \includegraphics[width=0.45\textwidth]{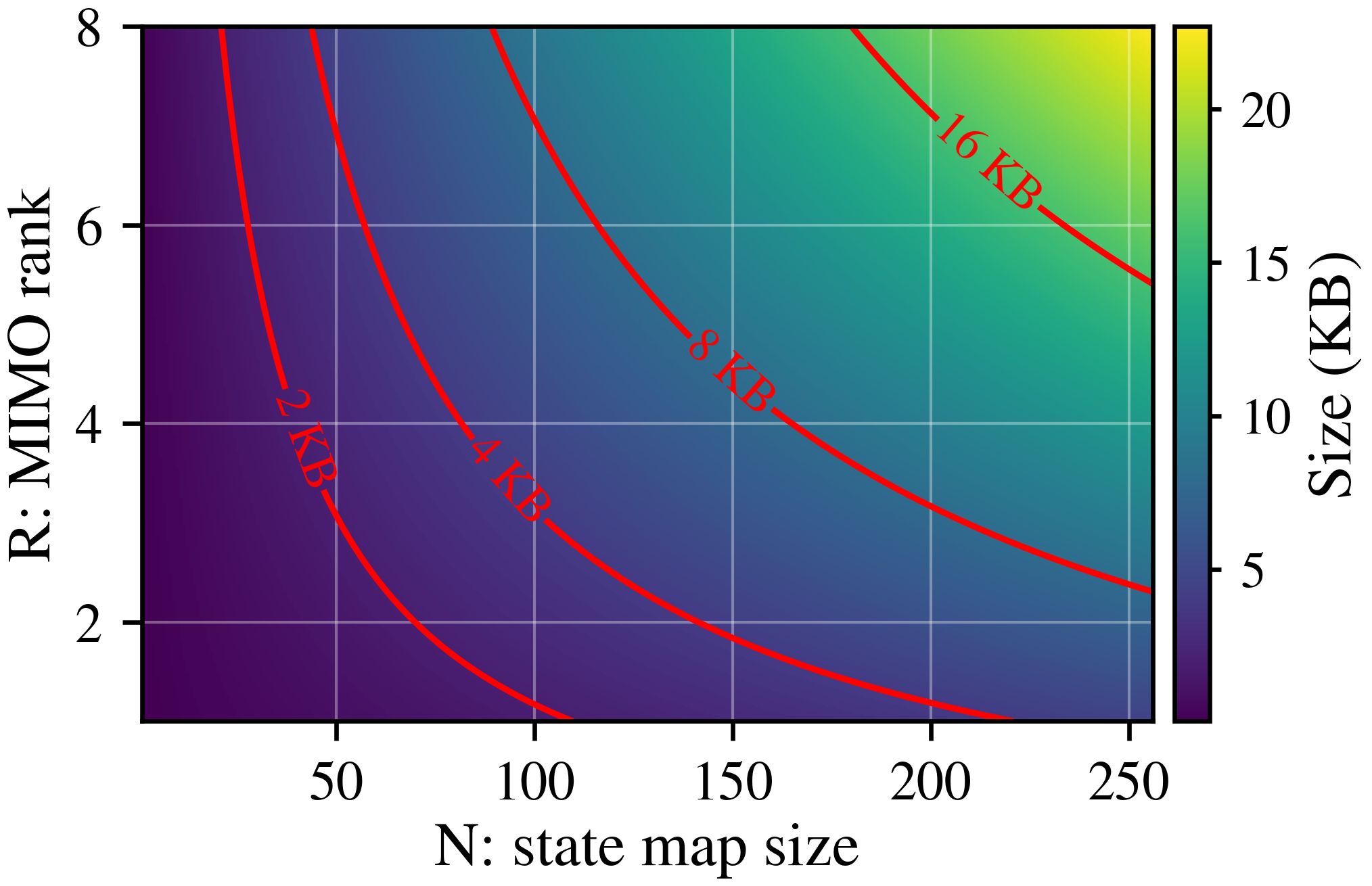}
    \caption{Total register file size per \acceleratorname with 16-bit data, across varying state map size, N, and MIMO-rank, R.}
    \label{fig:register_heamap}
\end{figure}

In addition to arithmetic logic, the storage required for recurrent states and intermediate values contributes to the hardware cost of \acceleratorname.
Figure~\ref{fig:register_heamap} illustrates how the register storage required by a single \acceleratorname unit scales with the state dimension $N$ and MIMO rank $R$, assuming 16-bit data.
The storage comprises two components: $4R$ values for the per-channel MIMO weights associated with $B$, $C$, $X$, and $Z$, and $5NR+3R+4N+4$ values for the recurrent state and intermediate results retained during pipeline execution.
The total storage requirement is therefore
\begin{equation}
S(N,R)=2(5NR+7R+4N+4)\ \text{bytes}.
\end{equation}
The dominant $NR$ term causes storage to increase approximately proportionally to the product of the state dimension and MIMO rank, resulting in approximately hyperbolic iso-storage contours in Figure~\ref{fig:register_heamap}.

Importantly, the storage requirement is independent of sequence length and batch size, since each \acceleratorname unit maintains only the state and intermediate values associated with the channel currently being processed, rather than a context-dependent cache.
Consequently, the register capacity is determined at design time and does not grow with the inference context length.

For the evaluated configurations, the required storage remains modest.
The $N=32$, $R=2$ timing configuration requires approximately $0.91$\,KiB, while the selected $N=32$, $R=4$ implementation requires $1.56$\,KiB per unit.
The $N=64$, $R=1$ configuration requires $1.15$\,KiB, and the larger $N=128$, $R=4$ configuration requires $6.06$\,KiB.
Even at $N=256$ and $R=8$, the estimated storage requirement is only $22.12$\,KiB per unit.

With four \acceleratorname units per SM, the selected $N=32$, $R=4$ configuration requires approximately $6.25$\,KiB of aggregate local register storage per SM, equivalent to $2.44\%$ of the A100 SM's 256\,KiB register-file capacity.
This comparison provides a reference for the magnitude of the added storage; the registers are implemented within \acceleratorname rather than allocated from the GPU's existing register file.
Overall, these results indicate that local storage requirements remain manageable across the evaluated design space, with arithmetic logic and the cost of GPU-wide replication representing the primary considerations when selecting the native \acceleratorname dimensions.

\subsection{Mapping Larger Models onto Fixed-Size \acceleratorname}\label{EVAL_SCALE_UP}

Although the selected \acceleratorname natively supports $N'=32$ and $R'=4$, the hybrid models in Table~\ref{tab:hyrid-model-params} include larger recurrent-state dimensions. 
We therefore evaluate how a fixed hardware configuration can execute models with dimensions $(N,R)$ exceeding its native capacity.

Figure~\ref{fig:scale_up} illustrates the scale-up mapping. 
The model computation is partitioned into operations supported by the fixed \acceleratorname dimensions $(N',R')$. 
When the model requires more state elements or MIMO contributions than a single unit can process natively, the additional work is executed through repeated passes or distributed across additional parallel resources.

Figure~\ref{fig:scale_out_heatmap} evaluates the effect of this mapping on the execution cycles required to decode a sequence of $L=1024$ tokens. 
For example, executing a model with $N=128$ and $R=4$ on a \acceleratorname supporting $N'=32$ and $R'=4$ requires four state-dimension partitions. 
Under the idealized sequential mapping, the corresponding computation incurs approximately four times the latency of a native implementation. 
Alternatively, additional parallel hardware can reduce the number of sequential passes at increased area and power cost.

The resulting trade-off is central to the fixed-size design. 
Provisioning a larger \acceleratorname reduces the number of execution passes required for large models but increases the cost of every unit replicated throughout the GPU. 
A smaller implementation limits that replicated hardware overhead while accommodating larger model dimensions through additional execution.

We retain the selected $N'=32$, $R'=4$ configuration for the full-system experiments and account for the additional execution required by each workload. 
This ensures that the evaluation reflects a common physical hardware design rather than separately optimized \acceleratorname implementations for different hybrid models.

\begin{figure}[t]
    \centering
    \includegraphics[width=\linewidth]{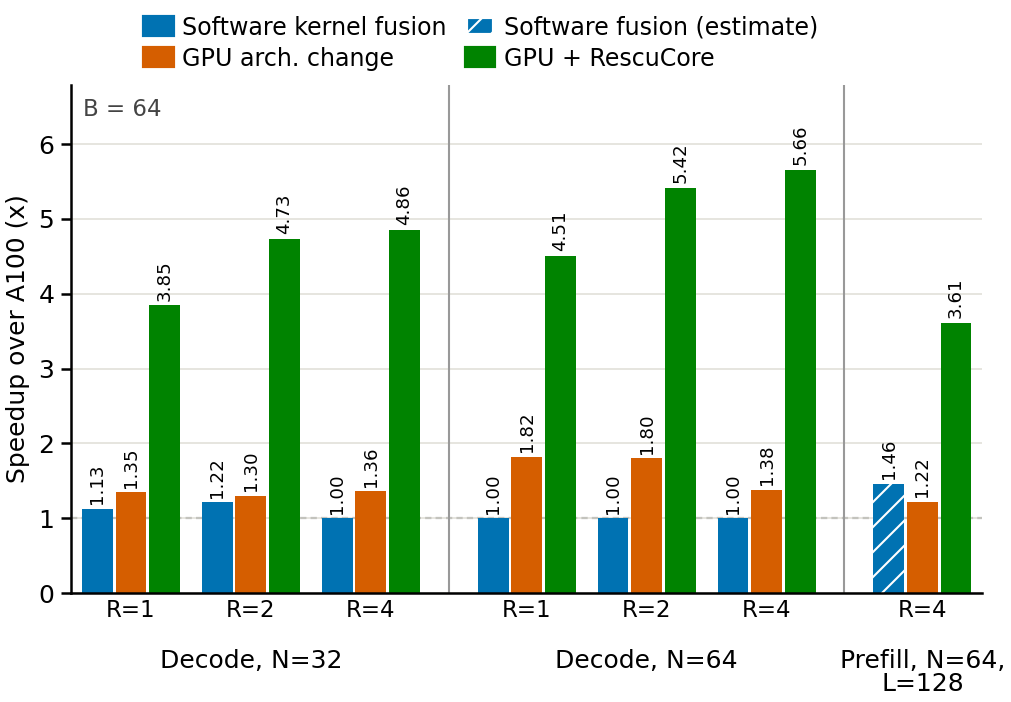}
    \caption{Mamba-3 layer speedup over the released A100 implementation at $B=64$, comparing software fusion, the best evaluated conventional GPU architectural modification, and GPU+\acceleratorname.
    Decode covers $N=32,64$ and $R=1,2,4$; prefill uses $N=64$, $R=4$, $L=128$.
    The software-fusion prefill result is estimated.
    \acceleratorname results combine measured GPU projection time with modeled specialized execution.}
    \label{fig:speedup_sw_arc_rescucore}
\end{figure}

\subsection{Comparison with GPU alternatives.}
At $B=64$, software fusion achieves at most $1.22\times$ decoding speedup, and conventional GPU architectural enhancements achieve $1.30$--$1.82\times$ (Fig.~\ref{fig:speedup_sw_arc_rescucore}).
In contrast, \acceleratorname achieves $3.85$--$5.66\times$ decoding and $3.61\times$ prefill speedup, outperforming the best alternatives by $2.5$--$4.1\times$.
These gains stem from specialized arithmetic, local state retention, and operand forwarding rather than kernel fusion or additional conventional GPU resources.

\begin{figure}[t]
    \centering
    \includegraphics[width=0.5\textwidth]{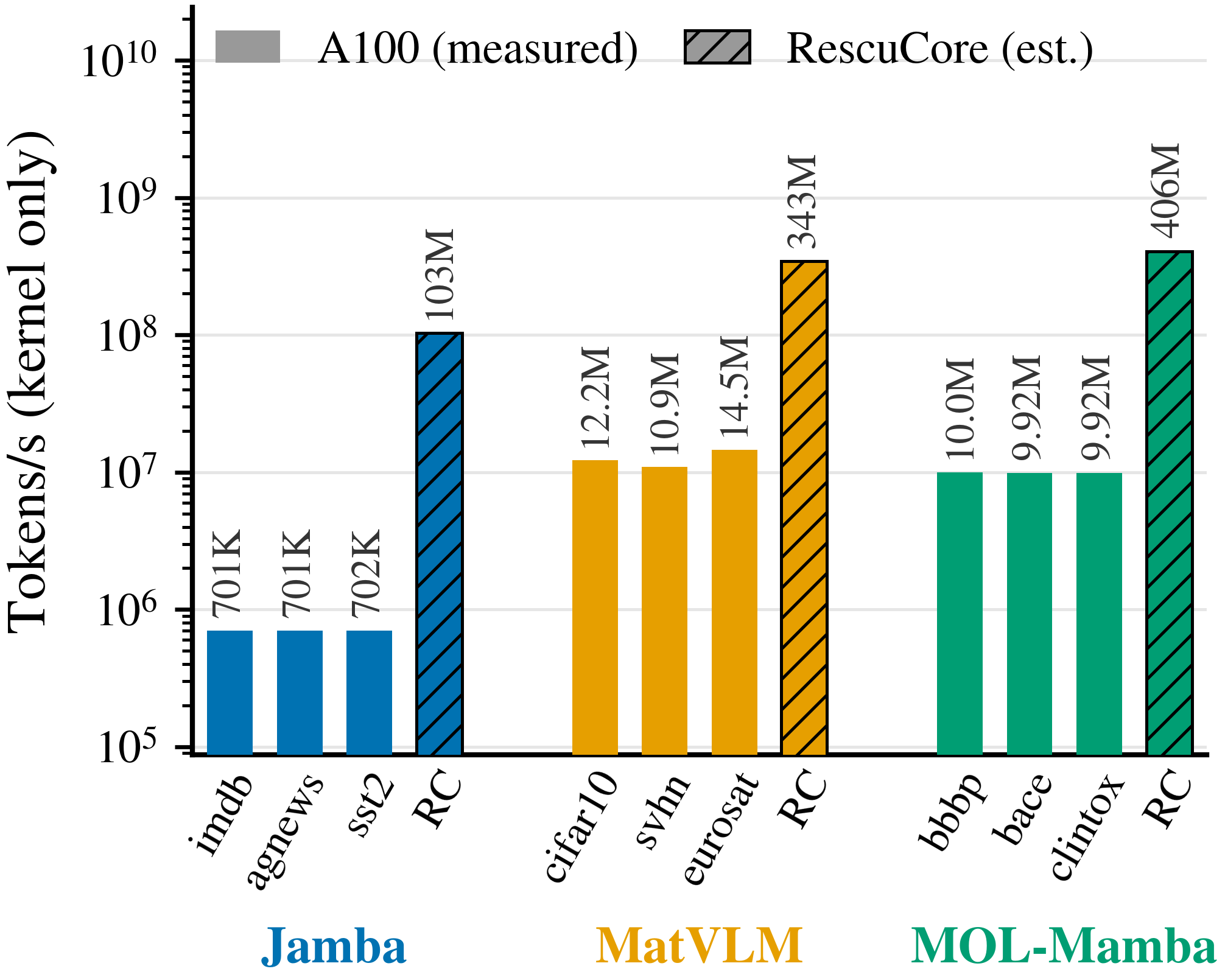}
    \caption{Mamba v3 kernel throughput comparison between NVIDIA A100 GPU with and without \acceleratorname using BF16 across various hybrid models. The selected \acceleratorname configuration uses \(N=32, R=4\) and scaled-out to support the dataset in the model.}
    \label{fig:tps_a100_vs_rc}
\end{figure}

\subsection{Full-System Performance and Energy Efficiency}

Having established the hardware cost, numerical representation, and scale-up behavior of the selected \acceleratorname, we now evaluate its impact on complete hybrid-model inference. 
All experiments use the same BF16 \acceleratorname configuration with $N'=32$ and $R'=4$, integrated at one unit per GPU SM subpartition. 
The evaluation includes the execution overhead required to support model dimensions larger than the native hardware configuration.

\textit{Mamba-3 execution throughput.}
Figure~\ref{fig:tps_a100_vs_rc} compares Mamba-3 execution throughput on the conventional NVIDIA A100 GPU and the \acceleratorname-augmented GPU across the evaluated hybrid models.
\acceleratorname accelerates the intermediate Mamba-3 computation, while the dense input and output projections continue to execute on conventional GPU resources.
All evaluations use the same fixed-size \acceleratorname configuration with $N'=32$ and $R'=4$.
For hybrid models whose state dimension $N$ or MIMO rank $R$ exceeds the native hardware capacity, we account for the additional execution cycles incurred by the scale-up mapping described in Section~\ref{EVAL_SCALE_UP}.
Thus, the reported throughput incorporates the cost of executing larger Mamba-3 configurations on the fixed-size datapath, rather than assuming a separately sized \acceleratorname for each model.
This enables a consistent comparison of the performance benefits delivered by a single practical hardware configuration across different hybrid workloads.

\begin{figure}[t]
    \centering
    \includegraphics[width=0.5\textwidth]{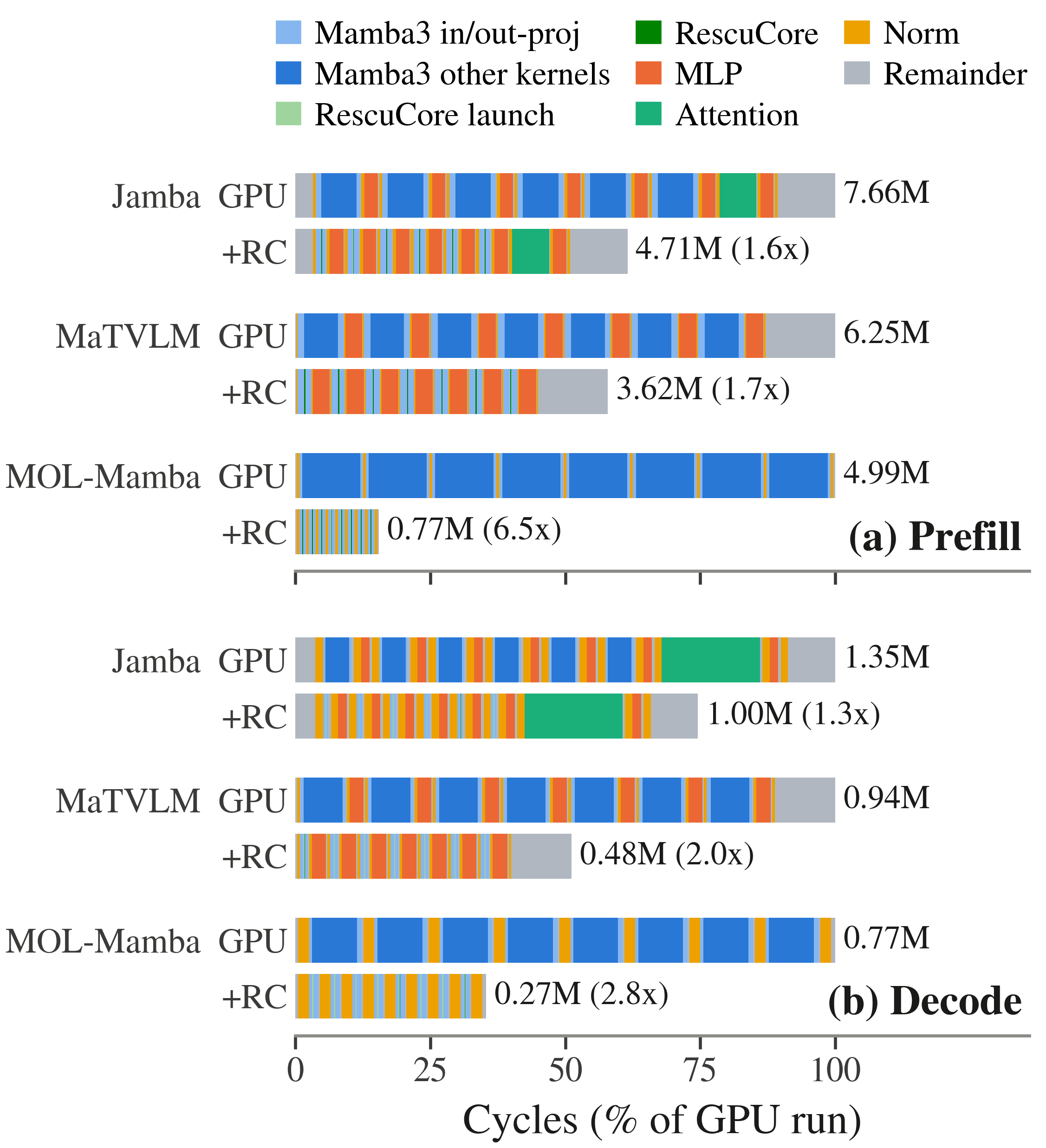}
    \caption{Accel-Sim GPU-cycle timeline with \acceleratorname. Replacing the Mamba v3 mixer with \acceleratorname's dedicated pipeline reduces the mixer's share of prefill runtime from $40-83\%$ to less than $5\%$ across the Jamba, MatVLM, and MOL-Mamba hybrid models.}
    \label{fig:hybrid_prefill_cycle_rescuore}
\end{figure}

\textit{Full-model performance.}
We next evaluate Jamba, MaTVLM, and MOL-Mamba to determine how accelerating Mamba-3 affects complete hybrid-model execution. These workloads contain different proportions of recurrent and non-recurrent computation, providing distinct opportunities for acceleration.

Across the evaluated configurations, the \acceleratorname-augmented GPU achieves reported prefill speedups of $1.4\times$--$9.4\times$ and autoregressive decoding speedups of $1.3\times$--$2.8\times$ relative to conventional GPU execution. 
The magnitude of improvement depends on the original fraction of execution spent in Mamba-3, the available recurrent parallelism, and the remaining GPU computation.

Figure~\ref{fig:hybrid_prefill_cycle_rescuore} presents the simulated GPU execution timelines before and after integrating \acceleratorname. In the characterized prefill workloads, Mamba-3 originally accounts for approximately $40\%$ of GPU execution cycles in Jamba, $54\%$ in MaTVLM, and $83\%$ in MOL-Mamba. With \acceleratorname, the Mamba-3 mixer's share of the modeled execution timeline decreases to below $5\%$.

This reduction reflects the substantially shorter execution interval of the specialized recurrent computation. 
However, the change in the mixer's runtime fraction is distinct from the full-model speedup, which depends on the total execution time of both recurrent and non-recurrent components. 
As Mamba-3 execution becomes faster, dense projections, attention, and other hybrid-model operations account for a larger fraction of the remaining execution time.

\begin{figure}[t]
    \centering
    \includegraphics[width=0.5\textwidth]{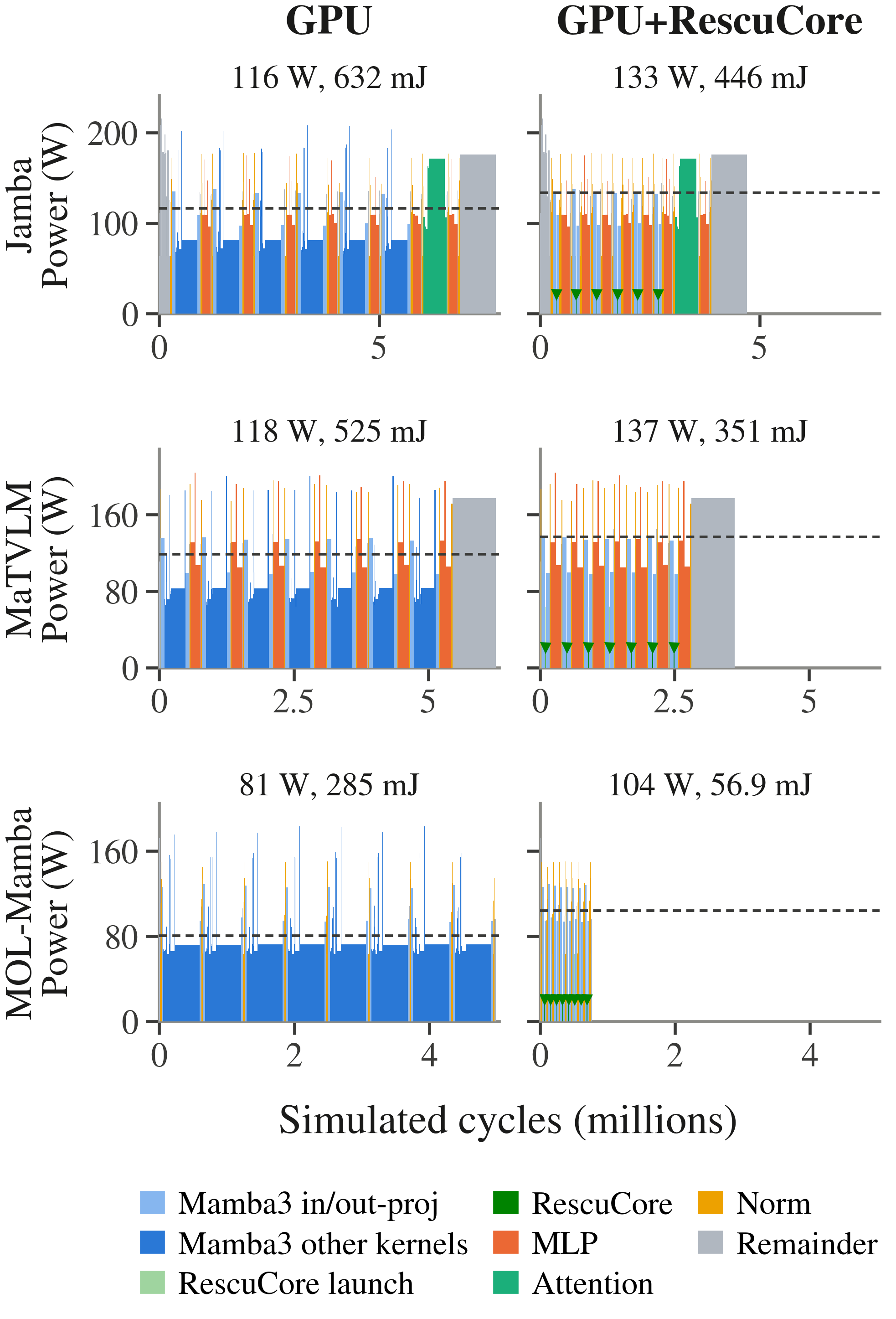}
    \caption{Prefill AccelWattch power timeline with \acceleratorname. Offloading the Mamba v3 mixer to \acceleratorname confines mixer execution to a short, low-power interval between GPU-executed layers.}
    \label{fig:hybrid_prefill_power_rescuore}
\end{figure}

\begin{figure}[t]
    \centering
    \includegraphics[width=0.5\textwidth]{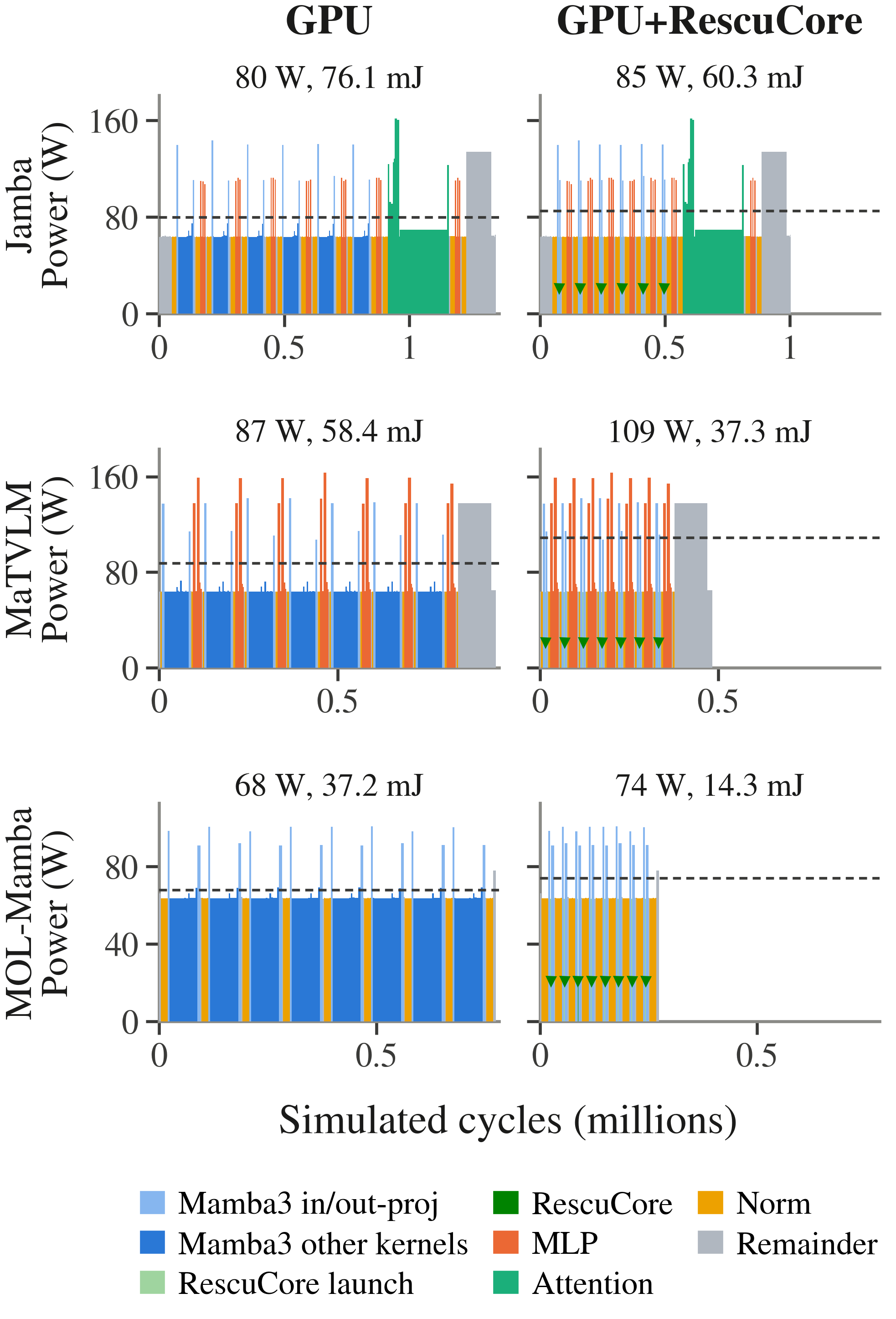}
    \caption{Decode AccelWattch power timeline with \acceleratorname. Offloading the Mamba v3 mixer to \acceleratorname confines mixer execution to a short, low-power interval between GPU-executed layers.}
    \label{fig:hybrid_decode_power_rescuore}
\end{figure}

\textit{Energy efficiency.}
We evaluate energy efficiency using AccelWattch augmented with the synthesized power characteristics of \acceleratorname. The reported energy reductions range from $1.4\times$ to $5.0\times$ during prefill and from $1.2\times$ to $2.6\times$ during decoding.

Figure~\ref{fig:hybrid_prefill_power_rescuore} and ~\ref{fig:hybrid_decode_power_rescuore} illustrate the simulated power timeline of a representative hybrid-model execution for prefill and decode operations.
Compared with conventional GPU execution, the \acceleratorname-augmented configuration confines the recurrent computation to a shorter execution interval, while the surrounding model components continue to execute on conventional GPU resources.

The energy savings arise primarily from reduced recurrent execution time and the specialized datapath's local handling of intermediate values. 
Although the additional \acceleratorname units contribute to GPU power consumption, the reduction in execution duration lowers the estimated energy required by the evaluated workloads.

Overall, the results demonstrate that the fixed BF16 \acceleratorname configuration can accelerate Mamba-3 inference across hybrid models with different recurrent dimensions while maintaining a common GPU-integrated hardware design. 
The hardware characterization establishes the cost of that design, the scale-up analysis accounts for larger model configurations, and the full-system evaluation quantifies the resulting performance and energy benefits.

\begin{figure}[t]
    \centering
    \includegraphics[width=0.5\textwidth]{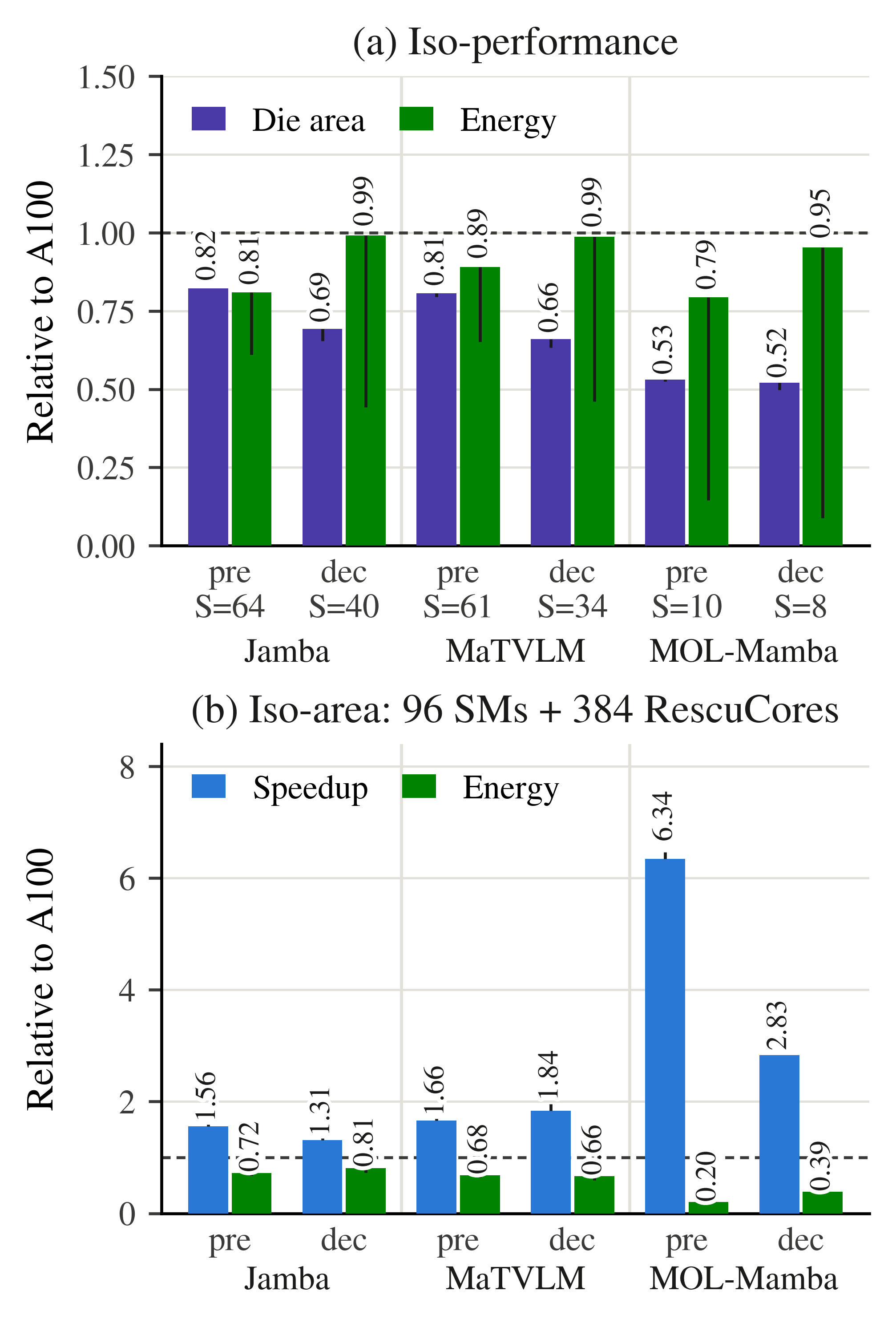}
    \caption{GPU + \acceleratorname vs a plain 108-SM A100 at \(L=2,048\). (a) Iso-performance: Fewest SMs matching the plain GPU's runtime, and the resulting area and energy. (b) Iso-area: Speedup and energy with 96 SMs + 384 \acceleratorname's. Bars: conservative models; black lines: range to the optimistic model}
    \label{fig:rescucore_iso_analysis}
\end{figure}

\subsection{Iso-Performance and Iso-Area Analysis}

To separate \acceleratorname's benefit from simply adding silicon, we compare GPU + \acceleratorname against the plain 108-SM A100 under two equal-cost constraints at $L=2,048$, as shown in Figure~\ref{fig:rescucore_iso_analysis}. 
At iso-performance, offloading the Mamba-3 recurrence lets the GPU match the plain A100's runtime with far fewer SMs: 64 and 40 SMs for Jamba prefill and decode, 61 and 34 for MaTVLM, and only 10 and 8 for MOL-Mamba. 
Including the \acceleratorname array, this reduces die area by \(18-31\%\) for Jamba and MaTVLM and by \(47-48\%\) for MOL-Mamba. 
Decode consistently needs fewer SMs than prefill, because the decode layer on the GPU is a long chain of small, latency-bound kernels, and once \acceleratorname absorbs that chain the remaining GPU work (mostly the projections) needs few SMs. 
Energy at iso-performance is \(0.79-0.99\times\) the plain GPU under our conservative assumption that static power does not shrink with the SM count, and falls to \(0.09-0.65\times\) if static power scales with the SMs removed; the gap between the two bounds is almost entirely this static-power assumption. 
At iso-area, trading 12 SMs for 384 \acceleratorname units (96 SMs + 384 \acceleratorname's in the plain GPU's SM-array area) speeds up Jamba by \(1.56\times\) (prefill) and \(1.31\times\) (decode), MaTVLM by \(1.66\times\) and \(1.84\times\), and MOL-Mamba by \(6.34\times\) and \(2.83\times\), while using \(0.20-0.81\times\) the energy. 
Here the two model bounds nearly coincide, since removing 12 of 108 SMs barely changes the GPU's share of the work. 
The benefit in both analyses tracks how much of each model's runtime is spent in Mamba-3 layers (\(40\%\), \(54\%\) and \(83\%\) of prefill time for Jamba, MaTVLM and MOL-Mamba): in this Amdahl-style trend, the more the workload is dominated by the recurrence, the more GPU area \acceleratorname can replace or the more speedup it delivers for the same area. 
These results come from an analytic model built on Accel-Sim and AccelWattch traces with synthesized \acceleratorname area and power; it does not model per-SM memory bandwidth limits, and the SM area is an order-of-magnitude estimate, so the iso-performance points at very small SM counts (MOL-Mamba) should be read as optimistic.

\begin{figure*}[t]
    \centering
    \includegraphics[width=0.85\textwidth]{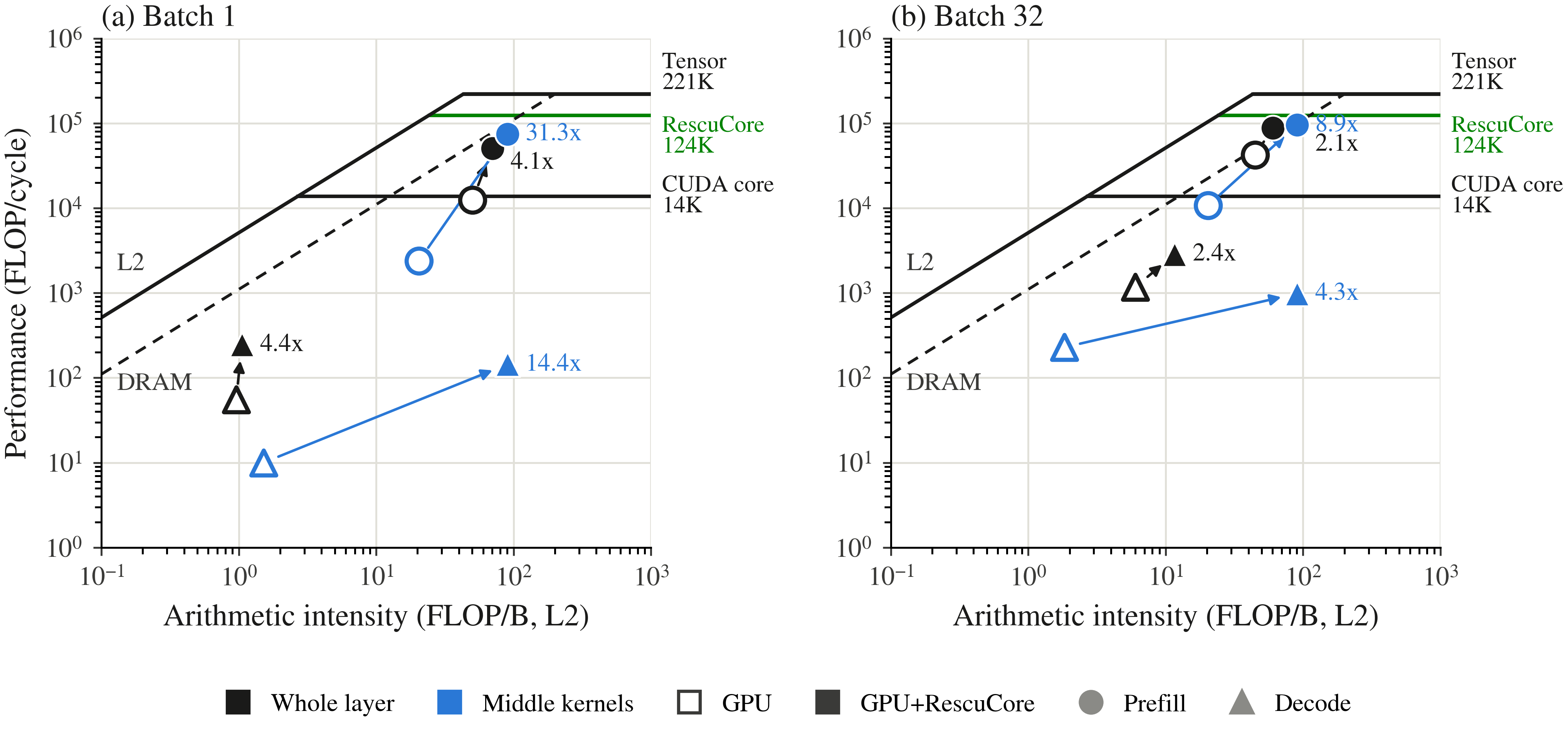}
    \caption{Roofline analysis of a Mamba-3 layer on an NVIDIA A100 (Accel-Sim) with GPU-only (hollow) and \acceleratorname augmented execution (filled, four units per SM) at (a) $B=1$ and (b) $B=32$. Model parameters are $d_{\mathrm{model}}=512$, 16 heads of 64 channels, $N=32$, MIMO rank $R=2$, chunk size 32, and BF16 precision. Prefill and decode use a sequence length and context length of 2,048 tokens, respectively. \acceleratorname replaces all middle kernels between \texttt{in\_proj} and \texttt{out\_proj} with a single kernel. In (a), \acceleratorname accelerates the latency-bound middle kernels by $31.3\times$/$14.4\times$ (prefill/decode). In (b), improved GPU utilization and near-linear \acceleratorname cycle scaling with batch size reduce these speedups to $8.9\times$/$4.3\times$. End-to-end layer speedups are $4.1\times$/$4.4\times$ in (a) and $2.1\times$/$2.4\times$ in (b).}
    \label{fig:roofline_gpu_vs_rescucore}
\end{figure*}

\subsection{Roofline Analysis}

Figure~\ref{fig:roofline_gpu_vs_rescucore} presents the roofline analysis of a Mamba-3 layer ($L=2{,}048$, $N=32$, $R=2$) on an A100 at (a) $B=1$ and (b) $B=32$, before (hollow) and after (filled) recurrence offloading to \acceleratorname{}.
Both configurations perform identical useful operations, with throughput measured in useful FLOPs per cycle.

\paragraph{Batch size 1}
The GPU middle kernels are latency-bound, achieving 2.4K~FLOP/cycle at 20.5~FLOP/B in prefill (17\% of the CUDA-core peak) and only 10~FLOP/cycle at 1.5~FLOP/B in decode.
This underutilization stems from limited thread-block parallelism in prefill (16 blocks) and sequential kernel dependencies and launch overheads in decode.
By retaining intermediate values within its pipeline, \acceleratorname{} increases arithmetic intensity to 90~FLOP/B and achieves 75K~FLOP/cycle in prefill (60\% of peak), yielding a $31.3\times$ speedup.
In decode, \acceleratorname{} achieves a $14.4\times$ speedup, despite kernel launch (1,800 cycles) and pipeline fill ($\sim$2,280 cycles) dominating its execution time.
These improvements translate into end-to-end layer speedups of $4.1\times$ and $4.4\times$ in prefill and decode, respectively.

\paragraph{Batch size 32}
Batching improves GPU utilization by increasing prefill thread-block parallelism to 512 blocks, raising throughput to 10.8K~FLOP/cycle (78\% of the CUDA-core peak).
Consequently, a $32\times$ increase in workload incurs only $7.1\times$ and $1.4\times$ increases in GPU cycles for prefill and decode, respectively.
In contrast, \acceleratorname{} already exploits channel-level parallelism at $B=1$, distributing 1,024 channels across 432 units and processing sequences sequentially within each channel.
Its execution time therefore scales nearly linearly with batch size.
Prefill throughput reaches 95.5K~FLOP/cycle (77\% of peak, compared to a mapping-imposed utilization ceiling of 79\%), yielding an $8.9\times$ speedup.
In decode, state save/restore dominates execution (80\% of cycles), requiring 160 cycles per sequence compared to two computation cycles and limiting the speedup to $4.3\times$.
The resulting end-to-end layer speedups are $2.1\times$ in prefill and $2.4\times$ in decode.

\begin{figure}[t]
\centering
\includegraphics[width=0.97\columnwidth]{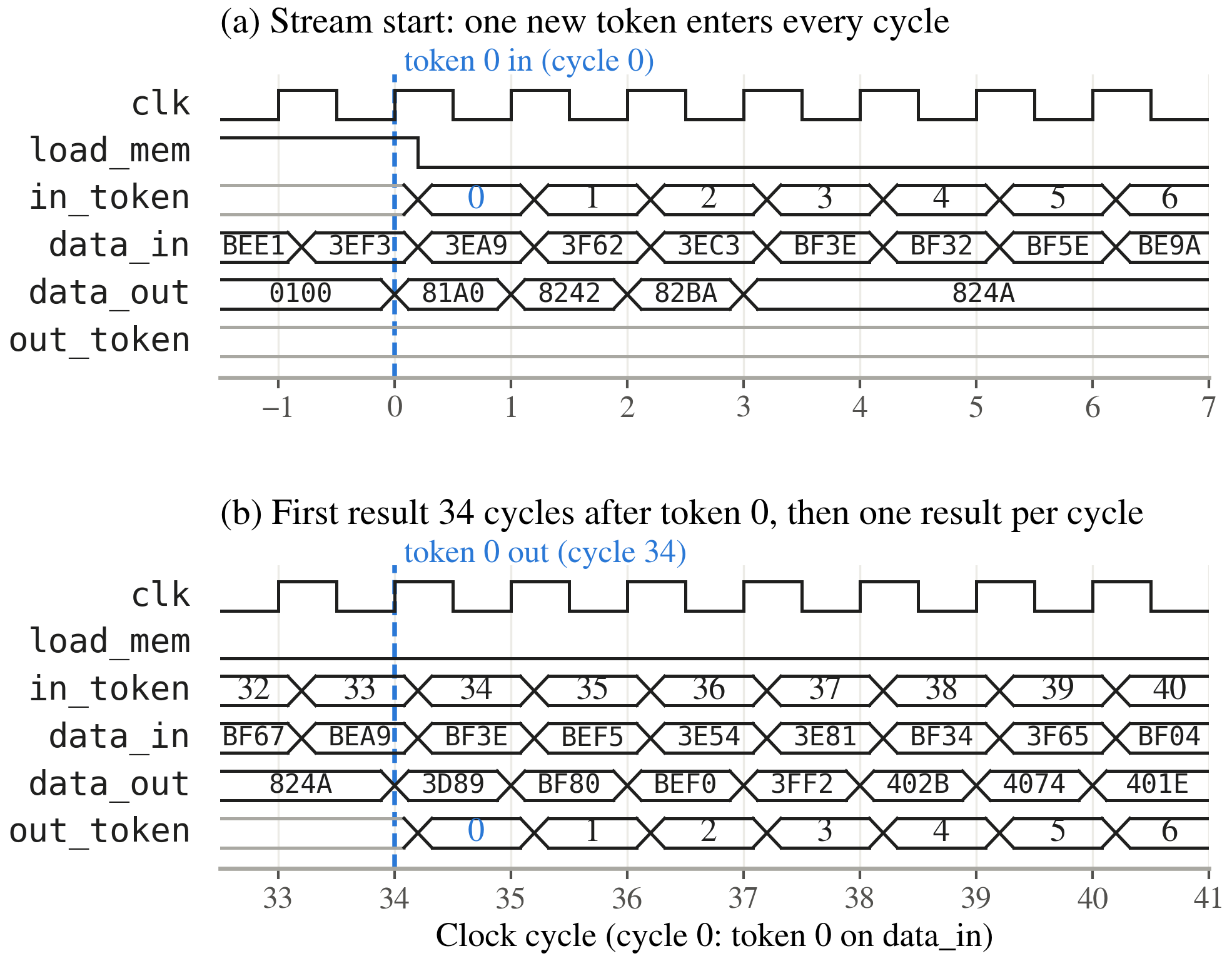}
\caption{Xcelium RTL waveform of one \acceleratorname{} unit ($N=32$, $R=4$, BF16).
(a)~After weight loading (\texttt{load\_mem}=0), the unit accepts one token per cycle.
(b)~The first valid result appears 34 cycles after token~0 enters, followed by one result per cycle, confirming $\mathrm{II}=1$.
Token indices on \texttt{data\_in} and \texttt{data\_out} are indicated by \texttt{in\_token} and \texttt{out\_token}, respectively.}
\label{fig:ii1_wave}
\end{figure}

\subsection{RTL Verification of Initiation Interval}
\label{sec:eval_ii1}

We validate the modeled pipeline depth (Eq.~\eqref{eq:rescucore_depth}) and initiation interval ($\mathrm{II}=1$) through cycle-accurate simulation of synthesizable \acceleratorname{} RTL using Cadence\textregistered Xcelium\texttrademark~\cite{cadence_xcelium}.
After loading random weights, we stream 400 BF16 tokens at one token per cycle.
Latency is measured by perturbing a single token in an identical RTL instance and identifying the first differing output.

For $(N,R)=(32,4)$, the measured latency is 34 cycles across tested token positions, with one output per cycle after pipeline fill (Figure~\ref{fig:ii1_wave}).
Configurations $(16,4)$, $(32,2)$, and $(64,4)$ yield 32, 31, and 36 cycles, respectively, matching Eq.~\eqref{eq:rescucore_depth}.
These results validate the modeled pipeline depth and sustained $\mathrm{II}=1$, with pipeline-fill latency incurred once per channel pass.

\begin{figure}[t]
\centering
\includegraphics[width=0.97\columnwidth]{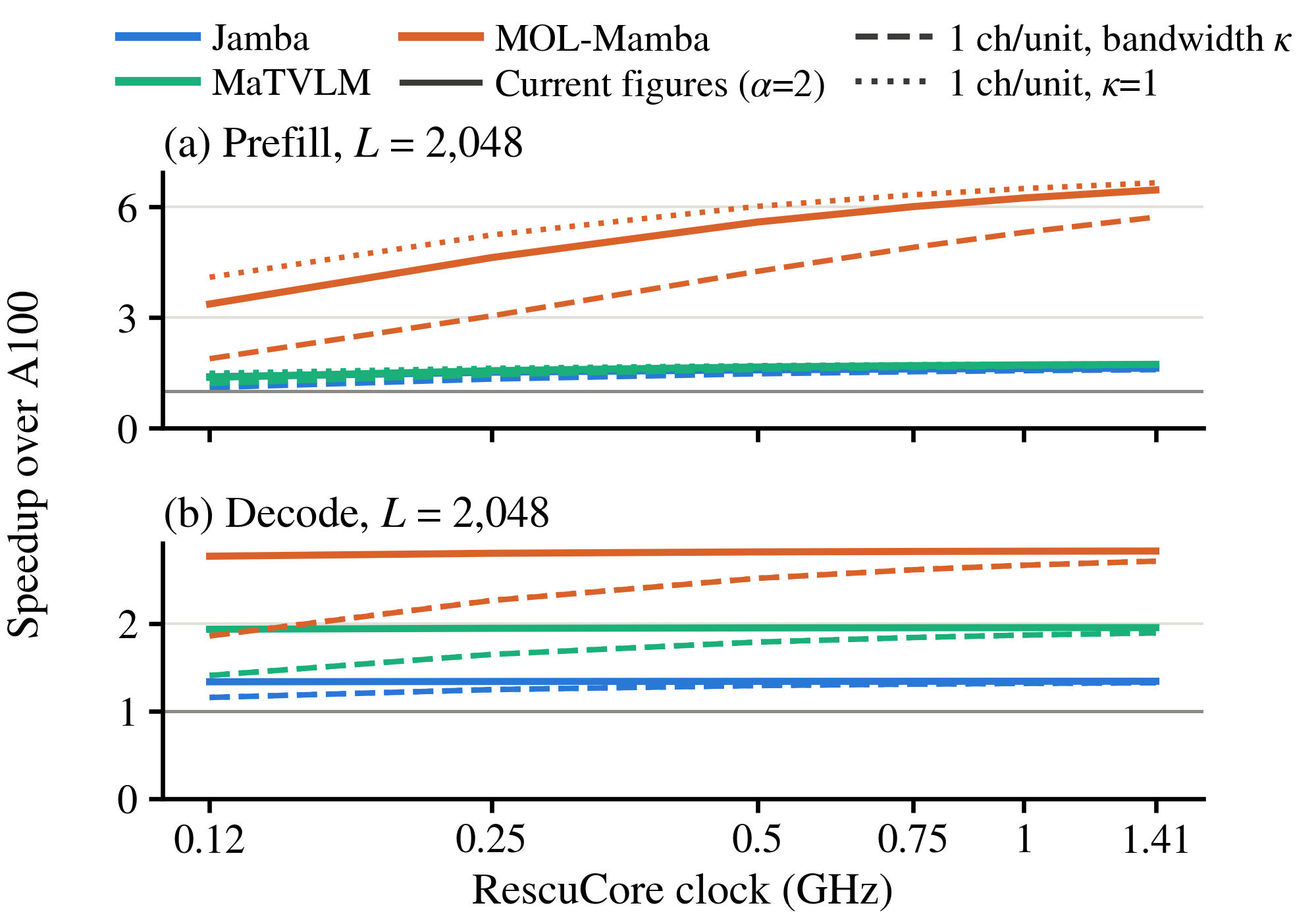}
\caption{
Projected Mamba-3 hybrid-model speedup over the A100 at $L=2{,}048$ as a function of the \acceleratorname clock (108 SMs, 432 \acceleratorname units, GPU at 1.41\,GHz). Solid: model used for the end-to-end figures. Dashed and dotted: Eq.~\eqref{eq:rescucore_cycles} with one channel per unit, using the operand-bandwidth-limited $\kappa$ (dashed) or the RTL-measured $\kappa=1$ (dotted). (a)~Prefill. (b)~Autoregressive decoding.}
\label{fig:freq_sens}
\end{figure}

\subsection{Sensitivity to \acceleratorname Clock Frequency}
\label{sec:eval_freq}

Because the 1.41\,GHz \acceleratorname clock is a target rather than a timing-closed result, Figure~\ref{fig:freq_sens} sweeps the \acceleratorname clock while \acceleratorname runs in its own clock domain and the rest of the model is unchanged.
Speedups depend only weakly on the clock.
At 0.5\,GHz, prefill speedups are $1.58\times$, $1.66\times$, and $5.59\times$ for Jamba, MaTVLM, and MOL-Mamba, compared with $1.62\times$, $1.73\times$, and $6.46\times$ at 1.41\,GHz, and decoding speedups change by less than 1\%.
Even at the 120\,MHz achieved at 45\,nm, prefill improves by $1.39\times$, $1.38\times$, and $3.36\times$.
The break-even clock, below which \acceleratorname would be slower than the GPU, is under 50\,MHz in every case.
The result is also insensitive to the cycle model: replacing $\kappa=2$ with the RTL-measured $\kappa=1$ or with the per-model operand-bandwidth bound changes the 1.41\,GHz speedups by at most 5\%.
Most of the benefit therefore comes from replacing the roughly 24 dependent GPU kernels between the input and output projections with a single \acceleratorname invocation, rather than from \acceleratorname's own arithmetic throughput.

\begin{figure*}[t]
    \centering
    \includegraphics[width=0.98\textwidth]{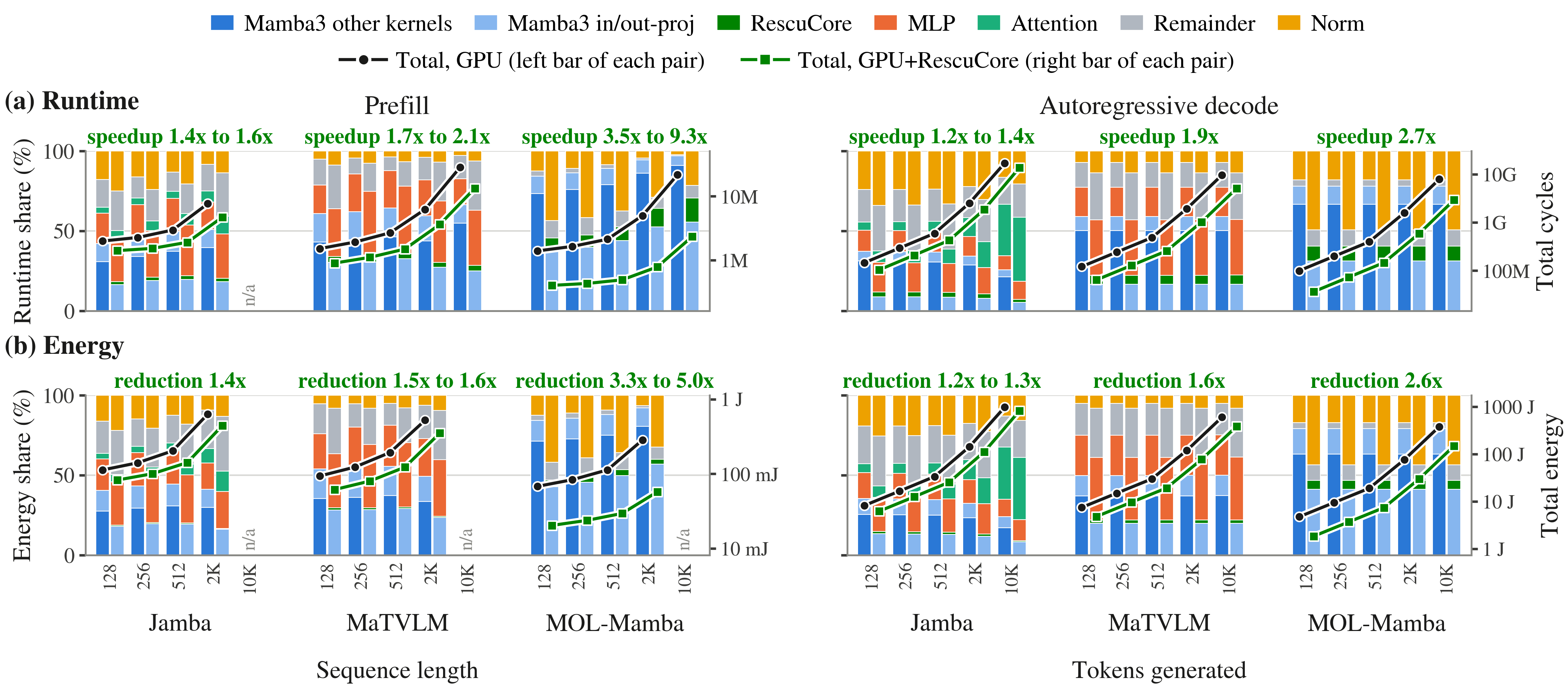}
    \caption{
    End-to-end (a) runtime and (b) energy breakdowns, along with absolute total execution cycles and energy, for prefill and autoregressive decoding across Jamba, MaTVLM, and MOL-Mamba. Stacked bars show the fraction of each configuration's total runtime or energy, normalized independently, while overlaid lines report absolute totals. All GPU+\acceleratorname results use the fixed BF16 $(N',R')=(32,4)$ implementation. Energy estimates combine AccelWattch GPU modeling with synthesized \acceleratorname power characteristics.}
    \label{fig:runtime_energy_rescucore}
\end{figure*}

\subsection{End-to-end Impact}
End-to-end impact across models and sequence lengths. 
Figure~\ref{fig:runtime_energy_rescucore} summarizes the end-to-end effect of \acceleratorname on all three hybrid models, for prefill of $L$ tokens and for autoregressive generation of $L_{gen}$ tokens, from 128 to 10K. 
Each pair of bars breaks down the plain GPU (left) and GPU + \acceleratorname (right) by kernel category, normalized to the design's own total, and the lines give total cycles (Figure~\ref{fig:runtime_energy_rescucore}a) and total energy (Figure~\ref{fig:runtime_energy_rescucore}b) on a log scale. 
Offloading the recurrence speeds up prefill by \(1.4-1.6\times\) for Jamba, \(1.7-2.1\times\) for MaTVLM and \(3.6-9.4\times\) for MOL-Mamba, and autoregressive decode by \(1.3-1.4\times\), \(2.0\times\) and \(2.8\times\) respectively. 
Energy falls by similar factors: \(1.4\times\), \(1.5-1.6\times\) and \(3.4-5.0\times\) in prefill, and \(1.2-1.3\times\), \(1.6\times\) and \(2.6\times\) in decode. 
Energy savings are slightly smaller than speedups, because the kernels that remain on the GPU keep their dynamic energy, and \acceleratorname adds its own. 
The ordering of the models follows directly from how much of their runtime the Mamba-3 recurrence occupies on the GPU. 
In MOL-Mamba, a Mamba-3 dominated hybrid model, the kernels between the projections take $73\%$ to $91\%$ of prefill time, and this share grows with sequence length, so its prefill speedup grows from \(3.6\times\) at 128 tokens to \(9.4\times\) at 10K. 
In Jamba, MLP, normalization and attention take a large fixed share, which caps the gain. 
Across every configuration, \acceleratorname's own slice is only a few percent of the remaining runtime and energy, so after offloading the recurrence is no longer the bottleneck. 
What remains is ordinary GPU work: the Mamba-3 input and output projections, the MLPs, and a large share of short normalization and element-wise kernels that are dominated by launch overhead. 
In Jamba decode, attention also grows with context length, reaching about a third of the GPU + \acceleratorname runtime at 10K tokens. 
\acceleratorname therefore removes the Mamba-3 recurrence as a bottleneck in both prefill and decode. 
Further end-to-end gains would have to come from the components that remain on the GPU, such as fusing the normalization and projection kernels around the \acceleratorname call, rather than from making the recurrence itself faster. 
All results use Accel-Sim and AccelWattch at batch size \(B=1\), ($N'=32$, $R'=4$) with the measured 1,800-cycle A100 launch cost; decode totals integrate the per-step cost over the generated tokens, and cells marked n/a were not simulated.

%% file: related_work.tex
\section{Related Work}\label{RELATED_WORK}

Prior work on SSM acceleration and GPU specialization can be broadly categorized into dedicated SSM accelerators, hybrid-model accelerators, and GPU-integrated execution units.

\subsection{Dedicated SSM and Mamba Accelerators}\label{RW_SSM_ACCEL}

PIMBA~\cite{kim2025pimba} employs processing-in-memory and quantization to reduce recurrent-state memory traffic, while MARCA~\cite{li2024marca} uses a reconfigurable array to accelerate Mamba's heterogeneous operations.
EpochCore~\cite{raja2025epochcore} exploits structured SSM computations through systolic execution, and FastMamba~\cite{FastMamba} combines FPGA acceleration with fine-grained quantization for Mamba-2.
More recently, RAIL~\cite{zhu2026rail} introduces a reconfigurable spatial accelerator for linear-time sequence models, combining inter-layer fusion, composable systolic arrays, and multi-tile scheduling to improve operand reuse and pipeline utilization.
Unlike these standalone accelerators, \acceleratorname{} integrates a dedicated recurrent datapath into GPU SM subpartitions and directly supports Mamba-3's exponential-trapezoidal recurrence, rotary transformations, and MIMO computation~\cite{lahoti2026mamba3}.

\subsection{Hybrid Transformer--Mamba Acceleration}\label{RW_HYBRID}

Hybrid models combine attention and SSM layers with different execution requirements~\cite{Lenz2025Jamba,Li2025MaTVLM}.
HLX~\cite{li2025hlx} proposes a unified pipelined accelerator for FlashAttention~\cite{dao2022flashattention} and Mamba-2's~\cite{dao2024mamba2} blockwise State Space Duality (SSD), exploiting operator fusion and intermediate-data reuse.
While \acceleratorname{} similarly employs pipelining and local forwarding, it targets Mamba-3's recurrent dependencies rather than matrix-oriented SSD execution.
Instead of implementing attention and SSM computation in a standalone accelerator, \acceleratorname{} retains existing GPU resources for dense operations and specializes recurrent execution for both prefill and autoregressive decoding.

\subsection{GPU-Integrated Specialized Execution}\label{RW_GPU_SPECIALIZATION}

Virgo~\cite{kim2025virgo} integrates a cluster-level matrix engine to improve GEMM execution, while FHECore~\cite{fhecore} introduces specialized modular-arithmetic hardware within GPU SMs for fully homomorphic encryption.
Better Call SoL~\cite{nag2026bettercallsol} integrates a reconfigurable Sea-of-LUTs fabric into GPU SMs to support custom operations and fine-grained dataflows.
Recent hybrid models, such as NVIDIA's Nemotron 3 Super~\cite{nemotron3super}, combine Mamba-2, Transformer attention, and mixture-of-experts layers, further motivating efficient execution of heterogeneous workloads on GPUs.
In contrast to prior GPU-integrated accelerators, \acceleratorname{} employs fixed-function Mamba-3 pipelines at SM subpartition granularity, with dedicated local registers for recurrent state and intermediate operands.
This organization accelerates dependent recurrent operations while preserving conventional GPU execution for dense projections, attention, and other model components, making it well suited to emerging hybrid SSM architectures.

%% file: conclusion.tex
\section{Conclusion}\label{CONCLUSION}

We presented \acceleratorname, a GPU-integrated execution unit designed to address the mismatch between Mamba-3's recurrent computation and conventional GPU execution. 
Our analysis identifies kernel fragmentation, intermediate data movement, and sequential dependencies as key bottlenecks that software fusion and conventional GPU optimizations cannot fully resolve. 
To address these limitations, \acceleratorname integrates specialized, pipelined computation into each SM subpartition, combining MIMO mapping, rotary transformations, recurrent updates, and output gating with dedicated local registers and operand forwarding to exploit parallelism across independent recurrent channels.

Our fixed BF16 implementation with native dimensions $(N',R')=(32,4)$ incurs an estimated $5.9\%$ A100 die-area overhead and supports larger configurations through additional execution passes. 
Across representative language, vision-language, and molecular hybrid models, simulation-based evaluation projects $1.4\times$--$9.4\times$ speedup and $1.4\times$--$5.0\times$ energy reduction during prefill, and $1.3\times$--$2.8\times$ speedup and $1.2\times$--$2.6\times$ energy reduction during autoregressive decoding. 
These results demonstrate the benefits of integrating workload-specific recurrent datapaths into general-purpose GPUs while preserving their flexibility for other model operations.

Future work will explore extending this approach to training, where fused recurrent-state computation and gradient propagation could further reduce data movement and synchronization overhead.